\documentclass[11pt]{article}
\usepackage{jheppub,graphicx,epsf,amssymb,amsbsy,amsfonts,amssymb,amsmath, empheq,mathrsfs,appendix}
\usepackage{hyperref}
\def\be{\begin{equation}}
\def\ee{\end{equation}}
\def\bea{\begin{eqnarray}}
\def\eea{\end{eqnarray}}

\def\s{\sigma}

\def\bg{\bar{g}}
\def\beq{\begin{eqnarray}}\def\eeq{\end{eqnarray}}
\def\ba#1\ea{\begin{align}#1\end{align}}
\def\bg#1\eg{\begin{gather}#1\end{gather}}
\def\bm#1\em{\begin{multline}#1\end{multline}}
\def\bmd#1\emd{\begin{multlined}#1\end{multlined}}

\def\D{\Delta}

\def\s{\sigma}

\def\({\left(}
\def\){\right)}
\def\[{\left[}
\def\]{\right]}

\def\s{\sigma}

\def\D{\Delta}

\title{MHV Gluon Scattering Amplitudes from Celestial Current Algebras }
\author[1]{Shamik Banerjee}
\author[2]{Sudip Ghosh}
\affiliation[1]{Institute of Physics, Sachivalaya Marg, Bhubaneshwar, India-751005 \\ and Homi Bhabha National Institute, Anushakti Nagar, Mumbai, India-400085}
\affiliation[2]{Okinawa Institute of Science and Technology,1919-1 Tancha, Onna-son, Okinawa 904-0495,Japan}
\emailAdd{banerjeeshamik.phy@gmail.com, sudip112phys@gmail.com}

\abstract{We show that the Mellin transform of an $n$-point tree level MHV gluon scattering amplitude, also known as the celestial amplitude in pure Yang-Mills theory, satisfies a system of $(n-2)$ linear first order partial differential equations corresponding to $(n-2)$ positive helicity gluons. Although these equations closely resemble Kniznik-Zamolodchikov equations for $SU(N)$ current algebra there is also an additional ``correction" term coming from the subleading soft gluon current algebra. These equations can be used to compute the leading term in the gluon-gluon OPE on the celestial sphere. Similar equations can also be written down for the momentum space tree level MHV scattering amplitudes. We also propose a way to deal with the non closure of subleading current algebra generators under commutation. This is then used to compute some subleading terms in the mixed helicity gluon OPE.}

\begin{document}
\maketitle
%\flushbottom

%%%%%%%%%%%%%%%%%%%%%%%%%%%%%%%%%%%%%%%%%%%%%%%%%%%%%%%%%%%%%%%%
\section{Introduction}

It is generally believed that any consistent theory of quantum gravity on a space-time with asymptotic boundary should have a holographic dual description in terms of a theory living on the boundary at infinity. The dual theory can compute all the observables which make sense in the bulk theory of quantum gravity. In the case of asymptotically flat space-time the observables are the $S$-matrix elements and it has been proposed that the dual theory is a conformal field theory \cite{Strominger:2017zoo,Kapec:2014opa,Kapec:2016jld,He:2017fsb,Ball:2019atb,Kapec:2017gsg,Banerjee:2019aoy,Banerjee:2019tam, Campiglia:2014yka,Donnay:2020guq, Barnich:2009se,Barnich:2011ct,Pasterski:2016qvg,Pasterski:2017kqt, Cheung:2016iub, deBoer:2003vf,Banerjee:2018gce,Banerjee:2018fgd, Banerjee:2019prz,Pasterski:2017ylz, Schreiber:2017jsr, Cardona:2017keg,Lam:2017ofc,Banerjee:2017jeg, Stieberger:2018edy, Stieberger:2018onx, Bagchi:2016bcd,Fotopoulos:2019tpe,Albayrak:2020saa,Casali:2020vuy,Law:2020xcf,Gonzalez:2020tpi,Muck:2020wtx,Narayanan:2020amh}, dubbed ``Celestial conformal field theory (CCFT)", which lives on the celestial sphere. The Lorentz group acts on the celestial sphere as the group of global conformal transformations and when the bulk space-time is four dimensional, one can show, starting from the subleading soft graviton theorem \cite{Cachazo:2014fwa} that CCFT should have a stress tensor \cite{Kapec:2014opa,Kapec:2016jld,He:2017fsb} which generates local conformal transformations on the two dimensional celestial sphere. So CCFT should have the full Virasoro symmetry just like a more conventional two dimensional CFT. But this is not the end of the story. Since the asymptotic symmetry group in four dimensions contains supertranslations \cite{Sachs:1962zza,Bondi:1962px}, CCFT should also have supertranslation symmetry \cite{Strominger:2013jfa,He,Strominger:2014pwa,Fotopoulos:2019vac,Banerjee:2020kaa,Banerjee:2020zlg}. On top of that, there are various other infinite dimensional current algebra symmetries, coming from soft factorisation theorems, under which CCFT should be invariant \cite{Strominger:2013lka,Pate:2019lpp,Banerjee:2020zlg,Campiglia:2014yka,Campiglia:2020qvc,Compere:2018ylh,Donnay:2020guq,He:2014cra,Kapec:2014zla,Kapec:2015ena,Campiglia:2015qka,Lysov:2014csa,Campiglia:2016hvg,Conde:2016csj,Himwich:2019dug,He:2020ifr,Nande:2017dba}. Now, the CCFT is supposed to compute bulk scattering amplitudes. So a natural question arises as to how the bulk scattering amplitudes in four dimensions are constrained by the infinite dimensional symmetries of the dual CCFT. For recent developments in this direction please see \cite{Pate:2019lpp,Law:2019glh,Banerjee:2020zlg,Himwich:2020rro,Fotopoulos:2019vac,Banerjee:2020kaa,Ebert:2020nqf}.

The study of CCFT is facilitated by going to the Mellin space \cite{Pasterski:2016qvg,Pasterski:2017kqt, Cheung:2016iub, deBoer:2003vf,Banerjee:2018gce,Banerjee:2018fgd, Banerjee:2019prz,Pasterski:2017ylz, Schreiber:2017jsr, Cardona:2017keg,Lam:2017ofc,Banerjee:2017jeg, Stieberger:2018edy, Stieberger:2018onx,Fotopoulos:2019tpe,Albayrak:2020saa,Casali:2020vuy,Law:2020xcf,Gonzalez:2020tpi,Muck:2020wtx,Narayanan:2020amh}. In Mellin space scattering amplitudes can be written as the correlation functions of conformal primaries \footnote{For a brief review of Mellin amplitudes for gluons please see the Appendix \ref{review}. In this appendix we also make some comments on the notation used in the paper.}. Conformal primaries are Mellin transform of Fock space creation (annihilation) operators which create (annihilate) asymptotic free particle states in a scattering event. They are called conformal primaries because under Lorentz transformations, which act on the celestial sphere as global conformal transformations, they transform like primary operators in a conformal field theory. Let us now briefly describe the main results of this paper. 

In this paper we focus on the Maximal Helicity Violating (MHV) gluon scattering amplitudes in pure Yang-Mills (YM) theory with gauge group $G= SU(N)$. These are amplitudes of the form $\langle{- - +++ \cdots}\rangle$, where two gluons have negative helicity and the rest have positive helicity. Their explicit expressions are given by the famous Parke-Taylor formula \cite{Parke:1986gb}. Now this is the first non-trivial helicity amplitude because at tree level in pure YM theory, amplitudes of the form $\langle{- +++ \cdots}\rangle$ with only one negative helicity gluon and $\langle{+++ \cdots}\rangle$ with all positive helicity gluons vanish. As a result of this at tree-level there is \textit{no negative helicity soft gluon} in the MHV-sector and also, the set of MHV amplitudes is \textit{closed under taking collinear limits}. This allows us to define, just like in the case of gravity \cite{Banerjee:2020zlg}, an \textit{autonomous} MHV-sector of the CCFT which computes the MHV gluon scattering amplitudes. The gluon MHV-sector is characterised by the fact that it is governed by the $G$ current algebra \cite{Strominger:2013lka,He:2014cra,Kapec:2014zla,Kapec:2015ena,Campiglia:2015qka,He:2020ifr,Nande:2017dba} at level zero which arises from the leading positive helicity soft gluon theorem. There is also the subleading positive helicity soft gluon theorem \cite{Pate:2019lpp,Lysov:2014csa,Campiglia:2016hvg,Conde:2016csj,Himwich:2019dug} which gives rise to another infinite set of currents which play an equally important role in the theory. The significance of the autonomous MHV sectors for gluons and gravitons \cite{Banerjee:2020zlg} is that they behave somewhat like Minimal models of two dimensional conformal field theories and hence are exactly solvable. 

Let us now consider an $n$-point MHV amplitude with two negative helicity gluons and $(n-2)$ positive helicity gluons. The main result of this paper is a system of $(n-2)$ coupled first order linear partial differential equations satisfied by the tree-level MHV amplitudes. In Mellin space they are given by, 
\begin{equation}
\label{ymdiffeq1}
\begin{split}
& \bigg[  \frac{C_{A}}{2}  \hspace{0.05cm} \frac{\partial}{\partial z_{i}}  -  h_{i} \hspace{0.05cm} \sum_{\substack{j=1 \\ j \ne i}}^{n}  \frac{T^{a}_{i} \hspace{0.05cm}T^{a}_{j}}{z_i - z_j}   \\  &+ \frac{1}{2}\sum_{\substack {j=1 \\ j\ne i }}^n    \frac{ \epsilon_{j}  \left(2\bar h_j -1 - (\bar z_i - \bar z_j) \frac{\partial}{\partial \bar z_j}\right)}{z_i - z_j} \hspace{0.05cm}T^a_j P^{-1}_j T^{a}_{i}    P_{-1,-1}(i) \bigg]   \bigg \langle  \prod_{k=1}^{n}\mathcal{O}^{a_{k}}_{h_{k},\bar{h}_{k}}(z_{k},\bar{z}_{k})   \bigg\rangle_{\text{MHV}}  =0
\end{split}
\end{equation} 

where $i \in \(1,2,\cdots, n-2\)$ runs over the $(n-2)$ \textit{positive helicity gluons} in the MHV amplitude. In the above equation, $\mathcal{O}^a_{h,\bar h}(z,\bar z)$ is a gluon conformal primary with scaling dimension $(h,\bar h)$ and $a$ is a Lie algebra index. $T^a$ is the Lie algebra generator in the adjoint representation and $C_A$ is the quadratic Casimir. We have also introduced the symbol $\epsilon_j$ which is $\pm 1$ depending on whether $\mathcal{O}^{a_j}_{h_j,\bar h_j}(z_j,\bar z_j)$ is outgoing or incoming. The operator $P_{j}^{-1}$ acts on a gluon conformal primary $\mathcal{O}^{a_k}_{h_k,\bar h_k}(z_k,\bar z_k)$ as 
\begin{equation}
\label{weightshift}
P_j^{-1} \mathcal{O}^{a_k}_{h_k,\bar{h}_k}(z_k, \bar{z}_k) = \mathcal{O}^{a_k}_{h_k-\frac{1}{2},\bar{h}_k- \frac{1}{2}}(z_k, \bar{z}_k) \delta_{jk}
\end{equation} 
Similarly the global time translation generator $P_{-1,-1}(i)$ acts on the $i$-th positive helicity gluon conformal primary according to
\be
P_{-1,-1}(i) \mathcal{O}^{a_i}_{h_i,\bar{h}_i}(z_i, \bar{z}_i) = \epsilon_i \mathcal{O}^{a_i}_{h_i+\frac{1}{2},\bar{h}_{i}+\frac{1}{2}}(z_i, \bar{z}_i)
\ee

Equation \eqref{ymdiffeq1} can be easily transformed to momentum space and for the momentum space MHV scattering amplitude it can be written as,
\begin{equation}
\label{defock1}
\begin{split}
& \bigg[  C_{A} \hspace{0.05cm}\frac{\partial}{\partial z_i}  + \left( \omega_{i} \frac{\partial}{\partial \omega_i} - 1\right)  \sum_{\substack{j=1 \\ j \ne i}}^{n}  \frac{T^{a}_{i} \hspace{0.05cm}T^{a}_{j}}{z_i - z_j}  \\ & +  \sum_{\substack{j=1 \\ j\ne i}}^n \frac{\epsilon_{i} \omega_{i}}{\epsilon_{j} \omega_{j}} \frac{ \left( \sigma_{j} + \omega_{j} \frac{\partial}{\partial\omega_j} + (\bar z_i - \bar z_j) \frac{\partial}{\partial\bar z_j } \right)}{z_i - z_j} T^a_i T^{a}_{j}     \bigg]  \bigg\langle  \prod_{k=1}^{n} A^{a_{k}}(\epsilon_{k} \omega_{k},z_{k},\bar{z}_{k}, \sigma_{k})  \bigg\rangle_{ \text{MHV}}  =0
\end{split}
\end{equation}

Here the null momentum of an on-shell gluon has been parametrised as
\be\nonumber
p = \epsilon\hspace{0.04cm}\omega(1+ z\bar z, z+\bar z, -i(z-\bar z), 1- z\bar z)
\ee

and $\epsilon = \pm 1$ for an outgoing (incoming) gluon.  We have also denoted the momentum space creation (annihilation) operators by $A^a(\epsilon\omega, z,\bar z, \sigma)$ where $\sigma$ is the helicity of the gluon. Note that the MHV amplitude in \eqref{defock1}  includes the overall momentum conserving delta function. 

Now \eqref{ymdiffeq1} or \eqref{defock1} are examples of (holographic) constraints on (hard) scattering amplitudes coming from the infinite dimensional symmetries of CCFT. They are obtained in the same way as differential equations for MHV graviton scattering amplitudes were obtained in \cite{Banerjee:2020zlg}. Presumably \eqref{ymdiffeq1} or \eqref{defock1}, together with the Ward identities coming from Poincare invariance, can be solved to obtain the MHV gluon scattering amplitudes. Along this line we make a preliminary check in which we determine the leading gluon-gluon OPE coefficients from the differential equations and our results match with those of \cite{Pate:2019lpp,Fan:2019emx}.

The origin of equations \eqref{ymdiffeq1} and \eqref{defock1} will be discussed in great detail in section \eqref{deymamp} but let us mention a few things before we close this section. 
\begin{enumerate}

\item The first two terms of \eqref{ymdiffeq1} closely resemble the Knizhnik-Zamolodchikov (KZ) equation \cite{Knizhnik:1984nr} satisfied by the correlation functions of current algebra primaries in WZW model. The only difference is the prefactor $h_i$ in the second term in \eqref{ymdiffeq1} and we also have to set the level $k$ of the current algebra to $0$. 

\item  In WZW model the scaling dimensions of the primaries are determined in terms of the level of the current algebra and the representation of the zero mode algebra under which the primary transforms. But, this is not the case here. \eqref{ymdiffeq1} holds for any value of the scaling dimension $(h,\bar h)$ of the gluon primary. This is consistent with the fact that in CCFT the scaling dimension $\D=h+\bar h$ of a (hard) primary is a continuously varying (complex) number and therefore should not be constrained in any way.

\item The third term in \eqref{ymdiffeq1} is an additional contribution coming from the (local) subleading soft gluon symmetry. This has no analog in the usual KZ equation and is most likely related to the fact that there is no Sugawara stress tensor in CCFT. This is also very different from pure gravity where the corresponding differential equations for MHV amplitudes do not have any contribution from the subsubleading soft graviton theorem. 

\end{enumerate}

It will be very interesting to understand the origin of the differential equations \eqref{ymdiffeq1} or \eqref{defock1} from the point of view of Twistor string theory \cite{Witten:2003nn,Nair:1988bq,Cachazo:2004kj,Cachazo:2004zb,Cachazo:2004by,Roiban:2004yf,Cachazo:2013hca,Cachazo:2013iea,Adamo:2014yya,Geyer:2014lca,Adamo:2015fwa}. This will also shed more light on the true nature of the CCFT.  We leave these questions for future research. \\

An outline of this paper is as follows. We begin in Section \eqref{poincareinv} by discussing the action of Poincare generators on gluon primary operators on the celestial sphere. The definition of a conformal primary operator is given explicitly in Section \eqref{poincareprim}. In Section \eqref{Kacmoody} we consider the leading conformal soft gluon theorem which is equivalent to the Ward identity for a level zero Kac-Moody algebra on the celestial sphere. We specify here the commutators involving modes of the Kac-Moody current and the Poincare generators. Using the current algebra Ward identity we also relate here the celestial correlators involving Kac-Moody descendants to correlation functions of gluon primary operators. Section \eqref{calgebprim} contains the definition of a primary operator under the current algebra.  In Section \eqref{subsoftth} we consider a set of currents $(J^{a},K^{a})$ on the celestial sphere which arise from the subleading conformal soft gluon theorem. In Section \eqref{subleadope} we derive the OPE between a subleading soft gluon and a hard gluon primary which yields an important constraint on the OPE of hard gluons in the subleading conformal soft limit. From this OPE we also extract the definition of descendants created by modes of $(J^{a},K^{a})$ and use the Ward identity, corresponding to the subleading soft gluon theorem, to obtain correlation functions with insertions of these descendants. In Section \eqref{subleadprim} the definition of a primary under the subleading soft symmetry algebra is provided. Section \eqref{comms} lists various useful commutation relations.  In Section \eqref{JKcomms}, we discuss the interpretation of the commutation relations between modes of the subleading soft symmetry generators in the light of the fact that these generators do not close to form a Lie algebra in the conventional sense. In Sections \eqref{deymamp} and \eqref{deymamp1} we derive the differential equations \eqref{ymdiffeq1} and \eqref{defock1} for tree-level MHV gluon amplitudes in Mellin space and Fock space respectively. In Section \eqref{leadopes} we use the differential equation \eqref{ymdiffeq1} to determine the structure of the leading OPE for gluon primaries in Yang-Mills theory. In particular, Section \eqref{opeoutout} deals with the case where both gluons in the OPE are either incoming or outgoing and in Section \eqref{opeinout} we consider the case where one of the gluons in the OPE is outgoing and the other is incoming. In Section \eqref{subleadopes}, we illustrate how some descendant OPE coefficients in the OPE between gluons of opposite helicities can be determined using the underlying infinite dimensional symmetry algebras. We end the paper with a set of Appendices. Appendix \ref{review} contains a brief review of celestial amplitudes and comments on some of the notation used in this paper. In Appendix \ref{ope5pt} we present a detailed calculation of the first subleading correction to the leading celestial OPE of positive helicity gluons using the Mellin transform of the $5$-point tree level MHV gluon amplitude in Yang-Mills theory. In Appendix \ref{pmope} we use the $4$-point MHV Mellin amplitude to extract the first set of subleading terms in the OPE between opposite helicity gluon primaries. 

%%%%%%%%%%%%%%%%%%%%%%%%%%%%%%%%%%%%%%%%%%%%%%%%%%%%%%%%%%%%%%%%

\section{Poincare invariance}
\label{poincareinv}
Since the scattering amplitudes are Poincare invariant, generators of the Poincare group act on the conformal primaries which live on the celestial sphere\footnote{For a brief review of conformal primaries and Mellin amplitudes for gluons and some comments on notations used in this paper, please see the Appendix \ref{review}.}. The Lorentz group $SL(2,\mathbb C)$ acts on the celestial sphere as the global conformal group and we denote its generators by $\(L_0,L_{\pm 1}, \bar L_0, \bar L_{\pm 1}\)$. Their commutation relations are given by,  
\be
\[L_m,L_n\] = (m-n)L_{m+n}, \quad \[ \bar L_m, \bar L_n\] = (m-n)\bar L_{m+n}, \quad \[ L_m, \bar L_n\] =0, \quad m,n= 0, \pm 1
\ee

They act on a gluon conformal primary $\mathcal O^a_{h,\bar h}(z,\bar z)$ as,
\be
\hspace{-0.5cm}\[ L_1, \mathcal{O}^a_{h,\bar h}(z,\bar z)\] = \( z^2 \partial + 2h z\) \mathcal{O}^a_{h,\bar h}(z,\bar z), \hspace{0.2cm} \[ L_0, \mathcal{O}^a_{h,\bar h}(z,\bar z)\] = h \mathcal{O}^a_{h,\bar h}(z,\bar z), \hspace{0.2cm} \[ L_{-1}, \mathcal{O}^a_{h,\bar h}(z,\bar z)\] = \partial \mathcal{O}^a_{h,\bar h}(z,\bar z)
\ee

\be
\hspace{-0.5cm}\[ \bar L_1, \mathcal{O}^a_{h,\bar h}(z,\bar z)\] = \( \bar z^2 \bar\partial + 2\bar h \bar z\) \mathcal{O}^a_{h,\bar h}(z,\bar z), \hspace{0.2cm} \[ \bar L_0, \mathcal{O}^a_{h,\bar h}(z,\bar z)\] = \bar h \mathcal{O}^a_{h,\bar h}(z,\bar z), \hspace{0.2cm} \[ \bar L_{-1}, \mathcal{O}^a_{h,\bar h}(z,\bar z)\] = \bar\partial \mathcal{O}^a_{h,\bar h}(z,\bar z)
\ee

The four global space-time translation generators will be denoted by $P_{m,n}$ where $m,n = 0,\pm 1$ and they are mutually commuting
\be
\[ P_{m,n}, P_{m',n'}\] =0 
\ee

The commutation relations between Lorentz and global space-time translation generators are given by,
\be
[L_n, P_{m',n'}] = \bigg(\frac{n-1}{2} - m'\bigg) P_{m'+n,n'} \qquad   [\bar L_n, P_{m',n'}] = \bigg(\frac{n-1}{2} - n'\bigg) P_{m',n'+n}
\ee

The translation generators $P_{m,n}$ act on conformal primaries according to
\be\label{poinc}
\[P_{m,n}, \mathcal O^a_{h,\bar h}(z,\bar z)\] = \epsilon z^{m+1}\bar z^{n+1} \mathcal O^a_{h+\frac{1}{2},\bar h + \frac{1}{2}}(z,\bar z)
\ee
where $\epsilon = \pm 1$ for an outgoing (incoming) gluon. 

\subsection{Poincare primary}
\label{poincareprim}

It follows from the definition of a conformal primary operator that the following standard relations hold
\begin{equation}
\label{lorentz} 
 L_{1} \mathcal{O}^a_{h,\bar h}(0) = \bar{L}_{1} \mathcal{O}^a_{h,\bar h}(0) = 0, \quad L_{0} \mathcal{O}^a_{h,\bar h}(0) =h \mathcal{O}^a_{h,\bar h}(0), \quad \bar{L}_{0} \mathcal{O}^a_{h,\bar h}(0)  = \bar{h}\mathcal{O}^a_{h,\bar h}(0) 
\end{equation}

but a Poincare primary \cite{Banerjee:2020kaa} must also satisfy the additional conditions
\be
P_{0,-1} \mathcal{O}^a_{h,\bar h}(0)  = P_{-1,0} \mathcal{O}^a_{h,\bar h}(0)  = P_{0,0}\mathcal O^a_{h,\bar h}(0) =0
\ee

which follow from \eqref{poinc}. 

\section{Leading soft gluon}
\label{Kacmoody}
The leading conformally soft \cite{Donnay:2018neh,Pate:2019mfs,Fan:2019emx,Nandan:2019jas,Adamo:2019ipt} \footnote{Conformally soft graviton theorems have been studied in \cite{Donnay:2018neh,Puhm:2019zbl,Guevara:2019ypd}.} gluon operator $j^a(z)$ with positive helicity is defined as,
\be
\label{leadsoftcurrent}
j^a(z) = \lim_{\D\rightarrow 1} (\D-1) \mathcal{O}^{a}_{\D,+}(z,\bar z)
\ee
where $\mathcal{O}^{a}_{\D,+}(z,\bar z)$ is a positive helicity gluon primary with scaling dimension $\D$. The soft operator $j^a(z)$ is a Kac-Moody current \cite{Strominger:2013lka,He:2014cra,Kapec:2014zla,Kapec:2015ena,Campiglia:2015qka,He:2020ifr,Nande:2017dba} whose correlation function with a collection of gluon primaries is given by the leading soft gluon theorem and has the standard form in Mellin space,
\be\label{leadsoftwi}
\bigg\langle{j^a(z) \prod_{i=1}^n \mathcal{O}^{a_i}_{h_i,\bar h_i}(z_i,\bar z_i)}\bigg\rangle 
= - \sum_{k=1}^n \frac{T^a_k}{z-z_k}\bigg\langle{\prod_{i=1}^n \mathcal{O}^{a_i}_{h_i,\bar h_i}(z_i,\bar z_i)}\bigg\rangle
\ee

where
\be
T^a_k O^{a_i}_{h_i,\bar h_i}(z,\bar z) = i f^{aa_i b} \mathcal{O}^b_{h_i,\bar h_i}(z_i,\bar z_i) \delta_{ik}
\ee

as gluons transform in the adjoint representation. Now let us consider the modes $j^a_n$ of the above current. They satisfy the algebra 
\be
\label{kacmoody}
\[j^a_m, j^b_n\] = - if^{abc} j^c_{m+n}
\ee

and act on a gluon primary as
\be
\[ j^a_m, \mathcal O^b_{h,\bar h}(z,\bar z)\] = -i f^{abc} z^n \mathcal O^c_{h,\bar h}(z,\bar z)
\ee

We note that the level of the current algebra here is zero which will be further justified by the form of the gluon-gluon OPE \cite{Ebert:2020nqf} in the MHV sector. 

\vspace{0.1cm}
The commutation relations with the (Lorentz) global conformal generators are given by,
\be
\[ L_m, j^a_n\] = -n j^a_{m+n}, \quad \[ \bar L_m, j^a_n\] = 0, \qquad m = 0, \pm 1
\ee

Similarly, the commutators with the generators $\{ P_{m,n}, m,n = 0, \pm 1 \}$ of global space-time translations are given by,
\be
\[ P_{m,n}, j^a_p\] = 0
\ee

For our purposes an important role is played by the correlation functions of the descendants $j^a_{-p}\mathcal{O}^b_{h,\bar h}(z,\bar z), p\ge 1$ with a collection of gluon primaries. These are given by
\be
\bigg\langle{j^{a}_{-p}\mathcal{O}^b_{h,\bar h}(z, \bar z) \prod_{i=1}^n \mathcal{O}^{a_i}_{h_i,\bar h_i}(z_i,\bar z_i)}\bigg\rangle = \mathscr{J}^{a}_{-p}(z)\bigg\langle \mathcal{O}^b_{h,\bar h}(z,\bar z) \prod_{i=1}^n \mathcal{O}^{a_i}_{h_i,\bar h_i}(z_i,\bar z_i) \bigg\rangle
\ee

where the operator $\mathscr{J}^a_{-p}(z)$ is defined as,
\be
\label{jamp}
\begin{gathered}
\mathscr{J}^{a}_{-p}(z)\bigg\langle \mathcal{O}^b_{h,\bar h}(z,\bar z) \prod_{i=1}^n \mathcal{O}^{a_i}_{h_i,\bar h_i}(z_i,\bar z_i)\bigg\rangle 
=  \sum_{k=1}^n \frac{T^a_k}{(z_k - z)^p} \bigg\langle \mathcal{O}^b_{h,\bar h}(z,\bar z) \prod_{i=1}^n \mathcal{O}^{a_i}_{h_i,\bar h_i}(z_i,\bar z_i) \bigg\rangle, \quad p\ge 1
\end{gathered}
\ee

\subsection{Leading current algebra primary}
\label{calgebprim}
A current algebra primary $\mathcal{O}^a_{h,\bar h}(z,\bar z)$ is defined by the standard conditions
\be
j^a_n \mathcal{O}^b_{h,\bar h}(0) = 0, \quad \forall n \ge 1
\ee

and 
\be\label{lcpj0}
j^a_0 \mathcal{O}^b_{h,\bar h}(0) = - T^a \mathcal{O}^b_{h,\bar h}(0) = -i f^{abc}\mathcal{O}^c_{h,\bar h}(0)
\ee

\section{Subleading soft gluon}
\label{subsoftth}
The subleading conformally soft \cite{Donnay:2018neh,Pate:2019mfs,Fan:2019emx,Nandan:2019jas,Adamo:2019ipt} gluon operator $S^{+a}_1(z,\bar z)$ with positive helicity is defined as, 
\be\label{ssg}
S^{+a}_1(z,\bar z) = \lim_{\D\rightarrow 0} \D \mathcal{O}^{a}_{\D,+}(z,\bar z)
\ee

where $\mathcal{O}^{a}_{\D,+}(z,\bar z)$ is a positive helicity gluon primary with scaling dimension $\D$. The correlation function of $S^{+a}_1(z,\bar z)$ with a collection of primary gluon operators is given by the subleading soft gluon theorem in Mellin space \cite{Pate:2019lpp},
\be\label{subl}
\begin{gathered}
\bigg\langle{S^{+a}_{1}(z,\bar z) \prod_{i=1}^n \mathcal{O}^{a_i}_{h_i,\bar h_i}(z_i,\bar z_i)}\bigg\rangle \\
= - \sum_{k=1}^n \frac{\epsilon_k}{z-z_k}\( -2 \bar h_k +1 + (\bar z - \bar z_k) \bar\partial_k\)T^a_k P_k^{-1} \bigg\langle \prod_{i=1}^n \mathcal{O}^{a_i}_{h_i,\bar h_i}(z_i,\bar z_i) \bigg\rangle
\end{gathered}
\ee

where 
\be
T^a_k \mathcal{O}^{a_i}_{h_i,\bar h_i}(z_i,\bar z_i) = i f^{aa_i b} \ \mathcal{O}^{b}_{h_i,\bar h_i}(z_i,\bar z_i) \ \delta_{ki}
\ee
and 
\be
P_k^{-1} \mathcal{O}^{a_i}_{h_i,\bar h_i}(z_i,\bar z_i) = \mathcal{O}^{a_i}_{h_i - \frac{1}{2}, \bar h_i - \frac{1}{2}}(z_i,\bar z_i) \ \delta_{ki}
\ee

Here $\epsilon_k = \pm 1$ depending on whether the gluon primary $\mathcal{O}^{a_k}_{h_k,\bar h_k}(z_k,\bar z_k)$ is outgoing or incoming. For  simplicity of notation we keep the additional label $\epsilon$ implicit when we write the correlation functions of the gluons. 

Now following \cite{Pate:2019lpp,Banerjee:2020zlg} we expand the R.H.S of \eqref{subl} in powers of the coordinate $\bar z$ of the subleading conformally soft gluon operator $S^{+a}_1(z,\bar z)$ and define two currents $J^a(z)$ and $K^a(z)$ whose Ward identities are given by,
\vspace{0.1cm}
\be\label{ward1}
\begin{gathered}
\bigg\langle{J^{a}(z) \prod_{i=1}^n \mathcal{O}^{a_i}_{h_i,\bar h_i}(z_i,\bar z_i)}\bigg\rangle = - \sum_{k=1}^n \frac{\epsilon_k}{z-z_k}\( -2 \bar h_k +1 - \bar z_k \bar\partial_k\)T^a_k P_k^{-1}\bigg\langle\prod_{i=1}^n \mathcal{O}^{a_i}_{h_i,\bar h_i}(z_i,\bar z_i)\bigg\rangle
\end{gathered}
\ee

and 
\be\label{ward2}
\begin{gathered}
\bigg\langle K^{a}(z) \prod_{i=1}^n \mathcal{O}^{a_i}_{h_i,\bar h_i}(z_i,\bar z_i)\bigg\rangle = - \sum_{k=1}^n \frac{\epsilon_k}{z-z_k} \bar\partial_k \ T^a_k P_k^{-1}\bigg\langle \prod_{i=1}^n \mathcal{O}^{a_i}_{h_i,\bar h_i}(z_i,\bar z_i)\bigg\rangle
\end{gathered}
\ee

This is equivalent to expanding the soft operator $S^{+a}_1(z,\bar z)$ as, 
\be
S^{+a}_1(z,\bar z) = J^a(z) + \bar{z} \hspace{0.05cm} K^a(z)
\ee 

We can now define the modes of the currents, $J^a_n(z)$ and $K^a_n(z)$, in the standard way. Their actions on a gluon primary are given by the following commutation relations,
\be
\label{Jacomm}
\begin{split}
\[ J^a_n, \mathcal{O}^b_{h,\bar h}(z,\bar z) \] &= - \epsilon z^n \(-2 \bar h +1 - \bar z \bar\partial\)T^a P^{-1} \mathcal{O}^b_{h,\bar h}(z,\bar z) \\
& = - i \epsilon f^{abc} z^n \(-2 \bar h +1 - \bar z \bar\partial\) \mathcal{O}^c_{h-\frac{1}{2},\bar h-\frac{1}{2}}(z,\bar z)
\end{split}
\ee

and 
\be
\label{Kacomm}
\begin{gathered}
\[ K^a_n, \mathcal{O}^b_{h,\bar h}(z,\bar z) \] = - \epsilon z^n \bar\partial \  T^a P^{-1} \mathcal{O}^b_{h,\bar h}(z,\bar z) 
= - i \epsilon f^{abc} z^n \bar\partial \ \mathcal{O}^c_{h-\frac{1}{2},\bar h-\frac{1}{2}}(z,\bar z)
\end{gathered}
\ee

Although the generators $\(J^a_n, K^a_n\)$ do not form a Lie algebra under commutation, we will see in section \eqref{JKcomms} that when the commutators $\[J^a_m, J^b_n\]$, $\[J^a_m, K^b_n\]$ and $\[K^a_m,K^b_n\]$ \textit{act on a gluon conformal primary or its descendants}, the results are given by simple expressions which look almost like closure. This is crucial for our purpose in this paper. 

\subsection{OPE between $S^{+a}_1(z,\bar z)$ and a hard gluon conformal primary }
\label{subleadope}

Suppose we want to compute the OPE between $S^{+a}_1(z,\bar z)$ and the gluon primary $\mathcal{O}^{a_1}_{h_1,\bar h_1}(z_1,\bar z_1)$. For this we have to expand the R.H.S of \eqref{subl} in powers of $(z-z_1)$ and $(\bar z - \bar z_1)$. So as $(z,\bar z)\rightarrow (z_1,\bar z_1)$ we can write,
\be\label{opecorr}
\begin{gathered}
\bigg\langle S^{+a}_{1}(z,\bar z) \prod_{i=1}^n \mathcal{O}^{a_i}_{h_i,\bar h_i}(z_i,\bar z_i)\bigg\rangle \\ 
= \(- \frac{\epsilon_1}{z-z_1} \( -2 \bar h_1 + 1\) T^a_1 P^{-1}_1 + \sum_{p=1}^{\infty} (z-z_1)^{p-1} \mathcal J^a_{-p}(1)\) \bigg\langle{\prod_{i=1}^n \mathcal{O}^{a_i}_{h_i,\bar h_i}(z_i,\bar z_i)}\bigg\rangle \\
+ (\bar z - \bar z_1) \( - \frac{\epsilon_1}{z-z_1}\bar\partial_1 T^a_1 P_1^{-1} + \sum_{p=1}^{\infty} (z-z_1)^{p-1} \mathcal K^a_{-p}(1)\) \bigg\langle{\prod_{i=1}^n \mathcal{O}^{a_i}_{h_i,\bar h_i}(z_i,\bar z_i)}\bigg\rangle  
\end{gathered}
\ee 

where the differential operators $\mathcal J^a_{-p}(1)$ and $\mathcal K^a_{-p}(1)$ are defined as, 
\be\label{1}
\begin{gathered}
\mathcal J^a_{-p}(1)\bigg\langle{\prod_{i=1}^n \mathcal{O}^{a_i}_{h_i,\bar h_i}(z_i,\bar z_i)}\bigg\rangle 
= - \( \sum_{j=2}^n \epsilon_j \frac{2\bar h_j -1 + (\bar z_j - \bar z_1)\bar\partial_j}{(z_j-z_1)^p} T^a_j P^{-1}_j \) \bigg\langle{\prod_{i=1}^n \mathcal{O}^{a_i}_{h_i,\bar h_i}(z_i,\bar z_i)}\bigg\rangle, \quad p\ge 1
\end{gathered}
\ee

and 
\be\label{2}
\begin{gathered}
\mathcal K^a_{-p}(1) \bigg\langle{\prod_{i=1}^n \mathcal{O}^{a_i}_{h_i,\bar h_i}(z_i,\bar z_i)}\bigg\rangle 
= \( \sum_{j=2}^n \epsilon_j \frac{\bar\partial_j}{(z_j-z_1)^p} T^a_j P^{-1}_j \) \bigg\langle{\prod_{i=1}^n \mathcal{O}^{a_i}_{h_i,\bar h_i}(z_i,\bar z_i)}\bigg\rangle, \quad p\ge 1 
\end{gathered}
\ee

From \eqref{opecorr} the OPE between $S^a_1(z,\bar z)$ and $\mathcal{O}^{a_1}_{h_1,\bar h_1}(z_1,\bar z_1)$ can be extracted to be, 
\be\label{ope1}
\begin{gathered}
S^{+a}_1(z,\bar z) \mathcal{O}^{a_1}_{h_1,\bar h_1}(z_1,\bar z_1) \\
= - \frac{\epsilon_1}{z-z_1} \( -2 \bar h_1 + 1\) T^a_1 P^{-1}_1 \mathcal{O}^{a_1}_{h_1,\bar h_1}(z_1,\bar z_1) + \sum_{p=1}^{\infty} (z-z_1)^{p-1} \(J^a_{-p} \mathcal{O}^{a_1}_{h_1,\bar h_1}\)(z_1,\bar z_1) \\
- (\bar z - \bar z_1) \(\epsilon_1 \frac{\bar\partial_1}{z-z_1} T^a_1 P_1^{-1} \mathcal{O}^{a_1}_{h_1,\bar h_1}(z_1,\bar z_1) + \sum_{p=1}^{\infty} (z-z_1)^{p-1} \(K^a_{-p}  \mathcal{O}^{a_1}_{h_1,\bar h_1}\)(z_1,\bar z_1)\)
\end{gathered} 
\ee

where the correlation functions with the insertion of the descendants $J^a_{-p} \mathcal{O}^{a_1}_{h_1,\bar h_1}(z_1,\bar z_1)$ and $K^a_{-p} \mathcal{O}^{a_1}_{h_1,\bar h_1}(z_1,\bar z_1)$ are given by \eqref{1} and \eqref{2}, respectively. 

Now \eqref{ope1} acts as a boundary condition on the OPE of two hard gluon primaries one which is positive helicity. Using the definition \eqref{ssg} of the subleading conformally soft gluon $S^{+a}_1(z,\bar z)$ we can write,
\be\label{con}
\boxed{
\lim_{\D\rightarrow 0} \D \mathcal{O}^{a}_{\D,+}(z,\bar z) \mathcal{O}^{a_1}_{h_1,\bar h_1}(z_1,\bar z_1) = \eqref{ope1}}
\ee

As we will see \eqref{con} is a nontrivial constraint on the OPE of two hard gluons.  Inside correlation functions \eqref{con} means that we first take the OPE limit $(z,\bar z)\rightarrow (z_1,\bar z_1)$ and then take the subleading conformal soft limit. In that limit we should always get back \eqref{opecorr}.

%\subsection{Subleading current algebra primary} 
%\label{subleadprim}

We also note that the operator product expansion \eqref{ope1} leads to the following conditions which are satisfied by any gluon primary $\mathcal{O}^{a}_{h,\bar h}(z,\bar z)$,
\be\label{subpri}
J^a_n \mathcal{O}^{a}_{h,\bar h}(0) =0, \quad J^a_0 \mathcal{O}^{b}_{h,\bar h}(0) = -i\epsilon f^{abc} (-2\bar h +1) \mathcal{O}^{c}_{h-\frac{1}{2},\bar h - \frac{1}{2}}(0), \quad \forall n>0 
\ee 

and 
\be\label{subpri2}
K^a_n \mathcal{O}^{a}_{h,\bar h}(0) =0, \quad K^a_0 \mathcal{O}^{b}_{h,\bar h}(0) = -i\epsilon f^{abc} \bar\partial \mathcal{O}^{c}_{h-\frac{1}{2},\bar h - \frac{1}{2}}(0), \quad \forall n>0 
\ee 

\subsection{Commutation relations involving subleading generators}
\label{comms}

In this section we collect some useful commutation relations. These are the "classical" commutators which can be easily obtained from the action of the generators on a primary operator. For the convenience of the reader we gather here the actions of the Poincare and current algebra generators on a gluon primary,
\be
\[ L_1, \mathcal{O}^a_{h,\bar h}(z,\bar z)\] = \( z^2 \partial + 2h z\) \mathcal{O}^a_{h,\bar h}(z,\bar z), \quad \[ L_{-1}, \mathcal{O}^a_{h,\bar h}(z,\bar z)\] = \partial \mathcal{O}^a_{h,\bar h}(z,\bar z)
\ee
\be
\[ \bar L_1, \mathcal{O}^a_{h,\bar h}(z,\bar z)\] = \( \bar z^2 \bar\partial + 2\bar h \bar z\) \mathcal{O}^a_{h,\bar h}(z,\bar z), \quad \[ \bar L_{-1}, \mathcal{O}^a_{h,\bar h}(z,\bar z)\] = \bar\partial \mathcal{O}^a_{h,\bar h}(z,\bar z)
\ee
\be
\[ P_{m,n}, \mathcal{O}^a_{h,\bar h}(z,\bar z)\] = \epsilon z^{m+1}\bar z^{n+1} \mathcal{O}^a_{h+\frac{1}{2}, \bar h+\frac{1}{2}}(z,\bar z)
\ee
\be
\[j^a_n , \mathcal{O}^b_{h,\bar h}(z,\bar z)\] = -i z^n f^{abc} \mathcal{O}^c_{h,\bar h}(z,\bar z)
\ee
\be\label{Jo}
\[ J^a_n, \mathcal{O}^b_{h,\bar h}(z,\bar z) \]
= -i \epsilon f^{abc} z^n \(-2 \bar h +1 - \bar z \bar\partial\) \mathcal{O}^c_{h-\frac{1}{2},\bar h-\frac{1}{2}}(z,\bar z)
\ee
\be
\[ K^a_n, \mathcal{O}^b_{h,\bar h}(z,\bar z) \]
= -i \epsilon f^{abc} z^n \bar\partial \mathcal{O}^c_{h-\frac{1}{2},\bar h-\frac{1}{2}}(z,\bar z)
\ee
where $\epsilon=\pm 1$ depending on whether the gluon is outgoing or incoming. Using these we arrive at the following commutation relations between generators,
\be
\[ L_1, J^a_n\] = - (n+1) J^a_{n+1}, \quad \[ L_0, J^a_n\] = \(n - \frac{1}{2}\)J^a_n, \quad \[ L_{-1}, J^a_n\] = - n J^a_{n-1}
\ee
\be
\[ \bar L_1, J^a_n\] = 0, \quad \[ \bar L_0 , J^a_n\] = - \frac{1}{2} J^a_n, \quad \[ \bar L_{-1}, J^a_n\] = - K^a_n
\ee
\be
\[ L_1, K^a_n\] = - (n+1) K^a_{n+1}, \quad \[ L_0, K^a_n\] = \(n - \frac{1}{2}\)K^a_n, \quad \[L_{-1}, K^a_n\] = - n K^a_{n-1}
\ee
\be
\[ \bar L_1, K^a_n\] = J^a_n , \quad \[ \bar L_0, K^a_n\] = \frac{1}{2} K^a_n, \quad \[ \bar L_{-1}, K^a_n\] = 0
\ee
\be
\[P_{m,-1}, J^a_n\] = j^a_{m+n+1}, \quad
\[ P_{m,0}, J^a_n\] = 0, \quad m = 0,-1
\ee
\be
\[P_{m,-1}, K^a_n\] = 0, \quad
\[ P_{m,0}, K^a_n\] = j^a_{m+n+1}, \quad m = 0,-1
\ee
\be
\[j^a_m, J^b_n\] = - if^{abc}J^c_{m+n}, \quad
\[j^a_m, K^b_n\] = - if^{abc}K^c_{m+n}
\ee

\section{How to interpret the commutators of subleading symmetry generators}
\label{JKcomms}

It is well known \cite{Lysov:2014csa} that the subleading symmetry generators $J^a_n$ and $K^a_n$ do not close under commutation. So they are not the generators of a Lie algebra symmetry in the ordinary sense. To see this we note that the scaling dimensions of $J^a_n$ and $K^a_n$ are given by $(-n - 1/2, -1/2)$ and $( -n - 1/2, 1/2)$, respectively. Therefore the antiholomorphic scaling dimension of the commutator amongst these generators must be an integer. But there is no generator with integer antiholomorphic scaling dimension that can appear here and so the generators $(J^a_m, K^a_m)$ cannot form a Lie algebra. 

Now, at least for the purposes of this paper, what we really need to know is how the commutator of two subleading generators acts on a gluon primary or its descendants. For example, in the OPE of two outgoing gluon primaries of opposite helicities given by, $\mathcal{O}^a_{\D_1, +}$ and $\mathcal{O}^b_{\D_1,-}$, one gets a subleading term of the form \eqref{glpmopeord1}, $J^a_{-1}\mathcal{O}^b_{\D_1+\D_2,-}$. In order to calculate the OPE coefficient multiplying this operator, one has to know the structure of the term  
\be\label{JJ}
J^c_{1}J^a_{-1}\mathcal{O}^b_{\D_1+\D_2,-} = \[ J^c_{1}, J^a_{-1}\] \mathcal{O}^b_{\D_1+\D_2,-}
\ee

In \eqref{JJ} we have used \eqref{subpri} which gives $J^c_{1}O^b_{\D_1 + \D_2, -}=0$. More generally, we will get terms like
\be
\(\[ J^a_m, J^b_n\], \[ J^a_m, K^b_n\], \[ K^a_m, K^b_n\]\) \prod_i J^{c_i}_{p_i} \prod_{j} K^{d_j}_{q_j} \prod_k j^{e_k}_{r_k} \mathcal{O}^f_{h,\bar h}(z,\bar z)
\ee

from the OPE and we need to get simplified expressions for them. In order to do this we start by computing the following commutators
\be
\[ \[J^a_m, J^b_n\], \mathcal{O}^c_{h,\bar h}(z,\bar z)\], \quad \[ \[J^a_m, K^b_n\], \mathcal{O}^c_{h,\bar h}(z,\bar z)\], \quad \[ \[K^a_m, K^b_n\], \mathcal{O}^c_{h,\bar h}(z,\bar z)\]
\ee

Let us focus on the first commutator. Using \eqref{Jo} and the Jacobi identity we get,
\be
\begin{gathered}
\[ \[J^a_m, J^b_n\], \mathcal{O}^c_{h,\bar h}(z,\bar z)\] \\ 
= - f^{abd} f^{dce} z^{m+n} \(-2 \bar h +1 - \bar z\bar\partial \) \(-2 \bar h +2 - \bar z\bar\partial \)\mathcal{O}^e_{h-1,\bar h-1}(z,\bar z) 
\end{gathered}
\ee 

Again using \eqref{Jo} we can write this as
\be\label{JJ1}
\[ \[J^a_m, J^b_n\], \mathcal{O}^c_{h,\bar h}(z,\bar z)\] = -i \epsilon f^{abd} \(-2 \bar h +1 - \bar z\bar\partial \) \[ J^d_{m+n}, \mathcal{O}^c_{h-\frac{1}{2},\bar h-\frac{1}{2}}(z,\bar z)\]
\ee

where we have used $\epsilon^2=1$. Now we take the limit $(z,\bar z)\rightarrow (0,0)$ and from \eqref{JJ1} we get
\be\label{subdes}
\[ \[J^a_m, J^b_n\], \mathcal{O}^c_{h,\bar h}(0)\] = -i \epsilon f^{abd} \(-2 \bar h +1 \) \[ J^d_{m+n}, \mathcal{O}^c_{h-\frac{1}{2},\bar h-\frac{1}{2}}(0)\]
\ee

Since the mode $J^a_n$ of the current $J^a$ is defined with respect to the point $(0,0)$, \eqref{subdes} can be interpreted as the relation between the descendants
\be\label{subint}
\boxed{
 \[J^a_m, J^b_n\]\mathcal{O}^c_{h,\bar h}(0) = -i \epsilon f^{abd} \(-2 \bar h +1 \) J^d_{m+n}\mathcal{O}^c_{h-\frac{1}{2},\bar h-\frac{1}{2}}(0)}
\ee

Now we can apply the same argument to get the other two relations
\be\label{subint2}
\boxed{
\[J^a_m, K^b_n\]\mathcal{O}^c_{h,\bar h}(0) = -i \epsilon f^{abd} \(-2 \bar h +1 \) K^d_{m+n}\mathcal{O}^c_{h-\frac{1}{2},\bar h-\frac{1}{2}}(0)}
\ee

and 
\be\label{subint3}
\boxed{
\[K^a_m, K^b_n\]\mathcal{O}^c_{h,\bar h}(0) = -i \epsilon f^{abd} K^d_{m+n}\bar\partial \mathcal{O}^c_{h-\frac{1}{2},\bar h-\frac{1}{2}}(0)}
\ee

These are the relations that will be used in the following sections to obtain recursion relations for subleading OPE coefficients and to define null states or primary descendants. 

%We would also like to point out that there is nothing special about the point $(0,0)$. 

%Identical relations \eqref{subint}, \eqref{subint2} and \eqref{subint3} hold at any arbitrary point $(z,\bar z)$ with the understanding that the modes $J^a_n$ and $K^a_n$ are now defined with respect to the point $(z,\bar z)$. 

We can see that the L.H.S of \eqref{subint}, \eqref{subint2} and \eqref{subint3} are linear in the subleading symmetry generators. But, they also depend on the scaling dimension of the gluon primary on which the commutators are acting and also the gluon primary appearing on the R.H.S has shifted dimension compared to the one appearing on the L.H.S. This is a signature of the fact that the generators do not form a Lie algebra.  This is perhaps the closest one can come towards forming a closed algebra out of $J^a_n$ and $K^a_n$ when they act on states in a Hilbert space and, as we will see, this is sufficient for finding (subleading) OPE coefficients as we will see in section \eqref{subleadopes}. 

Before we end this section we would like to mention the result when the commutator acts on a level-$1$ descendant. They are given by,
\be\label{example}
\[ J^a_m, J^b_n\] J^c_p \mathcal{O}^d_{h,\bar h}(0) = -i\epsilon f^{abe}(-2\bar h +1) J^e_{m+n}J^c_p \mathcal{O}^d_{h-\frac{1}{2}, \bar h - \frac{1}{2}}(0)
\ee

\be
\[ J^a_m, K^b_n\] J^c_p \mathcal{O}^d_{h,\bar h}(0) = -i\epsilon f^{abe}(-2\bar h +1) K^e_{m+n}J^c_p \mathcal{O}^d_{h-\frac{1}{2}, \bar h - \frac{1}{2}}(0)
\ee

\be
\[ K^a_m, K^b_n\] J^c_p \mathcal{O}^d_{h,\bar h}(0) = -i\epsilon f^{abe}(-2\bar h +1) K^e_{m+n}J^c_p \bar\partial \mathcal{O}^d_{h-\frac{1}{2}, \bar h - \frac{1}{2}}(0)
\ee

\be
\[ J^a_m, J^b_n\] K^c_p \mathcal{O}^d_{h,\bar h}(0) = -i\epsilon f^{abe}(-2\bar h +1) J^e_{m+n}K^c_p \mathcal{O}^d_{h-\frac{1}{2}, \bar h - \frac{1}{2}}(0)
\ee

\be
\[ J^a_m, K^b_n\] K^c_p \mathcal{O}^d_{h,\bar h}(0) = -i\epsilon f^{abe}(-2\bar h +1) K^e_{m+n}K^c_p \mathcal{O}^d_{h-\frac{1}{2}, \bar h - \frac{1}{2}}(0)
\ee

\be
\[ K^a_m, K^b_n\] K^c_p \mathcal{O}^d_{h,\bar h}(0) = -i\epsilon f^{abe}(-2\bar h +1) K^e_{m+n}K^c_p \bar\partial \mathcal{O}^d_{h-\frac{1}{2}, \bar h - \frac{1}{2}}(0)
\ee

\be
\[ J^a_m, J^b_n\] j^c_p \mathcal{O}^d_{h,\bar h}(0) = -i\epsilon f^{abe}(-2\bar h +1) J^e_{m+n}j^c_p \mathcal{O}^d_{h-\frac{1}{2}, \bar h - \frac{1}{2}}(0)
\ee

\be
\[ J^a_m, K^b_n\] j^c_p \mathcal{O}^d_{h,\bar h}(0) = -i\epsilon f^{abe}(-2\bar h +1) K^e_{m+n}j^c_p \mathcal{O}^d_{h-\frac{1}{2}, \bar h - \frac{1}{2}}(0)
\ee

\be
\[ K^a_m, K^b_n\] j^c_p \mathcal{O}^d_{h,\bar h}(0) = -i\epsilon f^{abe}(-2\bar h +1) K^e_{m+n}j^c_p \bar\partial \mathcal{O}^d_{h-\frac{1}{2}, \bar h - \frac{1}{2}}(0)
\ee

The above relations, say for example \eqref{example}, can be obtained by starting from the commutator
\be
\[ \[ J^a_m, J^b_n\], \[ J^c_p, \mathcal{O}^d_{h,\bar h}(z,\bar z)\]\]
\ee

The above relations have obvious generalisations to a general descendant and are given by,
\be\label{g1}
\[ J^a_m, J^b_n\] \prod_i J^{c_i}_{p_i} \prod_{j} K^{d_j}_{q_j} \prod_k j^{e_k}_{r_k} \mathcal{O}^f_{h,\bar h}(0) = -i\epsilon f^{abx}(-2\bar h +1) J^x_{m+n} \prod_i J^{c_i}_{p_i} \prod_{j} K^{d_j}_{q_j} \prod_k j^{e_k}_{r_k} \mathcal{O}^f_{h- \frac{1}{2},\bar h-\frac{1}{2}}(0)
\ee

\be\label{g2}
\[ J^a_m, K^b_n\] \prod_i J^{c_i}_{p_i} \prod_{j} K^{d_j}_{q_j} \prod_k j^{e_k}_{r_k} \mathcal{O}^f_{h,\bar h}(0) = -i\epsilon f^{abx}(-2\bar h +1) K^x_{m+n} \prod_i J^{c_i}_{p_i} \prod_{j} K^{d_j}_{q_j} \prod_k j^{e_k}_{r_k} \mathcal{O}^f_{h- \frac{1}{2},\bar h-\frac{1}{2}}(0)
\ee

\be\label{g3}
\[ K^a_m, K^b_n\] \prod_i J^{c_i}_{p_i} \prod_{j} K^{d_j}_{q_j} \prod_k j^{e_k}_{r_k} \mathcal{O}^f_{h,\bar h}(0) = -i\epsilon f^{abx}(-2\bar h +1) K^x_{m+n} \prod_i J^{c_i}_{p_i} \prod_{j} K^{d_j}_{q_j} \prod_k j^{e_k}_{r_k} \bar\partial \mathcal{O}^f_{h- \frac{1}{2},\bar h-\frac{1}{2}}(0)
\ee

We do not need these more general relations in this paper but, it will be interesting to check their consistency with explicit calculations performed using scattering amplitudes. For example, a good check of this will be to compute subleading OPE coefficients in the gluon-gluon OPE directly from the (MHV) scattering amplitude and compare them with the results from the recursion relations obtained using \eqref{g1}, \eqref{g2} and \eqref{g3}. Our derivation of these relations has been somewhat hand waving. We leave a more rigorous derivation to future work. 

%%%%%%%%%%%%%%%%%%%%%%%%%%%%%%%%%%%%%%%%%%%%%%%%%%%%%%%%%%%%%%

\section{Differential equation for MHV gluon amplitudes in Mellin space }
\label{deymamp}
In this section we will derive a differential equation for Mellin transformed tree level $n$-point MHV gluon amplitudes in Yang-Mills theory.

Consider the celestial OPE between two positive helicity outgoing gluon primaries in Yang-Mills theory. We will denote these operators  below as $\mathcal{O}^{ a}_{\Delta,+}(z,\bar{z})$ and $ \mathcal{O}^{a_{1}}_{\Delta_{1},+}(z_{1},\bar{z}_{1})$ where the subscript $(+)$ denotes that both operators have positive helicity.  In order to arrive at the proposed differential equation, we will be interested in the contribution from descendants which constitute the first subleading correction to the leading singular term in this OPE.  This was recently obtained in \cite{Ebert:2020nqf} and is given by
\begin{equation}
\label{glppope}
\begin{split}
  \mathcal{O}^{ a}_{\Delta,+}(z,\bar{z})   \mathcal{O}^{a_{1}}_{\Delta_{1},+}(z_{1},\bar{z}_{1}) & = -  i \hspace{0.05cm}B(\Delta-1,\Delta_{1}-1)\bigg[ \frac{f^{a a_{1}x}}{z-z_{1}} +  \frac{\Delta-1}{\Delta+\Delta_{1}-2}  \hspace{0.05cm} f^{a a_{1}x} L_{-1} \\
  & + i \left( \frac{ \Delta-1}{\Delta+\Delta_{1}-2}  \hspace{0.05cm} \delta^{a x}  \delta^{a_{1} y}+  \frac{ \Delta_{1}-1}{\Delta+\Delta_{1}-2}  \hspace{0.05cm} \delta^{a y} \delta^{a_{1} x} \right) j^{y}_{-1} \bigg] \mathcal{O}^{x}_{\Delta+\Delta_{1}-1,+}(z_{1},\bar{z}_{1}) + \cdots
\end{split}
\end{equation}

where $\mathcal{O}^{x}_{\Delta+\Delta_{1}-1,+}(z_{1},\bar{z}_{1})$ is a positive helicity (outgoing) gluon primary. The leading primary OPE coefficient is given by the Euler beta function, $B(\Delta-1,\Delta_{1}-1)$. The dots above denote contributions from descendants at further subleading orders in $(z-z_{1}), (\bar{z}-\bar{z}_{1})$.  In \cite{Ebert:2020nqf} the above OPE was extracted from the Mellin transform of the tree-level $4$-point MHV gluon amplitude. We refer the reader to Section \ref{ope5pt} of the Appendix in this paper for a derivation of \eqref{glppope} from the Mellin transform of the $5$-point MHV gluon amplitude. 

%\begin{equation}
%\label{ward2}
%\begin{gathered}
%\langle{J'^{a}(z) \prod_{i=1}^n O^{a_i}_{h_i,\bar h_i}(z_i,\bar z_i)}\rangle \\
%= \sum_{k=1}^n \frac{1}{z-z_k} \bar\partial_k \ T^a_k P_k^{-1}\langle{\prod_{i=1}^n O^{a_i}_{h_i,\bar h_i}(z_i,\bar z_i)}\rangle
%\end{gathered}
%\end{equation}

Now let us take the subleading conformal soft limit $\Delta\rightarrow 0$ in the above OPE. We then obtain
\begin{equation}
\label{opesubsoft1}
\begin{gathered}
 \lim_{\Delta \to 0} \Delta  \hspace{0.05cm} \mathcal{O}^{ a}_{\Delta,+}(z,\bar{z})  \mathcal{O}^{a_{1}}_{\Delta_{1},+}(z_{1},\bar{z}_{1}) \\
=   \bigg[ \frac{(\Delta_{1}-2)}{z-z_{1}} \hspace{0.05cm}  i f^{a a_{1}x} - i f^{a a_{1}x} L_{-1} + \left(   \delta^{a x}\delta^{a_{1}y} -  (\Delta_{1}-1) \delta^{a y}\delta^{a_{1}x} \right) j^{y}_{-1} \bigg] O^{x}_{\Delta_{1}-1,+}(z_{1},\bar{z}_{1}) +\cdots
\end{gathered}
\end{equation}

According to our discussion in Section \eqref{subleadope},  in the subleading conformal soft limit the OPE should obey the general constraint given by equation \eqref{con}. Therefore as $\Delta \rightarrow 0$ in \eqref{glppope} we should get
\begin{equation}
\label{opesubsoft2}
\begin{split}
& \lim_{\Delta \to 0} \Delta  \hspace{0.05cm} \mathcal{O}^{ a}_{\Delta,+ }(z,\bar{z}) \mathcal{O}^{a_{1}}_{\Delta_{1},+}(z_{1},\bar{z}_{1}) =  \bigg[   \frac{(\Delta_{1}-2)}{z-z_{1}} \hspace{0.05cm} T^{a }_{1} P_{1}^{-1} +   J^{a}_{-1} \bigg] \mathcal{O}^{a_{1}}_{\Delta_{1},+}(z_{1},\bar{z}_{1}) +\cdots
\end{split}
\end{equation}

where  
\begin{equation}
\label{adjrep}
\begin{split}
 T^{a} P_{1}^{-1}  \mathcal{O}^{a_{1}}_{\Delta_{1},+} (z_{1},\bar{z}_{1}) =  i f^{a a_{1}x}  \mathcal{O}^{ x}_{\Delta_{1}-1,+}(z_{1},\bar{z}_{1})
 \end{split}
\end{equation}

Then comparing \eqref{opesubsoft1} and \eqref{opesubsoft2}  we see that the leading singular terms in the OPE match. But the subleading terms appear to be different. Therefore in order for the OPE in \eqref{glppope} to be consistent with the subleading soft gluon theorem we must have the following relation
\begin{equation}
\label{nullst}
\begin{split}
& \left[  \delta^{a_{1}x} J^{a}_{-1} +  i f^{a a_{1}x } L_{-1} P_{1}^{-1} - \left(  \delta^{a x}\delta^{a_{1}y} -  (\Delta_{1}-1) \delta^{a y}\delta^{a_{1}x} \right) j^{y}_{-1} P_{1}^{-1} \right]  \mathcal{O}^{ x}_{\Delta_{1},+}(z_{1},\bar{z}_{1}) =0
\end{split}
\end{equation}

Multiplying the above by $i f^{aa_{1}b}$ and using the relation
\begin{equation}
\label{quadcas}
\begin{split}
& f^{aa_{1}b} f^{aa_{1}c} = C_{A} \hspace{0.05cm} \delta^{bc } 
\end{split}
\end{equation}

where $C_{A}$ is the quadratic Casimir of the adjoint representation, we can express \eqref{nullst} as 
\begin{equation}
\label{nullstKZform}
\begin{split}
& \bigg[  C_{A}  L_{-1} + \Delta_{1} \hspace{0.05cm} j^{a}_{-1}T^{a}_{1} + J^{a}_{-1} T^{a}_{1} P_{1} \bigg] \mathcal{O}^{ a_{1}}_{\Delta_{1}-1,+}(z_{1},\bar{z}_{1}) =0
\end{split}
\end{equation}

Further shifting $\Delta_{1} \rightarrow \Delta_{1}+1$ in \eqref{nullstKZform} we get 
\begin{equation}
\label{nullstKZform1}
\begin{split}
&  \bigg[  C_{A}  L_{-1} +  (\Delta_{1} +1)\hspace{0.05cm} j^{a}_{-1} T^{a}_{1}+ J^{a}_{-1}T^{a}_{1} P_{1} \bigg]  \mathcal{O}^{ a_{1}}_{\Delta_{1},+}(z_{1},\bar{z}_{1}) =0
\end{split}
\end{equation}

Up to this point we have been considering the gluon primary in \eqref{nullstKZform} to be outgoing. But this can be easily generalised to the case of an incoming positive helicity gluon. In that case we simply get an additional minus sign before the third term in \eqref{nullstKZform1}. Thus for an incoming positive helicity gluon we have
\begin{equation}
\label{nullstKZform2}
\begin{split}
&  \bigg[  C_{A}  L_{-1}  + (\Delta_{1} +1)\hspace{0.05cm} j^{a}_{-1} T^{a}_{1} - J^{a}_{-1}T^{a}_{1} P_{1}  \bigg]  \mathcal{O}^{ a_{1}}_{\Delta_{1},+}(z_{1},\bar{z}_{1}) =0
\end{split}
\end{equation}

Therefore in general we have the condition
\begin{equation}
\label{nullstKZform3}
\begin{split}
& \boxed{ \Psi^{a}(z,\bar{z}) =  \bigg[  C_{A}  L_{-1}  - (\Delta +1)\hspace{0.05cm} j^{b}_{-1} j^{b}_{0}  - J^{b}_{-1}  j^{b}_{0}P_{-1,-1} \bigg]  \mathcal{O}^{ a }_{\Delta ,+}(z,\bar{z}) =0 }
\end{split}
\end{equation}

where $P_{-1,-1}$ is the global time translation generator which acts on a gluon primary as $P_{-1,-1} \mathcal{O}^{ b }_{\Delta,+ } = \epsilon \hspace{0.04cm} \mathcal{O}^{ b }_{\Delta+1,+ }$. Here $\epsilon=\pm 1$ for an outgoing (incoming) gluon.  Note that in obtaining \eqref{nullstKZform3} we have used the definition \eqref{lcpj0} of a current algebra primary according to which
\begin{equation}
\begin{split}
 j^{a}_{0} \mathcal{O}^{ b }_{\Delta,+ }(z,\bar{z})  = - i f^{abc} \mathcal{O}^{ c }_{\Delta,+ }(z,\bar{z})  = - T^{a}  \mathcal{O}^{ b }_{\Delta,+ }(z,\bar{z}) 
\end{split}
\end{equation}

Now consider the linear combination of descendants denoted as $\Psi^{a}(z,\bar{z})$ in \eqref{nullstKZform3}. Applying  the definition of a Poincare and current algebra primary  from Sections \eqref{poincareprim} and \eqref{calgebprim}  and using the commutation relations given in Section \eqref{comms},  it can be easily checked that \footnote{In \eqref{nullst1} and \eqref{nullst2}, the modes of the symmetry generators have been defined with respect to the point $(z,\bar{z})$. } 
\begin{equation}
\label{nullst1} 
\begin{split}
 L_{1} \Psi^{a}(z,\bar{z}) = \bar{L}_{1} \Psi^{a}(z,\bar{z})= P_{0,-1} \Psi^{a}(z,\bar{z}) = P_{-1,0} \Psi^{a}(z,\bar{z}) =0
\end{split}
\end{equation}
\begin{equation}
\label{nullst2} 
\begin{split}
  j^{a}_{0} \Psi^{b}(z,\bar{z}) = - if^{abc}\Psi^{c}(z,\bar{z}) , \quad j^{a}_{m} \Psi^{b}(z,\bar{z}) = 0, \quad \forall m \ge 1
\end{split}
\end{equation}

%Using the results in Sections \eqref{subleadprim} and \eqref{JKcomms} we also get
%\begin{equation}
%\label{nullst3} 
%\begin{split}
%& J^{a}_{0} \Psi^{b}(z,\bar{z})= - i \epsilon f^{abc} (-2\bar{h}+1) \Psi^{c}(z,\bar{z}), \quad J^{a}_{m} \Psi^{b}(z,\bar{z})  =0,  \quad \forall m \ge 1 \\
%& K^{a}_{0} \Psi^{b}(z,\bar{z}) =  -  i \epsilon f^{abc} \bar{\partial}  \Psi^{c}(z,\bar{z}), \quad K^{a}_{m} \Psi^{b}(z,\bar{z})= 0, \quad \forall m \ge 1
%\end{split}
%\end{equation}

%where $2\bar{h} = (\Delta -1)$. 
Equations \eqref{nullst1} and \eqref{nullst2} together imply that $\Psi^{a}(z,\bar{z})$  is in fact a primary operator with respect to the Poincare group and the current algebra associated to the leading soft gluon theorem. \footnote{Since the subleading symmetry generators do not form a closed (Lie) algebra we do not impose ``primary-state'' condition under subleading soft symmetry. This will require further study and we hope to come back to this in future.} In fact, one can easily check that the null-state $\Psi^a$ is uniquely determined by the primary-state conditions under the Poincare group and leading soft gluon $SU(N)$ current algebra. Thus $\Psi^{a}(z,\bar{z})$ is a null field and we can consistently set $\Psi^{a}(z,\bar{z})$ to zero within celestial MHV gluon amplitudes. 

%without violating any of the underlying symmetries which characterise these amplitudes.  

Now let us insert \eqref{nullstKZform3} within Mellin transformed tree-level MHV gluon amplitudes. Below we will denote the gluon primaries as $\mathcal{O}^{a_{k}}_{h_{k},\bar{h}_{k}}(z_{k},\bar{z}_{k}) $. The negative helicity gluons in the MHV amplitude will be labelled by $(n-1)$  and $n$. Then for every positive helicity gluon $i \in (1,2,\cdots n-2)$ we get a decoupling relation
\begin{equation}
\label{decoup}
\begin{split}
& \bigg \langle \left[  C_{A}  L_{-1}(i) - 2 \hspace{0.05cm} h_{i} \hspace{0.05cm} j^{a}_{-1}(i) j^{a}_{0}(i)  -  J^{a}_{-1}(i) j^{a}_{0}(i) P_{-1,-1}(i) \right]  \prod_{k=1}^{n} \mathcal{O}^{a_{k}}_{h_{k},\bar{h}_{k}}(z_{k},\bar{z}_{k})   \bigg \rangle_{\text{MHV}}=0
\end{split}
\end{equation}

where we have used $2 \hspace{0.05cm} h_{i} = \Delta_{i}+1$ for a positive helicity gluon primary. The index $(i)$ accompanying $L_{-1}, j^{a}_{0},j^{a}_{-1},  J^{a}_{-1}$ and $P_{-1,-1}$ above denotes that these modes act on the $i$-th positive helicity gluon. Then using the representation of $L_{-1}, j^{a}_{-1}, J^{a}_{-1}$ in terms of differential operators, we obtain from \eqref{decoup}  with  $i \in (1, 2, \cdots, n-2) $
\begin{equation}
\label{ymdiffeq}
\begin{split}
&\bigg[  \frac{C_{A}}{2}  \hspace{0.05cm} \frac{\partial}{\partial z_{i}}  -  h_{i} \hspace{0.05cm} \sum_{\substack{j=1 \\ j \ne i}}^{n}  \frac{T^{a}_{i} \hspace{0.05cm}T^{a}_{j}}{z_{i}-z_{j}} \\ & +  \frac{1}{2} \sum_{\substack {j=1 \\ j\ne i }}^n    \frac{ \epsilon_{j}  \left(2\bar h_j -1 - (\bar z_{i}-\bar{z}_{j}) \frac{\partial}{\partial \bar z_j}\right)}{z_{i}-z_{j}} \hspace{0.05cm}T^a_j P^{-1}_j T^{a}_{i}    P_{-1,-1}(i) \bigg]   \bigg \langle  \prod_{k=1}^{n} \mathcal{O}^{a_{k}}_{h_{k},\bar{h}_{k}}(z_{k},\bar{z}_{k})   \bigg\rangle_{\text{MHV}}=0
\end{split}
\end{equation} 

We have thus obtained $(n-2)$ linear first order partial differential equations (PDEs) for tree-level $n$-point gluon MHV amplitudes in Mellin space. The $(n-2)$ PDEs correspond to the $(n-2)$ positive helicity gluons in the $n$-point MHV amplitude. 

Now let us note that the structure of \eqref{ymdiffeq} is similar to the Knizhnik-Zamolodchikov (KZ) equation \cite{Knizhnik:1984nr} obeyed by correlation functions of primary operators in WZW theory. The KZ equation is given by 
\begin{equation}
\label{KZeqn}
\begin{split}
& \bigg[  \left(k+ \frac{C_{A}}{2}\right)  \frac{\partial}{\partial z_{i}}  -  \sum_{\substack{j=1 \\ j \ne i}}^{n}  \frac{T^{a}_{i} \hspace{0.05cm}T^{a}_{j}}{z_{i}-z_{j}} \bigg]  \left \langle \phi^{a_{1}}_{h_{1},\bar{h}_{1}}(z_{1},\bar{z}_{1}) \cdots \phi^{a_{n}}_{h_{n},\bar{h}_{n}}(z_{n},\bar{z}_{n}) \right \rangle=0
\end{split}
\end{equation} 

where $\phi^{a_{i}}_{h_{i},\bar{h}_{i}}$ are the primary operators and  $k$ is the level of the current algebra. In the context of our paper, we will take these operators to transform in the adjoint representation of the zero mode algebra and so the superscript $a_{i}$ is a Lie algebra index.  Let us now compare the differential equation \eqref{ymdiffeq} and the KZ equation \eqref{KZeqn}.

First of all, for an $n$-point correlation function of primaries in WZW theory there are $n$ differential equations because every primary in WZW theory is degenerate. This should be contrasted with the case of MHV amplitudes where an $n$-point MHV amplitude satisfies $(n-2)$ differential equations corresponding to $(n-2)$ positive helicity gluons. This is related to the fact that \textit{within the MHV sector, governed by leading and subleading current algebras coming from positive helicity soft gluon, negative helicity gluons have no null states}. This is a major difference. 

Secondly, note that the coefficient of the $\partial_{z_{i}}$  term in \eqref{ymdiffeq} is $C_{A}/2$, where $C_{A}$ is the quadratic Casimir of the adjoint representation.  In the KZ equation \eqref{KZeqn},  this coefficient is given by $(k+ C_{A}/2)$.  At a superficial level this is consistent with the fact that in our case the $SU(N)$ current algebra has level $k=0$.

Now let us consider the second term within the square brackets in \eqref{ymdiffeq}. This arises from the  $ j^{a}_{-1}(i) j^{a}_{0}(i)$ piece in \eqref{decoup}. The analogous term is also present in the KZ equation \eqref{KZeqn} but, the coefficient of this term in our case depends on the holomorphic conformal weight $h_i$ of the primary operator $\mathcal O^{a_i}_{h_i,\bar h_i}(z_i,\bar z_i)$ whose null state gives rises to \eqref{ymdiffeq}. This is an important difference which plays a crucial role due to the following reason. 

In WZW theory, the KZ equation follows from the existence of a Sugawara stress tensor. From the expression of the Sugawara stress tensor it also follows that the holomorphic conformal weight of a current algebra primary is given by \cite{Knizhnik:1984nr,Ginsparg:1988ui}
\begin{equation}
\label{weightWZWprim}
\begin{split}
h_{r} = \frac{C_{r}}{2\hspace{0.04cm}k+C_{A}}
\end{split}
\end{equation} 

where $C_{r}$ is the quadratic Casimir of the representation $r$, under which the primary operator transforms. Here we are considering $r$ to be the adjoint representation and so $C_{r}=C_{A}$.  

The null state relation which gives rise to the usual KZ equation holds only when the primary operator, with respect to which the null state is defined, has (holomorphic) weight given by \eqref{weightWZWprim}. But \eqref{decoup} and consequently \eqref{ymdiffeq} hold for arbitrary values of the scaling dimensions for the positive helicity gluon primaries in the MHV amplitude\footnote{Here we are assuming that the dimensions $\Delta_{i}$ of primary operators can be analytically continued off the principal series where $\Delta_{i} =1+ i\lambda_{i}$ with $\lambda_{i} \in \mathbb{R}$. For tree-level amplitudes, the fact that such an analytic continuation is possible is evident from the explicit expressions of the corresponding Mellin amplitudes. See \cite{Donnay:2020guq} for a discussion on how conformal primaries with general dimensions can be expressed in terms of contour integrals over the principal series.}. The  coefficient $h_{i}$ in front of the second term in \eqref{decoup}, \eqref{ymdiffeq} plays a crucial role in ensuring this. 

Finally let us discuss the third term within the square brackets in \eqref{ymdiffeq}. This arises due to the subleading soft gluon symmetry. There is no counterpart of this term in the KZ equation \eqref{KZeqn}. Thus compared to the usual KZ equation,  this can be regarded as a correction term. Another consequence of this term is that unlike the KZ equation, the differential operators acting on the celestial MHV amplitude are not purely holomorphic.  

Recently in \cite{Fan:2020xjj} a Sugawara construction of the stress tensor was performed for celestial CFTs by studying Mellin transformed gluon amplitudes in Yang-Mills theory in the limit where a pair of gluons become conformally soft. However it was observed that within correlation functions, this stress tensor generates the correct conformal transformations only for the (leading) conformally soft gluons but fails to do so for the hard gluon primaries. This indicates that the Sugawara construction does not yield the full stress tensor in the celestial CFT putatively dual to Yang-Mills theory. The possibility that the full stress tensor may include contributions in addition to the Sugawara stress tensor was also pointed out in \cite{McLoughlin:2016uwa, Cheung:2016iub}. The  additional term coming from the subleading soft gluon symmetry in the differential equation \eqref{ymdiffeq} that we have obtained, further suggests that the standard form of the Sugawara construction involving only the leading current $j^a(z)$, may not apply to celestial CFTs and most likely the subleading currents $(J^a(z),K^a(z))$ play an important role in any such construction. 

In the following sections we will study the implications of this differential equation for the celestial OPE of gluon primary operators in Yang-Mills theory. In particular we will show that the leading celestial OPE of gluons can be determined using this equation. 

 %Now it will interesting to see if the KZ-like equation that we have obtained here 
 
 % If this were to happen for celestial CFTs, then the decoupling relation \eqref{decoup} and consequently \eqref{ymdiffeq} would not hold for arbitrary values of the conformal dimensions.  The presence of the coefficient $h_{i}$ avoids this issue.  \\

%%%%%%%%%%%%%%%%%%%%%%%%%%%%%%%%%%%%%%%%%%%%%%%%%%%%%%%%%%%%%%%%

\section{Differential equation for MHV gluon amplitudes in momentum space }
\label{deymamp1}

The differential equation \eqref{ymdiffeq} was derived for the Mellin transformed gluon amplitude. We can also write down an equivalent form of this equation for the amplitude in Fock space. Let us denote the tree-level Fock space MHV amplitude as
\begin{equation}
\label{famp}
\begin{split}
\bigg\langle  \prod_{k=1}^{n} A^{a_{k}} (\epsilon_{k} \omega_{k},z_{k},\bar{z}_{k}, \sigma_{k})   \bigg \rangle_{\text{MHV}}
\end{split}
\end{equation} 

where $\epsilon= \pm 1$. $ A^{a_{k}}( \omega_{k}, z_{k},\bar{z}_{k}, \sigma_{k}) $ and $A^{ a_{k}}( -\omega_{k}, z_{k},\bar{z}_{k}, \sigma_{k})$  are annihilation and creation operators respectively  for the external gluons with helicity $\sigma_{k}$ in the S-matrix. In \eqref{famp} we will take gluons $(1,2,\cdots, n-2)$ to have positive helicity. Then gluons $(n-1)$ and $n$ have negative helicity.  %This amplitude can be obtained by an inverse Mellin transform of the correlation function in \eqref{ymdiffeq}. 

Now  in order to recast \eqref{ymdiffeq} to momentum space, we make the following substitutions 
\begin{equation}
\begin{split}
& \Delta_{i} \rightarrow - \omega_{i} \partial_{\omega_{i}}, \quad P_{-1,-1}(i) \rightarrow \epsilon_{i} \omega_{i} , \quad P^{-1}_{i} \rightarrow \omega_{i}^{-1} 
\end{split}
\end{equation}

Applying an inverse Mellin transform we can replace the correlation function in \eqref{ymdiffeq} with the Fock space amplitude \eqref{famp}. Thus we obtain the differential equation
\begin{equation}
\label{defock}
\begin{split}
& \bigg[  C_{A} \hspace{0.05cm} \frac{\partial}{\partial z_{i}}  + \left( \omega_{i} \frac{\partial}{\partial \omega_{i}} - 1\right)  \sum_{\substack{j=1 \\ j \ne i}}^{n}  \frac{T^{a}_{i} \hspace{0.05cm}T^{a}_{j}}{z_{i}-z_{j}} \\ & +  \sum_{\substack{j=1 \\ j\ne i}}^n \frac{\epsilon_{i} \omega_{i}}{\epsilon_{j} \omega_{j}} \frac{ \left( \sigma_{j} + \omega_{j} \frac{\partial}{\partial\omega_j} + (\bar{z}_{i}-\bar{z}_{j}) \frac{\partial}{\partial\bar z_j} \right)}{z_{i}-z_{j}} T^a_i T^{a}_{j}     \bigg]  \bigg\langle  \prod_{k=1}^{n} A^{a_{k}} (\epsilon_{k} \omega_{k},z_{k},\bar{z}_{k}, \sigma_{k})   \bigg\rangle_{\text{MHV}}=0
\end{split}
\end{equation}

As in \eqref{ymdiffeq}, here we have $(n-2)$ partial differential equations labelled by the index $i$ in \eqref{defock} which runs over the $(n-2)$ positive helicity gluons in the MHV amplitude.  Also note that the amplitude appearing above is the full tree level scattering amplitude and so explicitly includes the delta function which imposes overall energy-momentum conservation. Therefore the partial derivative with respect to $\bar{z}_{j}$ in \eqref{defock} has a nontrivial action on the amplitude \eqref{famp}.

%%%%%%%%%%%%%%%%%%%%%%%%%%%%%%%%%%%%%%%%%%%%%%%%%%%%%%%%%%%%%%%%

\section{Leading OPE from differential equation }
\label{leadopes}

In this section we will show that the leading structure of the celestial OPE of gluons can be determined using the  differential equation \eqref{ymdiffeq}. 

\subsection{OPE for outgoing (incoming) gluons}
\label{opeoutout}

Let us first consider the celestial OPE between a positive helicity gluon primary $ \mathcal{O}^{a_{1}}_{\Delta_{1}, + } (z_{1},\bar{z}_{1}) $ and another gluon primary denoted by $\mathcal{O}^{a_{2} }_{\Delta_{2},\sigma_{2}}(z_{2},\bar{z}_{2}) $. The spin $\sigma_{2}$ of the second gluon primary will be left unspecified for now. We will also take both these gluons to be outgoing. Then let us assume that the leading OPE in this case takes the following general form 
\vspace{0.1cm}
\begin{equation}
\label{opegen}
\begin{split}
& \mathcal{O}^{a_{1}}_{\Delta_{1}, +} (z_{1},\bar{z}_{1}) \mathcal{O}^{a_{2} }_{\Delta_{2},\sigma_{2}}(z_{2},\bar{z}_{2}) = - i \hspace{0.05cm} z_{12}^{p} \bar{z}_{12}^{q} \hspace{0.05cm}   f^{a_{1}a_{2}x}  \hspace{0.05cm}  C_{p,q}  \left( \Delta_{1},\Delta_{2}, \sigma_{2}\right) \mathcal{O}^{x}_{\Delta,  \sigma} (z_{2},\bar{z}_{2}) + \cdots
\end{split}
\end{equation}

where $z_{12}=(z_{1}-z_{2}), \bar{z}_{12}=(\bar{z}_{1}-\bar{z}_{2})$. $ \mathcal{O}^{x}_{\Delta,  \sigma} (z_{2},\bar{z}_{2})$ is the leading primary operator that can appear in the OPE and $C_{p,q}  \left( \Delta_{1},\Delta_{2}, \sigma_{2}\right)$ is the associated OPE coefficient.  The dots denote possible contributions from descendants.  The conformal dimension and spin of $ \mathcal{O}^{x}_{\Delta,  \sigma} (z_{2},\bar{z}_{2})$ are given by
\begin{equation}
\label{dimspin}
\begin{split}
\Delta = \Delta_{1}+ \Delta_{2} + p +q, \quad \sigma = p-q+\sigma_{2}+1
\end{split}
\end{equation}

Our objective now is to determine the values of $p,q$ and the OPE coefficient $C_{p,q}  \left( \Delta_{1},\Delta_{2}, \sigma_{2}\right)$. In carrying out this analysis, we will assume that the structure of the OPE in \eqref{opegen} holds for arbitrary values of the dimensions $\Delta_{1}$ and $\Delta_{2}$ with $p, q$ and $\sigma_{2}$ fixed. This was also done in \cite{Banerjee:2020zlg} in the context of the celestial OPE of gravitons in Einstein gravity.  As in \cite{Banerjee:2020zlg}, our results below will further justify this assumption. We will see that the values of $p,q$ and $C_{p,q}  \left( \Delta_{1},\Delta_{2}, \sigma_{2}\right)$ obtained using the differential equation \eqref{ymdiffeq} precisely match with the corresponding results of \cite{Fan:2019emx}, where the leading celestial OPE was derived from the Mellin transform of the splitting function which appears in the leading collinear limit in gluon scattering amplitudes in Yang-Mills theory \cite{ Bern:1998sv}.   

\vspace{0.1cm}
Let us now write the differential equation \eqref{ymdiffeq} as 
 \begin{equation}
\label{ymde1}
\begin{split}
& \bigg[  \frac{C_{A}}{2} \frac{\partial}{\partial z_{1}}  -  h_{1} \sum_{j=2}^{n}  \frac{T^{a}_{j} \hspace{0.05cm}T^{a}_{1}}{z_{1}-z_{j}}  + \frac{1}{2} \sum_{j=2}^n \frac{\epsilon_{j}(2\bar h_j -1 - (\bar z_1 - \bar z_j) \frac{\partial}{\partial\bar z_j})}{(z_1-z_j)} \hspace{0.05cm} T^a_j  P^{-1}_j T^{a}_{1}    P_{-1,-1}(1) \bigg] \\ & \times \bigg \langle \mathcal{O}^{  a_{1}}_{\Delta_{1}, +}(z_{1},\bar{z}_{1})  \mathcal{O}^{a_{2} }_{\Delta_{2},\sigma_{2}}(z_{2},\bar{z}_{2})   \prod_{k=3}^{n} \mathcal{O}^{a_{k}}_{h_{k},\bar{h}_{k}}(z_{k},\bar{z}_{k})   \bigg \rangle_{\text{MHV}}=0
\end{split}
\end{equation}

where $\epsilon_{1}=\epsilon_{2}=1$. Using \eqref{ymde1} we can derive a recursion relation for the leading OPE coefficient  as follows. Let us substitute the leading OPE \eqref{opegen} inside the correlator in \eqref{ymde1}.  Then it is evident that at leading order in the OPE limit, the $z_{12},\bar{z}_{12}$ dependence on the L.H.S. of \eqref{ymde1} will be of the form $z_{12}^{p-1}\bar{z}_{12}^{q}$. Assuming that $\eqref{ymde1}$ is satisfied order by order in the OPE regime $(z_{12} \rightarrow 0, \bar{z}_{12}\rightarrow 0)$, we can then set the coefficient of the $z_{12}^{p-1}\bar{z}_{12}^{q}$ term to zero. Consequently we obtain the following recursion relation
\vspace{0.1cm}
\begin{equation}
\label{r0}
\begin{split}
& \left( p \hspace{0.05cm} C_{A} \hspace{0.05cm} f^{a_{1} a_{2}b } + 2 \hspace{0.05cm} h_{1} f^{a  a_{1} x }  f^{ x  b y }  f^{ y a_{2} a } \right)    C_{p,q} (\Delta_{1},\Delta_{2},\sigma_{2}) \\
& =    (2\bar{h}_{2}-1+q) f^{a  a_{1} x }  f^{ x  b y }  f^{ y a_{2} a } C_{p,q} (\Delta_{1}+1,\Delta_{2}-1,\sigma_{2})  
\end{split}
\end{equation}

where $2h_{1}= \Delta_{1}-1$ and  $2\bar{h}_{2} = \Delta_{2}-\sigma_{2}$. Then applying the identity
\begin{equation}
\label{fid}
\begin{split}
 f^{a  a_{1} x }  f^{ x  b y }  f^{ y a_{2} a } =  \frac{1}{2}  \hspace{0.05cm} C_{A} \hspace{0.05cm} f^{a_{1}  a_{2} b  }
\end{split}
\end{equation}

where $C_{A}$ is the quadratic Casimir of the adjoint representation,  we get from \eqref{r0} 
\begin{equation}
\label{r1}
\begin{split}
& \left( \Delta_{1}+ 2 p + 1 \right)    C_{p,q} (\Delta_{1},\Delta_{2},\sigma_{2})  = (\Delta_{2} - \sigma_{2} +q-1) C_{p,q} (\Delta_{1}+1,\Delta_{2}-1,\sigma_{2})  
\end{split}
\end{equation}

In the ensuing discussion, it will be convenient for us to express \eqref{r1} in another form. For this purpose, let us first note the following relation due to the invariance of the OPE under global time translations \cite{Pate:2019lpp}
\begin{equation}
\label{r6}
\begin{split}
& C_{p,q}(\Delta_{1},\Delta_{2},\sigma_{2})     =  C_{p,q}(\Delta_{1}+1,\Delta_{2},\sigma_{2}) + C_{p,q}(\Delta_{1},\Delta_{2}+1,\sigma_{2})
\end{split}
\end{equation}

Then shifting  $\Delta_{2}\rightarrow \Delta_{2}+1$ in \eqref{r1} and using \eqref{r6} we get 
\begin{equation}
\label{r7}
\begin{split}
& (\Delta_{1}+2p+1) C_{p,q}(\Delta_{1},\Delta_{2},\sigma_{2})     = \left(\Delta_{1} + \Delta_{2} + 2p +q- \sigma_{2} +1 \right)C_{p,q}(\Delta_{1}+1,\Delta_{2},\sigma_{2}) 
\end{split}
\end{equation}

We shall now derive another recursion relation for the leading OPE coefficient by appealing to the subleading soft gluon theorem in a similar fashion as in \cite{Pate:2019lpp}.  Consider the action of the subleading soft symmetry generator  $J^{a}_{0}$ on a gluon primary. This is given by \eqref{Jacomm} 
\begin{equation}
\label{subsoftglobal}
\begin{split}
& J^{a}_{0} \mathcal{O}^{b,\sigma}_{\Delta} (z,\bar{z}) = -  i  \epsilon f^{abc} (- 2 \bar{h}+ 1  - \bar{z}\partial_{\bar{z}} ) \mathcal{O}^{c,\sigma}_{\Delta-1} (z,\bar{z}) 
\end{split}
\end{equation}

where $\epsilon = \pm 1$ for an outgoing (incoming) gluon.  Then requiring both sides of the OPE \eqref{opegen} to transform in the same way under the action \eqref{subsoftglobal} we get\footnote{Note that in order to arrive at \eqref{r2} we have set $z_{2}=\bar{z}_{2}=0$ in the OPE \eqref{opegen}. }
\begin{equation}
\label{r2}
\begin{split}
&  f^{aa_{1}x} f^{x a_{2} y}(\Delta_{1}+q-2)  C_{p,q}(\Delta_{1}-1,\Delta_{2},\sigma_{2})  +  f^{a a_{2} x} f^{ a_{1} x  y}  (\Delta_{2} -\sigma_{2}-1) C_{p,q}(\Delta_{1},\Delta_{2}-1,\sigma_{2}) \\
& =   f^{a  x y  }f^{a_{1} a _{2} x} (\Delta_{1} +\Delta_{2} + 2q-\sigma_{2}-2 )  C_{p,q}(\Delta_{1},\Delta_{2},\sigma_{2})
\end{split}
\end{equation}

Multiplying both sides of \eqref{r2}  by $f^{yba}$ and using the identity \eqref{fid} we then obtain 
\begin{equation}
\label{r3}
\begin{split}
&  (\Delta_{1}+q-2)  C_{p,q}(\Delta_{1}-1,\Delta_{2},\sigma_{2})  +    (\Delta_{2} -\sigma_{2}-1) C_{p,q}(\Delta_{1},\Delta_{2}-1,\sigma_{2}) \\
& =  2  (\Delta_{1} +\Delta_{2} + 2q-\sigma_{2}-2 )  C_{p,q}(\Delta_{1},\Delta_{2},\sigma_{2})
\end{split}
\end{equation}

In order to easily determine the values of $p$ and $q$ let us bring \eqref{r3} into the same form as \eqref{r7}. To achieve this,  we shift $\Delta_{1}\rightarrow \Delta_{1}+1$ in \eqref{r3}. This gives,
\begin{equation}
\label{r4}
\begin{split}
&  (\Delta_{1}+q-1)  C_{p,q}(\Delta_{1},\Delta_{2},\sigma_{2})  +    (\Delta_{2} -\sigma_{2}-1) C_{p,q}(\Delta_{1}+1,\Delta_{2}-1,\sigma_{2}) \\
& =  2  (\Delta_{1} +\Delta_{2} + 2q-\sigma_{2}-1 )  C_{p,q}(\Delta_{1}+1,\Delta_{2},\sigma_{2})
\end{split}
\end{equation}

Then using \eqref{r1} we can eliminate $C_{p,q}(\Delta_{1}+1,\Delta_{2}-1, \sigma_{2}) $ from \eqref{r4} and obtain
\begin{equation}
\label{r5}
\begin{split}
& \bigg[  \Delta_{1}+q-1  +  \frac{(\Delta_{1}+2p+1)(\Delta_{2}-\sigma_{2}-1)}{\Delta_{2}-\sigma_{2}+q-1}  \bigg] C_{p,q}(\Delta_{1},\Delta_{2}, \sigma_{2})  \\
& =    2(\Delta_{1} +\Delta_{2} + 2q-\sigma_{2}-1 )  C_{p,q}(\Delta_{1}+1,\Delta_{2},\sigma_{2})
\end{split}
\end{equation}

Now we can solve for $p,q$ using equations \eqref{r7} and \eqref{r5}. These equations admit nontrivial solutions provided we have

\begin{equation}
\label{r8}
\begin{split}
&  \frac{\Delta_{1}+q-1}{2p+\Delta_{1}+1}  +   \frac{\Delta_{2} - \sigma_{2}-1}{\Delta_{2}- \sigma_{2}+q-1} = \frac{ 2 (\Delta_{1}+\Delta_{2}+ 2q -\sigma_{2}-1)}{\Delta_{1} + \Delta_{2} +2p +q- \sigma_{2}+1 } 
\end{split}
\end{equation}

Note that the differential equation \eqref{ymde1} holds for any value of $\Delta_{1}$.  Further as mentioned before, the values of  $p,q$ in the celestial OPE \eqref{opegen} do not depend on $\Delta_{1},\Delta_{2}$. Consequently we can vary $\Delta_{1}$ and $\Delta_{2}$ independently in \eqref{r8}.  Thereby the only non-trivial solution of the above equation is\footnote{A simple way to see this is by taking $\Delta_{1}$ or $\Delta_{2}$ to be large in \eqref{r8}. For example, keeping $\Delta_{2} $ fixed and taking $\Delta_{1} $ large in \eqref{r8} we immediately get $q$=0. Similarly in the limit where $\Delta_{2} $ becomes large we easily find that $p=-1$. The assumption that $\Delta_{1}, \Delta_{2}$ can be analytically continued off the principal series is again implicit here.} 
\begin{equation}
\label{pqvals}
\begin{split}
&  p=-1, \quad q=0
\end{split}
\end{equation}

This is precisely what we expect in pure Yang-Mills theory. Then substituting \eqref{pqvals} in \eqref{dimspin} we immediately get 
\begin{equation}
\label{dimspin1}
\begin{split}
\Delta = \Delta_{1}+ \Delta_{2} -1, \quad \sigma = \sigma_{2}
\end{split}
\end{equation}

The leading OPE \eqref{opegen} for outgoing gluon primaries then takes the form
\begin{equation}
\label{outope}
\begin{split}
& \mathcal{O}^{a_{1}}_{\Delta_{1}, +} (z_{1},\bar{z}_{1}) \mathcal{O}^{a_{2} }_{\Delta_{2},\sigma_{2}}(z_{2},\bar{z}_{2}) \sim - \frac{ i f^{a_{1}a_{2}x} }{ z_{12}} \hspace{0.05cm}  C_{-1,0}  \left( \Delta_{1},\Delta_{2}, \sigma_{2}\right) \mathcal{O}^{x}_{\Delta_{1}+\Delta_{2}-1,  \sigma_{2}} (z_{2},\bar{z}_{2}) 
\end{split}
\end{equation}

where $\sigma_{2}= \pm 1$. Now we can determine the OPE coefficient as follows. After substituting $p=-1,q=0$, equation \eqref{r7} as well as \eqref{r5} reduces to 
\begin{equation}
\label{r5a}
\begin{split}
&   C_{-1,0}(\Delta_{1}+1,\Delta_{2},\sigma_{2}) = \frac{\Delta_{1}-1}{\Delta_{1} +\Delta_{2} -\sigma_{2}-1 } \ C_{-1,0}(\Delta_{1},\Delta_{2}, \sigma_{2})    
\end{split}
\end{equation}

Then using the above in   the recursion relation \eqref{r6} we obtain 
\begin{equation}
\label{r5b}
\begin{split}
&   C_{-1,0}(\Delta_{1},\Delta_{2}+1,\sigma_{2}) = \frac{\Delta_{2}-\sigma_{2}}{\Delta_{1} +\Delta_{2} -\sigma_{2}-1} \ C_{-1,0}(\Delta_{1},\Delta_{2}, \sigma_{2})    
\end{split}
\end{equation}

Recursion relations of the form in \eqref{r5a} and \eqref{r5b} were also obtained using time translation invariance and the global subleading soft gluon symmetry in \cite{Pate:2019lpp}. The solution of these equations is given by 
\begin{equation}
\label{outopecoeff}
\begin{split}
& C_{-1,0}(\Delta_{1},\Delta_{2},\sigma_{2})  = \alpha \hspace{0.05cm} B(\Delta_{1}-1,\Delta_{2} - \sigma_{2})
\end{split}
\end{equation}

where $ B(x,y)$ is the Euler-Beta function. The constant $\alpha$ is as of yet undetermined. We can fix it by using the leading conformal soft limit. Consider $\Delta_{1} \rightarrow 1$ in \eqref{outope}. Then matching with the Ward identity \eqref{leadsoftwi} gives $\alpha =1$. Thus for outgoing gluons, we get 
\begin{equation}
\label{outopecoeff1}
\begin{split}
& C_{-1,0}(\Delta_{1},\Delta_{2},\sigma_{2})  =  B(\Delta_{1}-1,\Delta_{2} - \sigma_{2})
\end{split}
\end{equation}

In the case where both gluon primaries are incoming, an identical analysis again gives $p=-1,q=0$. The OPE coefficient also takes the same form as in \eqref{outopecoeff}. However in order to determine the overall constant we should note that the Kac-Moody current for an incoming gluon is given by 
\begin{equation}
\label{crossrel}
\begin{split}
& j^{a,-}(z,\bar{z})= -  \lim_{\Delta \to 1} (\Delta-1) \mathcal{O}^{a, - }_{\Delta, \sigma} (z,\bar{z})
\end{split}
\end{equation}

where the superscript $(-)$ above denotes an incoming gluon. Due to this minus sign in comparison to \eqref{leadsoftcurrent} for an outgoing gluon, the leading OPE coefficient for incoming gluons is given by
\begin{equation}
\label{ininopecoeff}
\begin{split}
& C_{-1,0}(\Delta_{1},\Delta_{2},\sigma_{2})  = -  B(\Delta_{1}-1,\Delta_{2} - \sigma_{2})
\end{split}
\end{equation}

The OPE coefficients in \eqref{outopecoeff1} and \eqref{ininopecoeff} precisely match with the corresponding results derived in \cite{Pate:2019lpp,Fan:2019emx}.

%%%%%%%%%%%%%%%%%%%%%%%%%%%%%%%%%%%%%%%%%%%%%%%%%%%%%%%%%%%%%%%%

\subsection{Outgoing-incoming OPE}
\label{opeinout}

We will now deal with the case where one of the gluon primaries in the celestial OPE is outgoing and the other is incoming. Here we will take the outgoing gluon primary to have positive helicity and denote it as $\mathcal{O}^{a_{1}}_{\Delta_{1}, +} (z_{1},\bar{z}_{1})$.  The incoming gluon primary will be denoted by $\mathcal{O}^{a_{2}, - }_{\Delta_{2},\sigma_{2}}(z_{2},\bar{z}_{2})$ where the superscript $(-)$ denotes that it is incoming. We will not fix the spin of the incoming gluon and so $\sigma_{2}=\pm 1$. 

In this case, both an outgoing and an incoming gluon primary can contribute to the OPE at leading order.  Then as in the  previous subsection we begin by assuming the following general form of the leading OPE
\begin{equation}
\label{ioopegen}
\begin{split}
& \mathcal{O}^{a_{1}}_{\Delta_{1},+} (z_{1},\bar{z}_{1}) \mathcal{O}^{a_{2}, - }_{\Delta_{2},\sigma_{2}}(z_{2},\bar{z}_{2}) \\
&= -  i f^{a_{1}a_{2}x}  z_{12}^{p} \bar{z}_{12}^{q} \bigg[ C^{+}_{p,q}  \left( \Delta_{1},\Delta_{2}, \sigma_{2}\right) \mathcal{O}^{x,+}_{\Delta,  \sigma} (z_{2},\bar{z}_{2}) + C^{-}_{p,q}  \left( \Delta_{1},\Delta_{2}, \sigma_{2}\right) \mathcal{O}^{x,-}_{\Delta,  \sigma} (z_{2},\bar{z}_{2}) \bigg] +\cdots
\end{split}
\end{equation}

where $\mathcal{O}^{x,+}_{\Delta,  \sigma} $ and $\mathcal{O}^{x,-}_{\Delta,  \sigma}$ in the R.H.S. of \eqref{ioopegen} respectively denote the outgoing and incoming primaries which contribute to the leading OPE.  Their conformal dimension and spin are 
\begin{equation}
\label{dimspin1}
\begin{split}
\Delta= \Delta_{1}+\Delta_{2}+p+q, \quad \sigma= p-q+ \sigma_{2}  +1
\end{split}
\end{equation}

In \eqref{ioopegen}  $C^{\pm}_{p,q}  \left( \Delta_{1},\Delta_{2}, \sigma_{2}\right)$ is the OPE coefficient corresponding to the outgoing (incoming) primary that appears in the OPE. The dots denote possible contributions from descendants.

Now following exactly the same steps as in the previous subsection \eqref{opeoutout} we can obtain a recursion relation for the leading OPE coefficients using the differential equation \eqref{ymde1}. Here we get 
\begin{equation}
\label{iooper1}
\begin{split}
(2 p + \Delta_{1}+1) C^{\pm}_{p,q} (\Delta_{1},\Delta_{2},\sigma_{2}) + (\Delta_{2} - \sigma_{2}-1+q)  C^{\pm}_{p,q} (\Delta_{1}+1,\Delta_{2}-1,\sigma_{2}) =0 
\end{split}
\end{equation}

Then applying the global subleading soft symmetry generator $J^{a}_{0}$ to the OPE \eqref{ioopegen} we get another  recursion relation analogous to \eqref{r3}
\begin{equation}
\label{iooper2}
\begin{split}
&( \Delta_{1}+q-2) C^{\pm}_{p,q} (\Delta_{1}-1,\Delta_{2},\sigma_{2}) -  (\Delta_{2} - \sigma_{2}-1)  C^{\pm}_{p,q} (\Delta_{1},\Delta_{2}-1,\sigma_{2}) \\
&= \pm 2 (\Delta_{1}+\Delta_{2}+ 2q -\sigma_{2}-2) C^{\pm}_{p,q} (\Delta_{1},\Delta_{2},\sigma_{2}) 
\end{split}
\end{equation}

Let us also note that invariance of the OPE \eqref{ioopegen} under global time translations yields 
\begin{equation}
\label{iooper3}
\begin{split}
  C^{ \pm }_{p,q}(\Delta_{1},\Delta_{2}, \sigma_{2}) = \pm\bigg[C^{ \pm}_{p,q}(\Delta_{1}+1,\Delta_{2},\sigma_{2}) -  C^{ \pm}_{p,q}(\Delta_{1},\Delta_{2}+1, \sigma_{2}) \bigg]
\end{split}
\end{equation}

Then using \eqref{iooper3} and performing similar manipulations as before we can rewrite \eqref{iooper1} and \eqref{iooper2} as follows 
\begin{equation}
\label{iooper4}
\begin{split}
 (2 p + \Delta_{1}+1) C^{\pm}_{p,q} (\Delta_{1},\Delta_{2},\sigma_{2}) = \pm (\Delta_{1} + \Delta_{2} +2p +q- \sigma_{2}+1)  C^{\pm}_{p,q} (\Delta_{1}+1,\Delta_{2},\sigma_{2}) 
\end{split}
\end{equation}

and
\begin{equation}
\label{iooper5}
\begin{split}
& \bigg[ \Delta_{1}+q-1  +   \frac{(\Delta_{1}+2p+1)(\Delta_{2} - \sigma_{2}-1)}{\Delta_{2}- \sigma_{2}+q-1} \bigg] C^{\pm}_{p,q} (\Delta_{1},\Delta_{2},\sigma_{2}) \\
&= \pm 2 (\Delta_{1}+\Delta_{2}+ 2q -\sigma_{2}-1) C^{\pm}_{p,q} (\Delta_{1}+1,\Delta_{2},\sigma_{2}) 
\end{split}
\end{equation}

The system of equations \eqref{iooper4} and \eqref{iooper5} have the same form as the analogous equations \eqref{r7} and \eqref{r5} obtained in the case of the OPE between two outgoing (incoming) gluons. It then follows by similar arguments that they admit non-trivial solutions iff
\begin{equation}
\label{pqval1}
\begin{split}
& p=-1, \quad q=0
\end{split}
\end{equation}

This is again the expected result in Yang-Mills theory.   The OPE coefficients can now be obtained by solving \eqref{iooper3} and \eqref{iooper4} with $p=-1, q=0$ in the same way as shown in the previous subsection \eqref{opeoutout}. We then get
\begin{equation}
\label{ioopecoeffs1}
\begin{split}
& C^{+}_{-1,0}(\Delta_{1},\Delta_{2},\sigma_{2})  =   \alpha \hspace{0.05cm} B( \Delta_{2} - \sigma_{2}, 2-\Delta_{1}-\Delta_{2} + \sigma_{2}) \\
& C^{-}_{-1,0}(\Delta_{1},\Delta_{2},\sigma_{2})  = \beta \hspace{0.05cm} B(\Delta_{1}-1,2-\Delta_{1}-\Delta_{2} +\sigma_{2})
\end{split}
\end{equation}

%%
%%
%\begin{split}
%%
%& C^{-}_{-1,0}(\Delta_{1},\Delta_{2},\sigma_{2}= - 1)  = \gamma \hspace{0.05cm} B(\Delta_{1}-1,1-\Delta_{1}-\Delta_{2} )
%\end{split}
%\end{equation}

where $\alpha,\beta$ are constants. These can be fixed by using the leading conformal soft limit. In order to determine $\beta$ we can put $p=-1,q=0$ in the OPE \eqref{ioopegen} and then take $\Delta_{1}\rightarrow 1$. The leading conformal soft theorem \eqref{leadsoftwi} then implies $\beta=1$.  Similarly considering the limit $\Delta_{2}\rightarrow 1$ in the OPE \eqref{ioopegen} and comparing with the Ward identity \eqref{leadsoftwi}  gives $\alpha=-1$. Thus finally the leading OPE coefficients in the case of outgoing-incoming gluon OPE are given by
\begin{equation}
\label{ioopes}
\begin{split}
& C^{+}_{-1,0}(\Delta_{1},\Delta_{2},\sigma_{2})  =  -  B( \Delta_{2} - \sigma_{2}, 2-\Delta_{1}-\Delta_{2} + \sigma_{2}) \\
& C^{-}_{-1,0}(\Delta_{1},\Delta_{2},\sigma_{2})  =  B(\Delta_{1}-1,2-\Delta_{1}-\Delta_{2} +\sigma_{2})
\end{split}
\end{equation}

Again the above results for the OPE coefficients are in perfect agreement with \cite{Pate:2019lpp}. 

%Shifting $\Delta_{1} \rightarrow \Delta_{1}+1$ in \eqref{iooper2} yields
%%
%\label{iooper3}
%%
%&( \Delta_{1}+q-1) C^{\pm}_{p,q} (\Delta_{1},\Delta_{2},\sigma_{2}) -  (\Delta_{2} - \sigma_{2}-1)  C^{\pm}_{p,q} (\Delta_{1}+1,\Delta_{2}-1,\sigma_{2}) \\
%&= \pm 2 (\Delta_{1}+\Delta_{2}+ 3q -\sigma_{2}-1) C^{\pm}_{p,q} (\Delta_{1}+1,\Delta_{2},\sigma_{2}) 
%\end{split}
%\end{equation}

%Note that the R.H.S. of \eqref{iooper7} is symmetric under $\Delta_{1} \leftrightarrow \Delta_{2}$. Therefore we must have
%\begin{equation}
%\label{iooper8}
%\begin{split}
%&  \frac{\Delta_{1}+q-1}{2p+\Delta_{1}+1}  -   \frac{\Delta_{1} - \sigma_{2}-1}{\Delta_{1}- \sigma_{2}+q-1} = \frac{\Delta_{2}+q-1}{2p+\Delta_{2}+1}  -   \frac{\Delta_{2} - \sigma_{2}-1}{\Delta_{2}- \sigma_{2}+q-1} 
%\end{split}
%\end{equation}

%Since $p,q$ do not depend on $\Delta_{1},\Delta_{2}$, we can set the L.H.S. and R.H.S. individually to zero. Thus we get
%\begin{equation}
%\label{iooper9}
%\begin{split}
%&  \frac{\Delta_{1}+q-1}{2p+\Delta_{1}+1}  =   \frac{\Delta_{1} - \sigma_{2}-1}{\Delta_{1}- \sigma_{2}+q-1} 
%\end{split}
%\end{equation}

%%%%%%%%%%%%%%%%%%%%%%%%%%%%%%%%%%%%%%%%%%%%%%%%%%%%%%%%%%%%%%%%

\section{Subleading OPE coefficients from symmetry}
\label{subleadopes}

In this section we will illustrate how the OPE coefficients of descendants in the celestial OPE between gluon primaries  can be determined using the underlying symmetries. For the OPE between positive helicity gluons, some of the descendant OPE coefficients were obtained in \cite{Ebert:2020nqf} using translation, global conformal and leading soft gluon current algebra symmetries. Here we will consider the mixed helicity case, i.e., the OPE between a positive helicity  and a negative helicity gluon primary.  We will see that the subleading soft gluon symmetry plays a crucial role here.

Let us denote the gluon primaries whose OPE we want to consider as $ \mathcal{O}^{ a }_{\Delta_{1},+}(z_{1},\bar{z}_{1})$ and $  \mathcal{O}^{b}_{\Delta_{2},-}(z_{2},\bar{z}_{2})$. We will also consider both of these to be outgoing. Then as shown in Section  \ref{pmope} of the Appendix in this paper,  upto the first subleading order this OPE is given by
\begin{equation}
\label{glpmope}
\begin{split}
&  \mathcal{O}^{ a }_{\Delta_{1},+}(z,\bar{z})  \mathcal{O}^{b}_{\Delta_{2},-}(z_{2},\bar{z}_{2})  \\
& \sim B(\Delta_{1}-1,\Delta_{2}+1)\bigg[ - \frac{ i f^{a b x}}{z_{12}} + \Delta_{1}  \hspace{0.05cm} \delta^{b x }\hspace{0.05cm} j^{a }_{-1}  + \frac{ (\Delta_{1}-1)}{(\Delta_{1}+\Delta_{2})} \hspace{0.04cm}   \delta^{ b x }  J^{a}_{-1}P_{-1,-1}   \bigg] \mathcal{O}^{x}_{\Delta_{1}+\Delta_{2}-1, - }(z_{2} ,\bar{z}_{2})
\end{split}
\end{equation}

where $P_{-1,-1}\mathcal{O}^{x}_{\Delta_{1}+\Delta_{2}-1,-}= \mathcal{O}^{x}_{\Delta_{1}+\Delta_{2},-} $. In the Appendix, Section \ref{pmope} we have derived this result from the Mellin transform of the $4$-point MHV gluon amplitude in Yang-Mills theory.  Although we have obtained this from the $4$-point Mellin amplitude, the above form of the mixed helicity OPE is  expected to hold within any $n$-point tree level MHV gluon amplitude in Yang-Mills. 

Now it is important to note that in the OPE \eqref{glpmope}, at order $\mathcal{O}(z_{12}^{0}\bar{z}_{12}^{0})$, we encounter descendants associated to both the leading soft gluon current algebra as well the subleading soft gluon symmetry algebra. These are given by the operators $j^{a }_{-1}\mathcal{O}^{x}_{\Delta_{1}+\Delta_{2}-1,-}$ and $J^{a}_{-1}\mathcal{O}^{x}_{\Delta_{1}+\Delta_{2},-}$ respectively in \eqref{glpmope}. We will now show that the OPE coefficients for these descendants can also determined using symmetries as follows.

In general, we can have the following descendants appearing at $\mathcal{O}(1)$ in the mixed-helicity OPE
\begin{equation}
\label{pmopeord1ops}
\begin{split}
 &  L_{-1} \mathcal{O}^{ a}_{\Delta,- }, \quad   j^{a}_{-1} \mathcal{O}^{ b}_{\Delta,- }, \quad  J^{a}_{-1}P_{-1,-1}  \mathcal{O}^{ b}_{\Delta,-}
\end{split}
\end{equation}

The above operators are linearly independent. This is because the vanishing condition \eqref{nullstKZform3} holds only for a positive helicity gluon primary. This will be further justified by our analysis below. Then the general form of the  $\mathcal{O}(1)$ term in the mixed helicity OPE can be written as
\begin{equation}
\label{pmopeord1gen}
\begin{split}
 & \mathcal{O}^{ a }_{\Delta_{1},+} (z ,\bar{z}) \mathcal{O}^{ b }_{\Delta_{2},- }(0) \supset B(\Delta_{1}- 1, \Delta_{2}+1) \bigg[ \alpha_{1}   f^{a b x}  \hspace{0.05cm} L_{-1} + \alpha_{2}  \hspace{0.05cm}\delta^{b x } j^{a}_{-1} + \alpha_{3}  \hspace{0.05cm} \delta^{bx} J^{a}_{-1}P_{-1,-1}\bigg]  \mathcal{O}^{ x}_{\Delta_{1}+\Delta_{2}-1,-} (0) 
\end{split}
\end{equation}

where $\alpha_{1},\alpha_{2}, \alpha_{3}$ are constants which we want to determine. The leading OPE coefficient $B(\Delta_{1}- 1, \Delta_{2}+1)$ can be obtained using the differential equation \eqref{ymdiffeq} as shown in Section \ref{leadopes}. Also note that we have placed the operator $\mathcal{O}^{ b }_{\Delta_{2},- }$ at the origin, without loss of generality. 

Now let us apply the subleading soft symmetry mode $J^{c}_{1}$  to the OPE in \eqref{pmopeord1gen}. Then using  \eqref{subpri} and applying the commutation relations listed in Sections \eqref{comms} and \eqref{JKcomms} we get the recursion relation
\begin{equation}
\label{pmoperec1}
\begin{split}
 &   i \hspace{0.04cm} \alpha_{1} f^{c x d}  f^{abx} +    \left( \alpha_{2} - (\Delta_{1}+\Delta_{2})\alpha_{3}-1\right)  f^{cax}  f^{x bd}  =0 
\end{split}
\end{equation}

The coefficients $\alpha_{1},\alpha_{2},\alpha_{3}$ in the above equation do not carry any Lie algebra indices. This equation should then hold for any allowed values of the free indices $(a,b,c,d)$. Thus we can set for example $a=c$ in \eqref{pmoperec1}. The structure constants being antisymmetric then immediately gives us
\begin{equation}
\label{rec1sola}
\begin{split}
 &  \alpha_{1} =0
\end{split}
\end{equation}

Similarly setting $a=b$ in \eqref{pmoperec1} we get
\begin{equation}
\label{rec1solb}
\begin{split}
 & \alpha_{2} - (\Delta_{1}+\Delta_{2})\alpha_{3}-1 =0
\end{split}
\end{equation}

Now it can be easily checked that applying the current algebra mode $j^{c}_{1}$ yields exactly the same recursion relation as in \eqref{pmoperec1}. Then in order to fix $\alpha_{2}$ we can apply $L_{1}$ to both sides of \eqref{pmopeord1gen}. Again using the relevant commutation relations from Section \ref{comms} we obtain
\begin{equation}
\label{pmoperec2}
\begin{split}
 &   i \hspace{0.04cm}\alpha_{1} (\Delta_{1}+ \Delta_{2})   f^{aby}  +  (\alpha_{2}- \Delta_{1}) f^{ab y} =0
\end{split}
\end{equation}

Substituting  $\alpha_{1}=0$ from \eqref{rec1sola} into the above equation we get  
\begin{equation}
\label{rec2sol}
\begin{split}
 &  \alpha_{2} = \Delta_{1}
\end{split}
\end{equation}

Finally we can solve for $\alpha_{3}$ from \eqref{pmoperec1} by putting in the value of $\alpha_{2}$ obtained above. This yields
\begin{equation}
\label{rec3sol}
\begin{split}
 &  \alpha_{3} = \frac{\alpha_{2}-1}{\Delta_{1}+\Delta_{2}} = \frac{\Delta_{1}-1}{\Delta_{1}+\Delta_{2}} 
\end{split}
\end{equation}

We thus find that the values of $\alpha_{1},\alpha_{2},\alpha_{3}$ obtained in \eqref{rec1sola}, \eqref{rec2sol} and \eqref{rec3sol} precisely agree with those extracted directly from the Mellin amplitude.  Now it is easy to see that \eqref{pmopeord1gen} is already invariant under the action of the translation generator $P_{-1,0}$.  It is  also straightforward to check that these values of the descendant OPE coefficients satisfy the recursion relations that follow from applying the translation generator $P_{0,-1}$ to the OPE \eqref{pmopeord1gen}. The fact that all these recursion relations are mutually consistent and admit a unique solution further justifies the absence of the null state relation \eqref{nullstKZform3} for a negative helicity gluon primary.

%Recursion from applying $P_{0,-1}$
%\begin{equation}
%\label{pmoperec4}
%\begin{split}
 %& \frac{(\Delta_{1}-1)}{(\Delta_{1}+\Delta_{2})}  \hspace{0.05cm}f^{a b d } = i  f^{abx} \alpha_{1} + f^{c x d} \alpha_{3}^{a b c x} 
%\end{split}
%\end{equation}

 %%%%%%%%%%%%%%%%%%%%%%%%%%%%%%%%%%%%%%%%%%%%%%%%%%%%%%%%%%%%%%%%

\section*{Acknowledgements}

The work of SB is partially supported by the Science and Engineering Research Board (SERB) grant MTR/2019/000937 (Soft-Theorems, S-matrix and Flat-Space Holography). S.G. would like to thank Yasha Neiman for useful discussions. The work of S.G. is supported by the Quantum Gravity Unit of the Okinawa Institute of Science and Technology Graduate University (OIST), Japan. 

%%%%%%%%%%%%%%%%%%%%%%%%%%%%%%%%%%%%%%%%%%%%%%%%%%%%%%%%%%%%%%%%

\section{Note Added}
In a previous version of the paper we demanded that the conditions \eqref{subpri} and \eqref{subpri2}, satisfied by gluon primaries, should also hold for any \textit{primary descendant} or null-state of a gluon primary. In other words, we demanded that any primary descendant should also be primary under the subleading soft gluon symmetry. We feel that a naive imposition of this condition is somewhat restrictive and in this version we have removed this. A more precise statement will appear elsewhere. 

The main results of the paper remain unchanged. For example, the null-state \eqref{nullstKZform3} is uniquely determined by the primary-state conditions with respect to the Poinacre group and the leading soft gluon $SU(N)$ current algebra.

%%%%%%%%%%%%%%%%%%%%%%%%%%%%%%%%%%%%%%%%%%%%%%%%%%%

%\begin{appendices} 

%\section{Notations and Conventions}
%\label{notconv}

%\section{Delta function representation for $n=5$ particles} 

%\label{appendix1}

%%%%%%%%%%%%%%%%%%%%%%%%%%%%%%%%%%%%%%%%%%%%%%%%%%%%%%%%%%%%%%%%
\begin{appendices} 

\section{Brief review of Celestial or Mellin amplitudes for massless particles}\label{review}

The Celestial or Mellin amplitude for $n$ gluons in four dimensions is defined as the Mellin transformation of the $n$-particle $S$-matrix element, given by \cite{Pasterski:2016qvg,Pasterski:2017kqt}
\be\label{mellin}
\mathcal M_n\big(\{z_i, \bar z_i, h_i, \bar h_i, a_i\}\big) = \prod_{i=1}^{n} \int_{0}^{\infty} d\omega_i \ \omega_i^{\D_i -1} S_n\big(\{\omega_i,z_i,\bar z_i, \sigma_i, a_i\}\big)
\ee 

where $\sigma_i= \pm 1$ denotes the helicity of the $i$-th gluon and the on-shell momenta are parametrized as,
\be\label{para}
p_i = \omega_i (1+z_i\bar z_i, z_i + \bar z_i , -i(z_i - \bar z_i), 1- z_i \bar z_i), \quad p_i^2 = 0
\ee

$a_i$ denotes the Lie algebra index carried by the $i$-th gluon.  The scaling dimensions $(h_i,\bar h_i)$ are defined as,
\be
h_i = \frac{\D_i + \sigma_i}{2}, \quad \bar h_i = \frac{\D_i - \sigma_i}{2}
\ee

The Lorentz group $SL(2,\mathbb C)$ acts on the celestial sphere as the group of global conformal transformations and the Mellin amplitude $\mathcal M_n$ transforms as,
\be
\mathcal M_n\big(\{z_i, \bar z_i, h_i, \bar h_i, a_i\}\big) = \prod_{i=1}^{n} \frac{1}{(cz_i + d)^{2h_i}} \frac{1}{(\bar c \bar z_i + \bar d)^{2\bar h_i}} \mathcal M_n\bigg(\frac{az_i+b}{cz_i+d} \ ,\frac{\bar a \bar z_i + \bar b}{\bar c \bar z_i + \bar d} \ , h_i,\bar h_i, a_i\bigg)
\ee

This is the familiar transformation law for the correlation function of primary operators of weight $(h_i,\bar h_i)$ in a $2$-D CFT under the global conformal group $SL(2,\mathbb C)$.

We can also define a modified Mellin amplitude \footnote{The exponentials in \eqref{mellinmod} can also be thought of as convergence factors. It is good to have them because, as discussed in \cite{Pate:2019mfs}, the Mellin amplitude for gluons, as defined in \eqref{mellin}, is only marginally convergent.} as in \cite{Banerjee:2018gce,Banerjee:2019prz}, 
\be\label{mellinmod}
\mathcal M_n\big(\{u_i,z_i, \bar z_i, h_i, \bar h_i, a_i\}\big) = \prod_{i=1}^{n} \int_{0}^{\infty} d\omega_i \ \omega_i^{\D_i -1} e^{-i\sum_{i=1}^n \epsilon_i \omega_i u_i} S_n\big(\{\omega_i,z_i,\bar z_i, \sigma_i, a_i\}\big)
\ee 

where $u$ can be thought of as a time coordinate and $\epsilon_i = \pm 1$ for an outgoing (incoming) particle. Under (Lorentz) conformal tranansformation the modified Mellin amplitude $\mathcal M_n$ transforms as,
\be
\mathcal M_n\big(\{u_i,z_i, \bar z_i, h_i, \bar h_i, a_i\}\big) = \prod_{i=1}^{n} \frac{1}{(cz_i + d)^{2h_i}} \frac{1}{(\bar c \bar z_i + \bar d)^{2\bar h_i}} \mathcal M_n\bigg(\frac{u_i}{|cz_i + d|^2} \ , \frac{az_i+b}{cz_i+d} \ ,\frac{\bar a \bar z_i + \bar b}{\bar c \bar z_i + \bar d} \ , h_i,\bar h_i, a_i\bigg)
\ee

Under global space-time translation, $u \rightarrow u + A + Bz + \bar B\bar z + C z\bar z$, the modified amplitude is invariant, i.e, 
\be
\mathcal M_n\big(\{u_i + A + Bz_i + \bar B\bar z_i + C z_i\bar z_i ,z_i, \bar z_i, h_i, \bar h_i, a_i\}\big) = \mathcal M_n\big(\{u_i,z_i, \bar z_i, h_i, \bar h_i, a_i\}\big)
\ee

Now in order to make manifest the conformal nature of the dual theory living on the celestial sphere it is useful to write the (modified) Mellin amplitude as a correlation function of conformal primary operators. So let us define a generic conformal primary operator as, 
\be
\label{confprim}
\mathcal O^{a,\epsilon}_{h,\bar h}(z,\bar z) = \int_{0}^{\infty} d\omega \  \omega^{\D-1} A^a(\epsilon\omega, z, \bar z, \sigma)
\ee

where $\epsilon=\pm 1$ for an annihilation (creation) operator of a massless gluon of helicity $\sigma$ and Lie algebra index $a$. Under (Lorentz) conformal transformation the conformal primary transforms like a primary operator of scaling dimension $(h,\bar h)$
\be
\mathcal O'^{a,\epsilon}_{h,\bar h}(z,\bar z) = \frac{1}{(cz + d)^{2h}} \frac{1}{(\bar c \bar z + \bar d)^{2\bar h}} \mathcal O^{a,\epsilon}_{h,\bar h}\bigg(\frac{az+b}{cz+d} \ ,\frac{\bar a \bar z + \bar b}{\bar c \bar z + \bar d}\bigg)
\ee

Similarly in the presence of the time coordinate $u$ we have,
\be
\label{confprimu}
\mathcal O'^{a,\epsilon}_{h,\bar h}(u,z,\bar z) = \int_{0}^{\infty} d\omega \ \omega^{\D-1} e^{-i \epsilon \omega u} A^a(\epsilon\omega, z, \bar z, \sigma)
\ee

Under (Lorentz) conformal transformations 
\be
\mathcal O'^{a,\epsilon}_{h,\bar h}(u,z,\bar z) = \frac{1}{(cz + d)^{2h}} \frac{1}{(\bar c \bar z + \bar d)^{2\bar h}} \mathcal O^{a,\epsilon}_{h,\bar h}\bigg(\frac{u}{|cz+d|^2},\frac{az+b}{cz+d} \ ,\frac{\bar a \bar z + \bar b}{\bar c \bar z + \bar d}\bigg)
\ee

In terms of \eqref{confprim},  the Mellin amplitude can be written as the correlation function of conformal primary operators
\be
\mathcal M_n = \bigg\langle{\prod_{i=1}^n \mathcal O^{a_i,\epsilon_i}_{h_i,\bar h_i}(z_i,\bar z_i)}\bigg\rangle
\ee

Similarly using \eqref{confprimu}, the modified Mellin amplitude can be written as,
\be
\mathcal M_n = \bigg\langle{\prod_{i=1}^n \mathcal O^{a_i,\epsilon_i}_{h_i,\bar h_i}(u_i,z_i,\bar z_i)}\bigg\rangle
\ee 

\subsection{Comments on notation in the paper}
Note that conformal primaries carry an additional index $\epsilon$ which distinguishes between an incoming and an outgoing particle. In the paper, for notational simplicity, we omit this additional index unless this plays an important role. So in most places we simply write the (modified) Mellin amplitude as,
\be
\mathcal M_n = \bigg\langle{\prod_{i=1}^n \mathcal O^{a_i}_{h_i,\bar h_i}(z_i,\bar z_i)}\bigg\rangle
\ee
or
\be
\mathcal M_n = \bigg\langle{\prod_{i=1}^n \mathcal O^{a_i}_{h_i,\bar h_i}(u_i,z_i,\bar z_i)}\bigg\rangle
\ee 

Similarly in many places in the paper we denote a gluon conformal primary of weight $\D = h+\bar h$ by $\mathcal O^{a}_{\D,\sigma}$ where $\sigma = \pm 1$ is the helicity (= $h-\bar h$). Since we are considering pure Yang-Mills, we can further simplify the notation to $\mathcal O^{a}_{\D,\pm}$ by omitting the $\sigma = \pm 1$. 

%%%%%%%%%%%%%%%%%%%%%%%%%%%%%%%%%%%%%%%%%%%%%%%%%%%%%%%

%%%%%%%%%%%%%%%%%%%%%%%%%%%%%%%%%%%%%%%%%%%%%%%%%%%%%%%%%%%%%%%%
\section{OPE from $5$-point MHV gluon amplitude} 
\label{ope5pt}

In this section of the Appendix we will consider the Mellin transform of the $5$-point tree-level MHV gluon amplitude in Yang-Mills theory.  Our main objective here is to show that the OPE in  \eqref{glppope} which was obtained in \cite{Ebert:2020nqf} from the $4$-point MHV amplitude also holds within the $5$-point Mellin amplitude. 

\subsection{$5$-point MHV gluon amplitude}
\label{5ptamp}

The tree-level $5$-point gluon amplitude in Yang-Mills theory can be expressed as  \cite{DelDuca:1999rs}
\begin{equation}
\label{glym5pt}
\begin{split}
\mathcal{A}(1^{a_{1}}, 2^{a_{2}}, 3^{a_{3}}, 4^{a_{4}}, 5^{a_{5}}) =  (i g)^{3} \sum_{\rho \in S_{3}} f^{a_{1}a_{\sigma_{2}}x_{1}} f^{x_{1}a_{\sigma_{3}}x_{2}} f^{x_{2}a_{\sigma_{4}}a_{5}} A(1, \rho(2), \rho(3), \rho(4), 5)
\end{split}
\end{equation}

where $g$ is the Yang-Mills coupling.  $f^{abc}$'s are Lie algebra structure constants . The sum above runs over a basis of $3!$ color-stripped partial amplitudes denoted by $A(1,\rho(2), \rho(3), \rho(4), 5)$.  This is an over-complete basis.  Owing to BCJ relations \cite{Bern:2008qj}, any $4$ of the sub-amplitudes in \eqref{glym5pt} can be written in terms of $2$ linearly independent sub-amplitudes. Let us choose this BCJ basis to be given by the sub-amplitudes
\begin{equation}
\label{bcjbasis5pt}
\begin{split}
A(1, 2, 3, 4, 5), \quad A(1, 4, 3, 2, 5)
\end{split}
\end{equation}

Then the remaining sub-amplitudes are given by \cite{Bern:2008qj}
\begin{equation}
\label{bcjbasis5pt1}
\begin{split}
& A(1, 2, 4, 3, 5) = \alpha_{1} A(1,2,3,4,5) + \alpha_{2}  A(1,4,3,2,5) \\
& A(1, 3, 2, 4, 5) = \alpha_{3} A(1,2,3,4,5) + \alpha_{4}  A(1,4,3,2,5) \\
& A(1, 3, 4, 2, 5) = \alpha_{5} A(1,2,3,4,5) + \alpha_{6}  A(1,4,3,2,5) \\
& A(1, 4, 2, 3, 5) = \alpha_{7} A(1,2,3,4,5) + \alpha_{8}  A(1,4,3,2,5) 
\end{split}
\end{equation}

where the $\alpha_{i}$'s are given by the following kinematic factors
\begin{equation}
\label{alphas}
\begin{split}
& \alpha_{1} = \frac{s_{45}(s_{12}+s_{24})}{s_{24}s_{35}}, \quad  \alpha_{2} = - \frac{s_{14} s_{25}}{s_{24}s_{35}}, \quad  \alpha_{3} = \frac{s_{12}(s_{24}+s_{45})}{s_{13}s_{24}}, \quad  \alpha_{4} = - \frac{s_{14} s_{25}}{s_{13}s_{24}}, \\
& \alpha_{5} =  - \frac{s_{12} s_{45}}{s_{13}s_{24}}, \quad  \alpha_{6} =  \frac{s_{14}( s_{24}+s_{25})}{s_{13}s_{24}}, \quad  \alpha_{7} = - \frac{s_{12} s_{45}}{s_{35}s_{24}}, \quad  \alpha_{8} =  \frac{s_{25}( s_{14}+s_{24})}{s_{35}s_{24}}
\end{split}
\end{equation}

Here $s_{ij} = (p_{i}+p_{j})^{2}$, with $p_{i}$ being the momenta of the external particles. Now our interest here will be in the MHV configuration. For this we will take the helicities of the gluons in \eqref{glym5pt} to be $\s_{1}=\s_{2} =-1, \s_{3}=\s_{4}=\s_{5} = 1$. Then using \eqref{bcjbasis5pt1}  we obtain
\begin{equation}
\label{glym5pt1}
\begin{split}
& \mathcal{A}(1^{-a_{1}}, 2^{-a_{2}}, 3^{+a_{3}}, 4^{+a_{4}}, 5^{+a_{5}}) \\
& =  (i g)^{3} \bigg[  \left( c_{1} + c_{2} \alpha_{1} + c_{3} \alpha_{3} +  c_{4} \alpha_{5} + c_{5} \alpha_{7}  \right) A(1^{-}, 2^{-}, 3^{+}, 4^{+}, 5^{+})  \\
& \hspace{1.2cm}+  \left( c_{6} + c_{2} \alpha_{2} + c_{3} \alpha_{4} +  c_{4} \alpha_{6} + c_{5} \alpha_{8}  \right) A(1^{-}, 4^{+}, 3^{+}, 2^{-}, 5^{+}) \bigg]
\end{split}
\end{equation}

where the $c_{i}$'s denote the following color structures
\begin{equation}
\label{colfacts}
\begin{split}
& c_{1} = f^{a_{1}a_{2}x_{1}} f^{x_{1}a_{3}x_{2}} f^{x_{2}a_{4}a_{5}} , \quad c_{2} = f^{a_{1}a_{2}x_{1}} f^{x_{1}a_{4}x_{2}} f^{x_{2}a_{3}a_{5}} , \quad  c_{3} = f^{a_{1}a_{3}x_{1}} f^{x_{1}a_{2}x_{2}} f^{x_{2}a_{4}a_{5}} \\
& c_{4} = f^{a_{1}a_{3}x_{1}} f^{x_{1}a_{4}x_{2}} f^{x_{2}a_{2}a_{5}} , \quad c_{5} = f^{a_{1}a_{4}x_{1}} f^{x_{1}a_{2}x_{2}} f^{x_{2}a_{3}a_{5}} , \quad  c_{6} = f^{a_{1}a_{4}x_{1}} f^{x_{1}a_{3}x_{2}} f^{x_{2}a_{2}a_{5}}
\end{split}
\end{equation}

Now parametrising the null momenta of the external gluons in the amplitude as 
\begin{equation}
\label{nullmomrep}
\begin{split}
& p^{\mu}_{i} =\epsilon_{i} \hspace{0.04cm} \omega_{i} q^{\mu}_{i}, \quad q^{\mu}_{i}=\left( 1+z_{i} \bar{z}_{i},  z_{i}+\bar{z}_{i} , - i(z_{i}-\bar{z}_{i}), 1- z_{i} \bar{z}_{i} \right), \quad i =1,2,3,4,5
\end{split}
\end{equation}

where $\epsilon_{i} =\pm 1$ and using the Parke-Taylor formula for MHV amplitudes \cite{Parke:1986gb}, we can write \eqref{glym5pt1} as
\begin{equation}
\label{glym5pt2}
\begin{split}
& \mathcal{A}(1^{-a_{1}}, 2^{-a_{2}}, 3^{+a_{3}}, 4^{+a_{4}}, 5^{+a_{5}}) \\
& = - \frac{(i g)^{3}}{2} \hspace{0.05cm} \frac{\omega_{1}\omega_{2}}{\omega_{3}\omega_{4}\omega_{5}} \hspace{0.05cm} \frac{z_{12}^{3}}{z_{15}z_{23}z_{35}} \bigg[  \frac{ \mathcal{X}_{1}}{z_{45}} + \frac{z_{12} }{z_{15}z_{25}} \left( 1- \frac{z_{45}}{z_{15}} \right)^{-1} \hspace{0.05cm} \mathcal{X}_{2} \bigg] \left( 1- \frac{z_{45}}{z_{35}} \right)^{-1}
\end{split}
\end{equation}

where $\mathcal{X}_{1}, \mathcal{X}_{2}$ are given by
\begin{equation}
\label{K1def}
\begin{split}
\mathcal{X}_{1} & =  c_{1} + c_{2} \ \frac{z_{45}\bar{z}_{45}}{z_{35}\bar{z}_{35}} \ \frac{\epsilon_{4}\omega_{4}}{\epsilon_{3}\omega_{3}} +  c_{3} \ \frac{z_{12}\bar{z}_{12}}{z_{13}\bar{z}_{13}} \  \frac{\epsilon_{2}\omega_{2}}{\epsilon_{3}\omega_{3}} \\
& +  \frac{z_{12}\bar{z}_{12}z_{45}\bar{z}_{45}}{z_{25}\bar{z}_{25}} \left( 1- \frac{z_{45}}{z_{25}} \right)^{-1}\left( 1- \frac{\bar{z}_{45}}{\bar{z}_{25}} \right)^{-1} \bigg[  (c_{3}-c_{4}) \ \frac{1}{z_{13}\bar{z}_{13}} \ \frac{\epsilon_{5}\omega_{5}}{\epsilon_{3}\omega_{3}} +  (c_{2}-c_{5}) \ \frac{1}{z_{35}\bar{z}_{35}} \ \frac{\epsilon_{1}\omega_{1}}{\epsilon_{3}\omega_{3}}\bigg]
\end{split}
\end{equation}

and
\begin{equation}
\label{K2def}
\begin{split}
\mathcal{X}_{2} & = c_{6} + c_{4} \ \frac{\epsilon_{4}\omega_{4}}{\epsilon_{3}\omega_{3}}  \ \frac{z_{15}\bar{z}_{15}}{z_{13}\bar{z}_{13}}  \left( 1- \frac{z_{45}}{z_{15}} \right)\left( 1- \frac{\bar{z}_{45}}{\bar{z}_{15}} \right) + c_{5} \ \frac{\epsilon_{2}\omega_{2}}{\epsilon_{3}\omega_{3}}  \ \frac{z_{25}\bar{z}_{25}}{z_{35}\bar{z}_{35}}   \\
& +   \bigg[  (c_{5}- c_{2}) \ \frac{\epsilon_{1}\omega_{1}}{\epsilon_{3}\omega_{3}} \ \frac{z_{15}\bar{z}_{15}}{z_{35}\bar{z}_{35}} + ( c_{4}  -  c_{3} ) \frac{\epsilon_{5}\omega_{5}}{\epsilon_{3}\omega_{3}} \  \frac{z_{15}\bar{z}_{15}}{z_{13}\bar{z}_{13}} \bigg]  \left( 1- \frac{z_{45}}{z_{15}} \right)\left( 1- \frac{\bar{z}_{45}}{\bar{z}_{15}} \right)  \left( 1- \frac{z_{45}}{z_{25}} \right)^{-1}\left( 1- \frac{\bar{z}_{45}}{\bar{z}_{25}} \right)^{-1}
\end{split}
\end{equation}

Note that we have written the amplitude in this particular form \eqref{glym5pt2} to facilitate the extraction of the celestial OPE between gluons $4^{+a_{4}}$ and $5^{+a_{5}}$ in the next subsection. 

%%%%%%%%%%%%%%%%%%%%%%%%%%%%%%%%%%%%%%%%%%%%%%%%%%%%%%%%%%%%%%%%
%\subsection{$4$-point color dressed MHV amplitude}

%%%%%%%%%%%%%%%%%%%%%%%%%%%%%%%%%%%%%%%%%%%%%%%%%%%%%%%%%%%%%%%%
\subsection{Mellin transform of $5$-point gluon MHV amplitude }

Let us now consider the Mellin transform of the $5$-point gluon MHV amplitude in \eqref{glym5pt1}. Here we will use the modified Mellin transform prescription which is reviewed in Section \ref{review}\footnote{ The Mellin transform prescription of \cite{Pasterski:2017kqt} can also be used here since it is convergent for tree-level gluon amplitudes. The celestial OPE finally will be the same irrespective of wether it is obtained from the usual Mellin amplitude or its modified version involving the time coordinates, $u_{i}$.}. For convenience in what follows we will set the Yang-Mills coupling constant $g=2$. The modified Mellin amplitude is given by
\begin{equation}
\label{mym5pt}
\begin{split}
& \mathcal{M}(1^{-a_{1}}, 2^{-a_{2}} ,3^{+a_{3}}, 4^{+a_{4}}, 5^{+a_{5}}) \\
& = \left \langle  \mathcal{O}^{a_{1}}_{\Delta_{1},-}(1)  \mathcal{O}^{a_{2}}_{\Delta_{2},-}(2)  \mathcal{O}^{a_{3}}_{\Delta_{3},+}(3)   \mathcal{O}^{a_{4}}_{\Delta_{4},+}(4) \mathcal{O}^{a_{5}}_{\Delta_{5},+}(5)  \right\rangle  \\
 &  = \int_{0}^{\infty}  \prod_{i=1}^{5} d\omega_{i} \hspace{0.05cm} \omega_{i}^{\Delta_{i}-1}  \hspace{0.05cm} e^{- i\sum\limits_{i=1}^{5} \epsilon_{i}\omega_{i}u_{i}} \hspace{0.05cm}  \mathcal{A}1^{-a_{1}}, 2^{-a_{2}} ,3^{+a_{3}}, 4^{+a_{4}}, 5^{+a_{5}}) \hspace{0.05cm} \delta^{(4)} \left( \sum_{i=1}^{5} \epsilon_{i} \omega_{i}q_{i}^{\mu}\right)%, \quad \Delta_i = 1 + i\lambda_i
\end{split}
\end{equation}

where $ \mathcal{O}^{a_{i}}_{\Delta_{i},\pm}(i) $ denotes a gluon primary operator with dimension $\Delta_{i}=1+i\lambda_{i}$. The subscript $(\pm)$ here denotes the spin of the operator. The label $(i)$ collectively denotes the coordinates $(z_{i},\bar{z}_{i},u_{i})$ at null infinity where the $i$-th gluon primary is inserted.

Now we are interested in extracting the celestial OPE between the gluons $(4^{+a_{4}},5^{+a_{5}})$. We will further take them to  outgoing and so $\epsilon_{4}=\epsilon_{5}=1$. Then let us define
\begin{equation}
\label{colparam}
\begin{split}
 \omega_{4} =\omega_{p} t , \quad \omega_{5} = (1-t)\omega_{p}
\end{split}
\end{equation}

where $t\in [0,1]$. The  delta function imposing overall energy-momentum conservation in \eqref{mym5pt} can be expressed in the following form
\begin{equation}
\label{delta5pt}
\begin{split}
& \delta^{(4)} \left( \sum_{i=1}^{5} \varepsilon_{i} \omega_{i} q_{i}^{\mu}\right)  \\
& = \frac{1}{4\ \omega_{P}} \delta \left( x- \bar{x} -t \hspace{0.05cm}   z_{45} \left(\frac{x}{z_{35}}-\frac{\bar{x}}{z_{25}} \right) -t\hspace{0.05cm}  \bar{z}_{45} \left(\frac{x}{\bar{z}_{25}}-\frac{\bar{x}}{\bar{z}_{35}}\right) +t \hspace{0.05cm} z_{45}\bar{z}_{45} \left( \frac{x}{z_{35}\bar{z}_{25}}-\frac{\bar{x}}{z_{25}\bar{z}_{35}}\right)\right)  \prod_{i=1}^{3}  \delta(\omega_{i} -\omega_{i}^{*})  
\end{split}
\end{equation}

where we have defined
\begin{equation}
\label{omegast}
\begin{split}
\omega_{i}^{*} = \varepsilon_{i}  \hspace{0.04cm} \omega_{P} (\sigma_{i,1} + t \ z_{45} \ \sigma_{i,2}+ t \ \bar{z}_{45} \ \sigma_{i,3}+  t \ z_{45}\bar{z}_{45} \ \sigma_{i,4}), \quad i \in (1,2,3)
\end{split}
\end{equation}

and
\begin{equation}
\label{xxbar}
\begin{split}
x=z_{12}z_{35} \bar{z}_{13}\bar{z}_{25}, \quad  \bar{x} =  z_{13}z_{25} \bar{z}_{12}\bar{z}_{35}
\end{split}
\end{equation}
 
The $\sigma_{i,j}$'s in \eqref{omegast} are given by
\begin{equation}
\label{5ptsigma1to3}
\begin{split}
& \sigma_{1,1} = - \frac{z_{25} \bar{z}_{35}}{z_{12}\bar{z}_{13}}, \quad \sigma_{1,2} = \frac{ \bar{z}_{35}}{z_{12}\bar{z}_{13}}, \quad \sigma_{1,3} = \frac{ z_{25}}{z_{12}\bar{z}_{13}}, \quad \sigma_{1,4} = -  \frac{ 1} {z_{12}\bar{z}_{13}} \\
&  \sigma_{2,1} =  \frac{z_{15} \bar{z}_{35}}{z_{12}\bar{z}_{23}}, \quad \sigma_{2,2} = - \frac{ \bar{z}_{35}}{z_{12}\bar{z}_{23}}, \quad \sigma_{2,3} = - \frac{ z_{15}}{z_{12}\bar{z}_{23}}, \quad \sigma_{2,4} =    \frac{ 1} {z_{12}\bar{z}_{23}} \\
& \sigma_{3,1} = -  \frac{z_{25} \bar{z}_{15}}{z_{23}\bar{z}_{13}}, \quad \sigma_{3,2} =  \frac{ \bar{z}_{15}}{z_{23}\bar{z}_{13}}, \quad \sigma_{3,3} =  \frac{ z_{25}}{z_{23}\bar{z}_{13}}, \quad \sigma_{3,4} =   -  \frac{ 1} {z_{23}\bar{z}_{13}} 
\end{split}
\end{equation}

%Note that they depend only on $z_{ij},\bar{z}_{ij}$ with $(i,j) \in (1,2,3,5)$. 
The representation of the delta function in \eqref{delta5pt} enables us to localise the integrals with respect to $\omega_{1},\omega_{2},\omega_{3}$ in \eqref{mym5pt}. This leaves us with integrals over $\omega_{P}$ and $t$. The $\omega_{P}$-integral can be easily done and we get
\begin{equation}
\label{omegapint}
\begin{split}
&   \int_{0}^{\infty}   d\omega_{P} \hspace{0.1cm} \omega_{P}^{i \Lambda -1} \exp\left[ -i \hspace{0.03cm} \omega_{P} \left( \mathcal{U}_{1} + z_{45} \hspace{0.03cm}t \hspace{0.04cm}  \mathcal{U}_{2}  + \bar{z}_{45} \hspace{0.03cm}t \hspace{0.04cm}  \mathcal{U}_{3} + z_{45} \bar{z}_{45}\hspace{0.03cm}t  \hspace{0.04cm}  \mathcal{U}_{4}  + t \hspace{0.04cm} u_{45}\right)\right] \\
& = \frac{\Gamma(i\Lambda)}{( i \hspace{0.05cm} \mathcal{U}_{1})^{i\Lambda}} \left[ 1+\frac{t}{\mathcal{U}_{1}} \left( z_{45} \hspace{0.04cm}\mathcal{U}_{2} + \bar{z}_{45} \hspace{0.04cm}\mathcal{U}_{3} + z_{45} \bar{z}_{45} \hspace{0.04cm}\mathcal{U}_{4}  + u_{45}\right) \right]^{-i\Lambda}
\end{split}
\end{equation}

where 
\begin{equation}
\label{Uidef}
\begin{split}
& \Lambda = \sum_{i=1}^{5}\lambda_{i}, \quad \mathcal{U}_{1} = \sum_{i=1}^{3} \sigma_{i,1} u_{i5}, \quad   \mathcal{U}_{2}= \sum_{i=1}^{3} \sigma_{i,2} u_{i5}, \quad \mathcal{U}_{3}= \sum_{i=1}^{3} \sigma_{i,3} u_{i5}, \quad \mathcal{U}_{4}= \sum_{i=1}^{3} \sigma_{i,4} u_{i5}
\end{split}
\end{equation}

Then using \eqref{omegapint} and the expression of the $5$-point MHV amplitude obtained in \eqref{glym5pt2} we finally get
\begin{equation}
\label{mym5pt1}
\begin{split}
& \mathcal{M}(1^{-a_{1}}, 2^{-a_{2}} ,3^{+a_{3}}, 4^{+a_{4}}, 5^{+a_{5}})\\
&= - i  \hspace{0.05cm} \mathcal{N} \int_{0}^{1} dt \hspace{0.1cm} t^{i\lambda_{4}-1}(1-t)^{i\lambda_{5}-1} \ \prod_{i=1}^{3} \Theta(\epsilon_{i}(\sigma_{i,1} +  z_{45} \hspace{0.03cm} t \hspace{0.03cm} \sigma_{i,2} + \bar{z}_{45} \hspace{0.03cm} t \hspace{0.03cm} \sigma_{i,3} + z_{45} \bar{z}_{45} \hspace{0.03cm} t \hspace{0.03cm} \sigma_{i,4} )) \ \mathcal{I}_{1}(t) \hspace{0.06cm} \mathcal{I}_{2}(t)
\end{split}
\end{equation}

In the above integral  the theta functions 
\begin{equation}
\label{thetafunc5pt}
\begin{split}
 \Theta(\epsilon_{i}(\sigma_{i,1} +  z_{45} \hspace{0.03cm} t \hspace{0.03cm} \sigma_{i,2} + \bar{z}_{45} \hspace{0.03cm} t \hspace{0.03cm} \sigma_{i,3} + z_{45} \bar{z}_{45} \hspace{0.03cm} t \hspace{0.03cm} \sigma_{i,4} )), \quad i=1,2,3
\end{split}
\end{equation}

simply impose the condition that $(\frac{\omega_{i}^{*}}{\omega_{P}}) \ge 0$. This is required because in the original integral \eqref{mym5pt}, the energy variables satisfy $\omega_{i}\ge 0$. The prefactor $\mathcal{N}$ in \eqref{mym5pt1} is given by
\vspace{0.2cm}
\begin{equation}
\label{Ngldef5pt}
\begin{split}
\mathcal{N} &=    \frac{z_{12}^{3}}{z_{15}z_{23}z_{35}} \ (\epsilon_{1}\sigma_{1,1})^{1+i\lambda_{1}} (\epsilon_{2}\sigma_{2,1})^{1+i\lambda_{2}}  (\epsilon_{3}\sigma_{3,1})^{i\lambda_{3}-1}\ \frac{\Gamma(i\Lambda)}{( i\hspace{0.04cm} \mathcal{U}_{1})^{i\Lambda}}  
\end{split}
\end{equation}

\vspace{0.1cm}
and $\mathcal{I}_{1}(t),  \mathcal{I}_{2}(t)$ in the integrand in \eqref{mym5pt1} are 
\begin{equation}
\label{I1gldef5pt}
\begin{split}
\mathcal{I}_{1}(t) =  & \left( 1+ z_{45} \hspace{0.03cm} t \hspace{0.03cm} \frac{\sigma_{1,2}}{\sigma_{1,1}} + \bar{z}_{45} \hspace{0.03cm} t \hspace{0.03cm} \frac{\sigma_{1,3}}{\sigma_{1,1}}  + z_{45}\bar{z}_{45} \hspace{0.03cm} t \hspace{0.03cm} \frac{\sigma_{1,4}}{\sigma_{1,1}}   \right)^{1+i\lambda_{1}}   \left( 1+ z_{45} \hspace{0.03cm} t \hspace{0.03cm} \frac{\sigma_{2,2}}{\sigma_{2,1}} + \bar{z}_{45} \hspace{0.03cm} t \hspace{0.03cm} \frac{\sigma_{2,3}}{\sigma_{2,1}}  + z_{45}\bar{z}_{45} \hspace{0.03cm} t \hspace{0.03cm} \frac{\sigma_{2,4}}{\sigma_{2,1}}   \right)^{1+i\lambda_{2}} \times \\
& \left( 1+ z_{45} \hspace{0.03cm} t \hspace{0.03cm} \frac{\sigma_{3,2}}{\sigma_{3,1}} + \bar{z}_{45} \hspace{0.03cm} t \hspace{0.03cm} \frac{\sigma_{3,3}}{\sigma_{3,1}}  + z_{45}\bar{z}_{45} \hspace{0.03cm} t \hspace{0.03cm} \frac{\sigma_{3,4}}{\sigma_{3,1}}   \right)^{i\lambda_{3}-1}   \left[ 1+ \frac{t}{\mathcal{U}_{1}} \left( z_{45} \ \mathcal{U}_{2} + \bar{z}_{45} \ \mathcal{U}_{3}+z_{45}\bar{z}_{45} \ \mathcal{U}_{4}+ u_{45}\right)\right]^{-i\Lambda}  \times \\
&  \delta \left( x- \bar{x} -t \   z_{45} \left(\frac{x}{z_{35}}-\frac{\bar{x}}{z_{25}} \right) -t\  \bar{z}_{45} \left(\frac{x}{\bar{z}_{25}}-\frac{\bar{x}}{\bar{z}_{35}}\right) +t \ z_{45}\bar{z}_{45} \left( \frac{x}{z_{35}\bar{z}_{25}}-\frac{\bar{x}}{z_{25}\bar{z}_{35}}\right)\right)
\end{split}
\end{equation}

and
\begin{equation}
\label{I2gldef5pt}
\begin{split}
\mathcal{I}_{2}(t) = \bigg[\frac{\mathcal{X}_{1}(t)}{z_{45}} + \frac{z_{12}}{z_{15}z_{25}} \left( 1- \frac{z_{45}}{z_{15}} \right)^{-1} \mathcal{X}_{2}(t)\bigg]\left( 1- \frac{z_{45}}{z_{35}} \right)^{-1}
\end{split}
\end{equation}

In \eqref{I2gldef5pt},  $\mathcal{X}_{1}(t), \mathcal{X}_{2}(t)$ are given by \eqref{K1def} and  \eqref{K2def} respectively after substituting the parametrisation of $\omega_{4},\omega_{5}$ in \eqref{colparam} and setting $\omega_{i} =\omega_{i}^{*}, i =1,2,3$.  \\

%%%%%%%%%%%%%%%%%%%%%%%%%%%%%%%%%%%%%%%%%%%%%%%%%%%%%%%%%%%%%%%%
\subsection{Mellin transform of $4$-point gluon MHV amplitude }

Here we will obtain the modified Mellin transform of the tree-level $4$-point MHV gluon amplitude which is required in order to extract the OPE from the $5$-point Mellin amplitude \eqref{mym5pt1}.  Since we shall be interested in taking the OPE between gluons $(4^{+a_{4}} ,5^{+a_{5}})$ in \eqref{mym5pt1}, it will be convenient to denote the momentum space gluon amplitude here as $\mathcal{A}(1^{-a_{1}} , 2^{-a_{2}}, 3^{+a_{3}}, 5^{+x} )$. We will also take the gluon labelled by $5^{+x}$ to be outgoing.   

The color-dressed $4$-point MHV amplitude is given by
\begin{equation}
\label{ym4pt}
\begin{split}
& \mathcal{A}(1^{-a_{1}} , 2^{-a_{2}}, 3^{+a_{3}}, 5^{+x} ) \\
& = (i g)^{2} \left( f^{a_{1}a_{2}y} f^{y a_{3}x} A(1^{-},2^{-},3^{+},5^{+})+  f^{a_{1}a_{3}y} f^{y a_{2}x} A(1^{-},3^{+},2^{-},5^{+}) \right)
\end{split}
\end{equation}

where $A(1^{-},2^{-},3^{+},5^{+})$ and $A(1^{-},3^{+},2^{-},5^{+}) $ are color-stripped partial amplitudes. Now applying the BCJ relation \cite{Bern:2008qj} 
\begin{equation}
\label{bcj4pt}
\begin{split}
&  A(1^{-} , 3^{+}, 2^{-}, 5^{+} ) =  \frac{s_{12}}{s_{13}} A(1^{-} , 2^{-}, 3^{+}, 5^{+} ) 
\end{split}
\end{equation}

we can express \eqref{ym4pt} as 
\begin{equation}
\label{ym4pt1}
\begin{split}
& \mathcal{A}(1^{-a_{1}} , 2^{-a_{2}}, 3^{+a_{3}}, 5^{+x}  ) = (i g)^{2} \left( f^{a_{1}a_{2}y} f^{y a_{3} x} +  f^{a_{1}a_{3}y} f^{y a_{2}x} \ \frac{s_{12}}{s_{13}}  \right) A(1^{-} , 2^{-}, 3^{+}, 5^{+} ) 
\end{split}
\end{equation}

Then parametrising the null momenta as
\begin{equation}
\label{nullmomrep}
\begin{split}
& p^{\mu}_{i} =\epsilon_{i} \hspace{0.04cm} \omega_{i} q^{\mu}_{i}, \quad q^{\mu}_{i}=\left( 1+z_{i} \bar{z}_{i},  z_{i}+\bar{z}_{i} , - i(z_{i}-\bar{z}_{i}), 1- z_{i} \bar{z}_{i} \right), \quad i =1,2,3,5
\end{split}
\end{equation}

%and using the Parke-Taylor formula for the MHV amplitude $A(1^{-} , 2^{-}, 3^{-}, 5^{+} ) $
%\begin{equation}
%\label{PTmhv4pt}
%\begin{split}
%& A(1^{-} , 2^{-}, 3^{+}, 5^{+} ) = \frac{\langle 1 2 \rangle^{3}}{\langle 2 3 \rangle \langle 3 5 \rangle \langle 5 1 \rangle}
%\end{split}
%\end{equation}

we get from \eqref{ym4pt1}
\begin{equation}
\label{ym4pt2}
\begin{split}
& \mathcal{A}(1^{-a_{1}} , 2^{-a_{2}}, 3^{+a_{3}}, 5^{+x} ) \\
&= - (i g)^{2} \left( f^{a_{1}a_{2}y} f^{y a_{3}x} +  f^{a_{1}a_{3}y} f^{y a_{2}x} \ \frac{\epsilon_{2}\omega_{2}}{\epsilon_{3}\omega_{3}} \ \frac{z_{12}\bar{z}_{12}}{z_{13}\bar{z}_{13}} \right) \frac{\omega_{1}\omega_{2}}{\omega_{3} \omega_{5}} \ \frac{z_{12}^{3}}{z_{15}z_{23}z_{35}}
\end{split}
\end{equation}

Now let us compute the modified Mellin transform of \eqref{ym4pt2}. In order to relate this to the factorisation of the $5$-point Mellin amplitude in the $(4^{+},5^{+})$ OPE channel we will take the dimension of the primary corresponding to the gluon $5^{+x}$ in \eqref{ym4pt2} to be $\Delta_{4}+\Delta_{5}-1$. Then the Mellin amplitude is 
\begin{equation}
\label{mellinym4pt}
\begin{split}
& \mathcal{M}(1^{- a_{1}}, 2^{-a_{2}} ,3^{+a_{3}}, 5^{+x}) \\
& = \left \langle  \mathcal{O}^{a_{1}}_{\Delta_{1},-}(1)  \mathcal{O}^{a_{2}}_{\Delta_{2},-}(2)  \mathcal{O}^{a_{3}}_{\Delta_{3},+}(3)   \mathcal{O}^{ x}_{\Delta_{4}+\Delta_{5}-1,+}(5)  \right\rangle  \\
 &  = \int_{0}^{\infty}  d\omega_{5} \hspace{0.05cm} \omega_{5}^{\Delta_{5}-1} e^{- i \omega_{5}u_{5}} \prod_{i=1}^{3} d\omega_{i} \hspace{0.05cm} \omega_{i}^{\Delta_{i}-1}  \hspace{0.05cm} e^{- i\sum\limits_{i=1}^{3} \epsilon_{i}\omega_{i}u_{i}} \hspace{0.05cm}  \mathcal{A}(1^{-a_{1}} , 2^{-a_{2}}, 3^{+a_{3}}, 5^{+x} ) \hspace{0.05cm} \delta^{(4)} \left( \sum_{i=1}^{3} \epsilon_{i} \omega_{i}q_{i}^{\mu}+\omega_{5}q^{\mu}_{5} \right)%, \quad \Delta_i = 1 + i\lambda_i
\end{split}
\end{equation}

The momentum conservation imposing delta function in the above integral can be represented as
\begin{equation}
\label{delta4pt}
\begin{split}
 &\delta^{(4)} \left( \sum_{i=1}^{3} \epsilon_{i} \omega_{i}q_{i}^{\mu}+\omega_{5}q^{\mu}_{5} \right) = \frac{1}{4\ \omega_{5}}   \prod_{i=1}^{3} \delta(\omega_{i} -\omega_{i}^{*}) \hspace{0.04cm}  \delta \left( x- \bar{x} \right)
\end{split}
\end{equation}

where 
\begin{equation}
\label{omegast4pt}
\begin{split}
 & \omega_{i}^{*} = \epsilon_{i}  \hspace{0.05cm} \omega_{5}   \hspace{0.05cm} \sigma_{i,1}, \quad i =1,2,3. 
\end{split}
\end{equation}

The $\sigma_{i,1}$'s in \eqref{omegast4pt} were specified in  \eqref{5ptsigma1to3}. Note that using the definition of $x,\bar{x}$ given in \eqref{xxbar} it follows that the delta function $\delta(x-\bar{x})$ above imposes the constraint 
\begin{equation}
\label{creq}
\begin{split}
&  r =\bar{r} 
\end{split}
\end{equation}

where $r$ and $\bar{r}$ are the cross-ratios
\begin{equation}
\label{cr}
\begin{split}
 r = \frac{z_{12}z_{35}}{z_{13}z_{25}}, \quad  \bar{r} = \frac{\bar{z}_{12}\bar{z}_{35}}{\bar{z}_{13}\bar{z}_{25}}
\end{split}
\end{equation}

Now using \eqref{delta4pt} in \eqref{mellinym4pt}  localises the integrals with respect to $\omega_{i}, i=1,2,3$ . Then doing the remaining integral over $\omega_{5}$ we get
\begin{equation}
\label{omegapint4pt}
\begin{split}
&   \int_{0}^{\infty}   d\omega_{5} \hspace{0.1cm} \omega_{5}^{i \Lambda -1} \exp\left( -i \hspace{0.03cm} \omega_{5} \hspace{0.03cm}\mathcal{U}_{1} \right)  = \frac{\Gamma(i\Lambda)}{( i \hspace{0.05cm} \mathcal{U}_{1})^{i\Lambda}}
\end{split}
\end{equation}

where 
\begin{equation}
\label{Uidef}
\begin{split}
& \Lambda = \sum_{i=1}^{5}\lambda_{i}, \quad \mathcal{U}_{1} = \sum_{i=1}^{3} \sigma_{i,1} u_{i5}
\end{split}
\end{equation}

Therefore the $4$-pt Mellin amplitude takes the form (with $g=2$)
\begin{equation}
\label{ymgl4ptmellin}
\begin{split}
 \mathcal{M}(1^{- a_{1}}, 2^{-a_{2}} ,3^{+a_{3}}, 5^{+x}) &=  \frac{z_{12}^{3}}{z_{14}z_{23}z_{34}} \ (\epsilon_{1}\sigma_{1,1})^{1+i\lambda_{1}} (\epsilon_{2}\sigma_{2,1})^{1+i\lambda_{2}} (\epsilon_{3}\sigma_{3,1})^{i\lambda_{3}-1}  \frac{\Gamma(i\Lambda)}{( i\hspace{0.04cm} \mathcal{U}_{1})^{i\Lambda}}  \times \\
 & \hspace{1.0cm} \left( f^{a_{1}a_{2}x} f^{x a_{3}a_{4}} - r \hspace{0.06cm}  f^{a_{1}a_{3}x} f^{x a_{2}a_{4}} \right) \delta(x-\bar{x})  \ \prod_{i=1}^{3} \Theta(\epsilon_{i}\sigma_{i,1}) 
\end{split}
\end{equation}

Let us also note that  $\mathcal{N}$ defined in \eqref{Ngldef5pt} which appears as a prefactor in the $5$-point Mellin amplitude in \eqref{mym5pt1} is simply related to the $4$-point Mellin amplitude as
\begin{equation}
\label{ymgl4ptmellin1}
\begin{split}
 \mathcal{M}(1^{- a_{1}}, 2^{-a_{2}} ,3^{+a_{3}}, 5^{+x})  &= \left( f^{a_{1}a_{2}x} f^{x a_{3}a_{4}} - r \hspace{0.06cm}  f^{a_{1}a_{3}x} f^{x a_{2}a_{4}} \right) \mathcal{N} \ \delta(x-\bar{x})  \ \prod_{i=1}^{3} \Theta(\epsilon_{i}\sigma_{i,1}) 
\end{split}
\end{equation}

%%%%%%%%%%%%%%%%%%%%%%%%%%%%%%%%%%%%%%%%%%%%%%%%%%%%%%%%%%%%%%%%

\subsection{OPE decomposition of $5$-point Mellin amplitude}

We shall now extract the celestial OPE between gluon primaries $\mathcal{O}^{a_{4}}_{\Delta_{4},+}$ and $\mathcal{O}^{a_{5}}_{\Delta_{5},+}$ from the $5$-point Mellin amplitude. For this purpose we will set $u_{45}=0$ and expand \eqref{mym5pt1} around $z_{45}= 0,\bar{z}_{45} = 0$. Now note that expanding the theta functions in  \eqref{mym5pt1} generates delta function contributions whose arguments are functions of $z_{ij}, \bar{z}_{ij}$ with $ (i,j)\in (1,2,3,5)$.  We will assume that the operators which do not participate in the OPE are all inserted at separated points at null infnity and thereby such contact terms can be ignored. The contributions of interest here will then only come from expanding the integrands $\mathcal{I}_{1}(t), \mathcal{I}_{2}(t)$ in \eqref{mym5pt1}. 

Here we will restrict attention to the leading and the first subleading terms in the OPE regime. The leading term corresponds to the Mellin transform of the collinear splitting function \cite{Bern:1998sv}. The subleading terms in the Mellin amplitude can likely be related to the subleading corrections to the leading collinear limit in the momentum space amplitude which has been studied in \cite{Nandan:2016ohb}.

%%%%%%%%%%%%%%%%%%%%%%%%%%%%%%%%%%%%%%%%%%%%%%%%%%%%%%%%%%%%%%%%

\subsubsection{Leading term}

Using \eqref{I1gldef5pt} and \eqref{I2gldef5pt}, it is easy to see that the  terms from $\mathcal{I}_{1}(t)$ and $\mathcal{I}_{2}(t)$ which contribute at leading order in the OPE limit $(z_{45} \rightarrow 0,\bar{z}_{45} \rightarrow 0)$ are given by

\begin{equation}
\label{leadord}
\begin{split}
\mathcal{I}_{1}(t)\Big|_{\mathcal{O}(1)} = \delta(x-\bar{x})
\end{split}
\end{equation}
and
\begin{equation}
\label{leadord1}
\begin{split}
\mathcal{I}_{2}(t)\Big|_{\mathcal{O}(z_{45}^{-1})}& = \frac{1}{z_{45}} \ \mathcal{X}_{1}(t)\Big|_{\mathcal{O}(1)}  = \frac{1}{z_{45}}\left( c_{1}  -  \frac{\bar{r}(1-r)}{(1-\bar{r})} \hspace{0.05cm} c_{3}\right) = (c_{1} - r c_{3})
\end{split}
\end{equation}

where in obtaining the last equality in \eqref{leadord1} we have used the fact on the support of $\delta(x-\bar{x})$ we get $r=\bar{r}$. Then doing the $t$-integral and using \eqref{ymgl4ptmellin1}, we find that the $5$-point Mellin amplitude factorises as
\begin{equation}
\label{leadord2}
\begin{split}
& \mathcal{M}(1^{- a_{1}}, 2^{-a_{2}} ,3^{+a_{3}}, 4^{+a_{4}}, 5^{+a_{5}}) \\
& = - \frac{i f^{a_{4}a_{5}x}}{z_{45}} \hspace{0.05cm} B(\Delta_{4}-1,\Delta_{5}-1)  \mathcal{M}(1^{- a_{1}}, 2^{-a_{2}} ,3^{+a_{3}},  5^{+x}) + \cdots
\end{split}
\end{equation}

where $B(x,y)$ is the Euler Beta function and  the dots denote subleading terms. Then \eqref{leadord2} implies that the leading celestial OPE is 
\begin{equation}
\label{leadord3}
\begin{split}
& \mathcal{O}^{a_{4}}_{\Delta_{4},+}(z_{4},\bar{z}_{4})  \mathcal{O}^{a_{5}}_{\Delta_{5},+} (z_{5},\bar{z}_{5})  \sim - \frac{i f^{a_{4}a_{5}x}}{z_{45}} \hspace{0.05cm} B(\Delta_{4}-1,\Delta_{5}-1) \hspace{0.05cm}  \mathcal{O}^{x}_{\Delta_{4}+\Delta_{5}-1,+} (z_{5},\bar{z}_{5}) 
\end{split}
\end{equation}

This of course agrees with the expected result \cite{Pate:2019lpp,Fan:2019emx}. 

%%%%%%%%%%%%%%%%%%%%%%%%%%%%%%%%%%%%%%%%%%%%%%%%%%%%%%%%%%%%%%%%
\subsubsection{Subleading terms: $\mathcal{O}(z_{45}^{0}\bar{z}_{45}^{0})$}

We now want to consider the subleading term of order $\mathcal{O}(z_{45}^{0}\bar{z}_{45}^{0})$  in the OPE decomposition. This will correspond to the contribution of descendants in the celestial OPE. Let us first gather the relevant terms from $\mathcal{I}_{1}(t)$ and $\mathcal{I}_{2}(t)$ which can contribute at this order. From \eqref{I1gldef5pt} we get
\begin{equation}
\label{ord15pt}
\begin{split}
\mathcal{I}_{1}(t)  \Big|_{\mathcal{O}(z_{45})}  = z_{45}\  t \bigg[ & \mathcal{Z} \hspace{0.05cm} \delta(x-\bar{x}) + \partial_{z_{5}} \delta(x-\bar{x}) \bigg] 
\end{split}
\end{equation}

where $\mathcal{Z}$ is given by
\begin{equation}
\label{Zdef}
\begin{split}
\mathcal{Z} =  \bigg(  (1+i\lambda_{1}) \frac{\sigma_{1,2}}{\sigma_{1,1}} + (1+i\lambda_{2}) \frac{\sigma_{2,2}}{\sigma_{2,1}} + (i\lambda_{3}-1) \frac{\sigma_{3,2}}{\sigma_{3,1}} -  \frac{ i \Lambda}{\mathcal{U}_{1}} \hspace{0.05cm} \mathcal{U}_{2}   \bigg) 
\end{split}
\end{equation}

Then from \eqref{I2gldef5pt} we have
\begin{equation}
\label{ord15pt1}
\begin{split}
\mathcal{I}_{2}(t)  \Big|_{\mathcal{O}(z_{45}^{-1})} & = \frac{1}{z_{45}} \ \mathcal{X}_{1}(t)\Big|_{\mathcal{O}(1)}  = \frac{1}{z_{45}}\left( c_{1}  -  \frac{\bar{r}(1-r)}{(1-\bar{r})} \hspace{0.05cm} c_{3}\right)% +  c_{3} \ \frac{z_{12}\bar{z}_{12}}{z_{13}\bar{z}_{13}} \  \frac{\sigma_{2,1}}{\sigma_{3,1}} \right)
\end{split}
\end{equation}

and
\begin{equation}
\label{ord15pt2}
\begin{split}
\mathcal{I}_{2}(t)  \Big|_{\mathcal{O}(1)} & = \frac{1}{z_{45}} \ \mathcal{X}_{1}(t)\Big|_{\mathcal{O}(z_{45})}  + \frac{1}{z_{35}} \ \mathcal{X}_{1}(t)\Big|_{\mathcal{O}(1)} + \frac{z_{12}}{z_{15}z_{25}} \ \mathcal{X}_{2}(t)\Big|_{\mathcal{O}(1)} \\
& =  \frac{ (c_{1} - r \ c_{3})}{z_{35}} + \left( -\frac{1}{z_{15}} + \frac{1}{z_{25}}\right) (  c_{6} -  c_{5}  + (1-r) (c_{3}-c_{4}))  + c_{2} \left( \frac{1}{z_{25}} - \frac{1}{z_{35}}\right) \\
& + t \left( \frac{1}{z_{15}} - \frac{1}{z_{25}}\right) c_{3} %\bigg]  \delta(x-\bar{x}) \ \mathcal{N } 
%& = j^{a_{4}}_{-1} \mathcal{M}^{a_{1}a_{2}a_{3}a_{5}} +  t \left( \frac{1}{z_{15}} - \frac{1}{z_{25}}\right) c_{3} \  \delta(x-\bar{x}) \ \mathcal{N }
\end{split}
\end{equation}

Using \eqref{ord15pt}, \eqref{ord15pt1}, \eqref{ord15pt2} and performing the $t$-integral in \eqref{mym5pt1}, we find the $\mathcal{O}(z_{45}^{0}\bar{z}_{45})$ contribution to the $5$-point Mellin amplitude in the OPE regime to be given by

\begin{equation}
\label{ord15pt3}
\begin{split}
& \mathcal{M}(1^{- a_{1}}, 2^{-a_{2}} ,3^{+a_{3}}, 4^{+a_{4}}, 5^{+a_{5}})\Big|_{\mathcal{O}(z_{45}^{0}\bar{z}_{45}^{0})}\\
& = - i\hspace{0.05cm} B(\Delta_{4}-1, \Delta_{5}-1)  \hspace{0.05cm} \mathcal{N} \bigg[ \frac{\Delta_{4}-1}{(\Delta_{4}+\Delta_{5}-2)} \left( \mathcal{Z} \hspace{0.05cm}  \delta(x-\bar{x})+ \partial_{z_{5}} \delta(x-\bar{x})  \right) \left( c_{1}  -  \frac{\bar{r}(1-r)}{(1-\bar{r})} \hspace{0.05cm} c_{3}\right) +  \\
& \left( \frac{ (c_{1} - r \ c_{3})}{z_{35}} + \left( -\frac{1}{z_{15}} + \frac{1}{z_{25}}\right) (  c_{6} -  c_{5}  + (1-r) (c_{3}-c_{4}))  + c_{2} \left( \frac{1}{z_{25}} - \frac{1}{z_{35}}\right) \right) \delta(x-\bar{x}) \\
& + \frac{\Delta_{4}-1}{(\Delta_{4}+\Delta_{5}-2)}  \left( \frac{1}{z_{15}} - \frac{1}{z_{25}}\right) c_{3}  \hspace{0.05cm} \delta(x-\bar{x})\bigg] \prod_{i=1}^{3} \Theta(\epsilon_{i}\sigma_{i,1}) 
\end{split}
\end{equation}

We can simplify the above further by noting the following identity
\begin{equation}
\label{deldervid}
\begin{split}
&   \frac{\bar{r}(1-r)}{(1-\bar{r})}  \ \partial_{z_{5}} \delta(x-\bar{x})  = r \hspace{0.05cm} \partial_{z_{5}} \delta(x-\bar{x}) +  \left( \frac{1}{z_{15}} - \frac{1}{z_{25}} \right) \delta(x-\bar{x})
\end{split}
\end{equation}

Applying \eqref{deldervid} in \eqref{ord15pt3} then gives
\begin{equation}
\label{ord15pt4}
\begin{split}
& \mathcal{M}(1^{- a_{1}}, 2^{-a_{2}} ,3^{+a_{3}}, 4^{+a_{4}}, 5^{+a_{5}}) \Big|_{\mathcal{O}(z_{45}^{0}\bar{z}_{45}^{0})}\\
& = - i  \hspace{0.05cm} B(\Delta_{4}-1, \Delta_{5}-1)  \hspace{0.05cm} \mathcal{N} \bigg[\frac{\Delta_{4}-1}{(\Delta_{4}+\Delta_{5}-2)} \left( \mathcal{Z} \hspace{0.05cm}  \delta(x-\bar{x})+ \partial_{z_{5}} \delta(x-\bar{x})  \right) \left( c_{1}  - r \hspace{0.05cm} c_{3}\right)  \\
& + \left( \frac{ (c_{1} - r \ c_{3})}{z_{35}} + \left( -\frac{1}{z_{15}} + \frac{1}{z_{25}}\right) (  c_{6} -  c_{5}  + (1-r) (c_{3}-c_{4}))  + c_{2} \left( \frac{1}{z_{25}} - \frac{1}{z_{35}}\right) \right) \delta(x-\bar{x}) \bigg]  \prod_{i=1}^{3} \Theta(\epsilon_{i}\sigma_{i,1}) 
\end{split}
\end{equation}

Now at this order in the OPE we expect to find descendants created by the action of $L_{-1}$ and the current algebra mode $j^{a}_{-1}$. In order to demonstrate this we need to know the action of these modes on the $4$-point Mellin amplitude. We will  now evaluate each of these in turn. First let us consider the correlator
\vspace{0.1cm}
\begin{equation}
\label{Lm14pt}
\begin{split}
&  \left \langle  \mathcal{O}^{a_{1}}_{\Delta_{1},-}(1)  \mathcal{O}^{a_{2}}_{\Delta_{2},-}(2)  \mathcal{O}^{a_{3}}_{\Delta_{3},+}(3)   \left( L_{-1}\mathcal{O}^{x}_{\Delta_{4}+\Delta_{5}-1,+}(5) \right)  \right\rangle \\ 
&  = \mathcal{L}_{-1} (5) \mathcal{M}(1^{- a_{1}}, 2^{-a_{2}} ,3^{+a_{3}},  5^{+x }) \\
& = \partial_{z_{5}} \mathcal{M}(1^{- a_{1}}, 2^{-a_{2}} ,3^{+a_{3}},  5^{+x }) 
\end{split}
\end{equation}

Using the expression of the $4$-point Mellin amplitude obtained in \eqref{ymgl4ptmellin1} the above can be straightforwardly evaluated and we obtain\footnote{Since we are interested in the $4$-point correlator at separated points on the celestial sphere, we do not differentiate the theta functions in \eqref{ymgl4ptmellin1} with respect to $z_{5}$ while evaluating \eqref{Lm14pt}.}
\begin{equation}
\label{Lm14pt1}
\begin{split}
&  f^{a_{4}a_{5}x} \mathcal{L}_{-1} \mathcal{M}(1^{- a_{1}}, 2^{-a_{2}} ,3^{+a_{3}},  5^{+x }) \\ 
& =  \bigg[ (c_{1}-r \hspace{0.04cm} c_{3}) \mathcal{N}  \left[ \mathcal{Z} \hspace{0.05cm} \delta(x-\bar{x}) + \partial_{z_{5}} \delta(x-\bar{x}) \right] + (c_{1}-r \hspace{0.04cm} c_{3}) \left( \frac{1}{z_{15}}+ \frac{1}{z_{35}} \right) \mathcal{N} \ \delta(x-\bar{x})  \\
& - r\left( \frac{1}{z_{25}} - \frac{1}{z_{35}}\right) c_{3} \hspace{0.05cm} \mathcal{N}\  \delta(x-\bar{x})\bigg]  \prod_{i=1}^{3} \Theta(\epsilon_{i}\sigma_{i,1}) 
\end{split}
\end{equation}

where $\mathcal{Z}$ is given by \eqref{Zdef}.  The other correlation function of interest here is 
\vspace{0.1cm}
\begin{equation}
\label{ja4m14pt}
\begin{split}
& \left \langle  \mathcal{O}^{a_{1}}_{\Delta_{1},-}(1)  \mathcal{O}^{a_{2}}_{\Delta_{2},-}(2)  \mathcal{O}^{a_{3}}_{\Delta_{3},+}(3)   \left( j^{a_{4}}_{-1}\mathcal{O}^{a_{5}}_{\Delta_{4}+\Delta_{5}-1,+}(5) \right)  \right\rangle \\
 & =  \mathscr{J}^{a_{4}}_{-1}(5) \mathcal{M}(1^{- a_{1}}, 2^{-a_{2}} ,3^{+a_{3}},  5^{+a_{5} }) 
 \end{split}
\end{equation}

This can be easily calculated using \eqref{jamp}. We then get
\begin{equation}
\label{ja4m14pt1}
\begin{split}
& \mathscr{J}^{a_{4}}_{-1}(5) \mathcal{M}(1^{- a_{1}}, 2^{-a_{2}} ,3^{+a_{3}},  5^{+a_{5} }) \\
 & =  - i\bigg[ \frac{1}{z_{15}} \  f^{a_{4} a_{1} x} (f^{x a_{2} y} f^{ y a_{3} a_{5}} - r \hspace{0.05cm} f^{x  a_{3} y} f^{y a_{2} a_{5}} ) + \frac{1}{z_{25}} \  f^{a_{4} a_{2} x} (f^{ a_{1} x  y} f^{ y a_{3} a_{5}} - r \hspace{0.05cm} f^{a_{1}  a_{3} y} f^{y x a_{5}} ) \\
&   \hspace{1.2cm} + \frac{1}{z_{35}}  f^{a_{4} a_{3} x} (f^{a_{1} a_{2} y} f^{ y x a_{5}} - r \hspace{0.05cm} f^{ a_{1} x  y} f^{y a_{2} a_{5}} ) \bigg] \mathcal{N} \ \delta(x-\bar{x}) \ \prod_{i=1}^{3} \Theta(\epsilon_{i}\sigma_{i,1}) 
\end{split}
\end{equation}

For our purposes it is more convinient to write the above in terms of the independent color structures $c_{i}$ given in \eqref{colfacts}. We then arrive at
\begin{equation}
\label{ja4m14pt2}
\begin{split}
 & \mathscr{J}^{a_{4}}_{-1}(5)  \mathcal{M}(1^{- a_{1}}, 2^{-a_{2}} ,3^{+a_{3}},  5^{+a_{5} }) \\
 & =   i \bigg[ \frac{1}{z_{15}}  (- c_{5} + r \hspace{0.05cm} c_{6} ) + \frac{1}{z_{25}} (c_{5}-c_{2} - r \hspace{0.05cm} (c_{4}-c_{3}) ) + \frac{1}{z_{35}}  (c_{2}-c_{1} - r \hspace{0.05cm} (c_{6}-c_{4})) \bigg] \mathcal{N} \ \delta(x-\bar{x}) \ \prod_{i=1}^{3} \Theta(\epsilon_{i}\sigma_{i,1}) 
\end{split}
\end{equation}

Now it can be easily shown that the term appearing in the second line of \eqref{ord15pt4} is simply
\begin{equation}
\begin{split}
& \bigg[ \frac{ (c_{1} - r \ c_{3})}{z_{35}} + \left( -\frac{1}{z_{15}} + \frac{1}{z_{25}}\right) (  c_{6} -  c_{5}  + (1-r) (c_{3}-c_{4}))  + c_{2} \left( \frac{1}{z_{25}} - \frac{1}{z_{35}}\right) \bigg] \mathcal{N}\  \delta(x-\bar{x})  \ \prod_{i=1}^{3} \Theta(\epsilon_{i}\sigma_{i,1})  \\
 & = -  \bigg[ \frac{1}{z_{15}}  (- c_{5} + r \hspace{0.05cm} c_{6} ) + \frac{1}{z_{25}} (c_{5}-c_{2} - r \hspace{0.05cm} (c_{4}-c_{3}) ) + \frac{1}{z_{35}}  (c_{2}-c_{1} - r \hspace{0.05cm} (c_{6}-c_{4})) \bigg]\mathcal{N} \  \delta(x-\bar{x}) \ \prod_{i=1}^{3} \Theta(\epsilon_{i}\sigma_{i,1})  \\
 & =  i \mathscr{J}^{a_{4}}_{-1}(5)  \mathcal{M}(1^{- a_{1}}, 2^{-a_{2}} ,3^{+a_{3}},  5^{+a_{5} })
\end{split}
\end{equation}

In order to arrive at our desired result let us also note the following useful relation 
\begin{equation}
\label{ja5a4m1diff}
\begin{split}
&  i \left( \mathscr{J}^{a_{5}}_{-1}(5)\mathcal{M}(1^{- a_{1}}, 2^{-a_{2}} ,3^{+a_{3}},  5^{+a_{4} })-   \mathscr{J}^{a_{4}}_{-1} (5)\mathcal{M}(1^{- a_{1}}, 2^{-a_{2}} ,3^{+a_{3}},  5^{+a_{5} })\right) \\
  & =  \left[  - ( c_{1} -r \hspace{0.04cm} c_{3} ) \left(  \frac{1}{z_{15}}+  \frac{1}{z_{35}} \right) + r \ c_{3} \left(  \frac{1}{z_{25}} -  \frac{1}{z_{35}} \right)   \right] \mathcal{N} \ \delta(x-\bar{x}) \ \prod_{i=1}^{3} \Theta(\epsilon_{i}\sigma_{i,1}) 
\end{split}
\end{equation}

Then applying  \eqref{Lm14pt1}, \eqref{ja4m14pt2} and \eqref{ja5a4m1diff} in equation \eqref{ord15pt4} we finally get
\begin{equation}
\label{ord15ptfinal}
\begin{split}
&  \mathcal{M}(1^{- a_{1}}, 2^{-a_{2}} ,3^{+a_{3}},  4^{+a_{4} }, 5^{+a_{5} } )\Big|_{\mathcal{O}(z_{45}^{0}\bar{z}_{45}^{0})}   \\
& = -  i \hspace{0.05cm} B(\Delta_{4}-1, \Delta_{5}-1) \bigg[  \frac{\Delta_{4}-1}{(\Delta_{4}+\Delta_{5}-2)} \ f^{a_{4}a_{5}x}\mathcal{L}_{-1}(5) \mathcal{M}(1^{- a_{1}}, 2^{-a_{2}} ,3^{+a_{3}},  5^{+x })  \\
&  +  \frac{\Delta_{4}-1}{(\Delta_{4}+\Delta_{5}-2)} \   i \mathscr{J}^{a_{5}}_{-1}(5)\mathcal{M}(1^{- a_{1}}, 2^{-a_{2}} ,3^{+a_{3}},  5^{+a_{5} })  +  \frac{\Delta_{5}-1}{(\Delta_{4}+\Delta_{5}-2)}\    i \mathscr{J}^{a_{4}}_{-1}(5)\mathcal{M}(1^{- a_{1}}, 2^{-a_{2}} ,3^{+a_{3}},  5^{+a_{5} }) \bigg]
\end{split}
\end{equation}

The above result implies that upto the order under consideration here, the celestial OPE for positive helicity outgoing gluon primaries is given by
\begin{equation}
\label{glppopeord1}
\begin{split}
  \mathcal{O}^{ a_{4}}_{\Delta_{4},+}(z_{4},\bar{z}_{4})  \mathcal{O}^{a_{5}}_{\Delta_{5},+}(z_{5},\bar{z}_{5}) & \sim -  i \hspace{0.05cm}B(\Delta_{4}-1,\Delta_{5}-1)\bigg[ \frac{f^{a_{4} a_{5}x}}{z_{45}} +  \frac{(\Delta_{4}-1)}{(\Delta_{4}+\Delta_{5}-2)} f^{a_{4}a_{5}x} L_{-1} \\
  & \hspace{-0.5cm}+ i \left( \frac{ (\Delta_{4}-1)}{(\Delta_{4}+\Delta_{5}-2)} \delta^{a_{4} x}  \delta^{a_{5} y}+  \frac{ (\Delta_{4}-1)}{(\Delta_{4}+\Delta_{5}-2)} \delta^{a_{4} y} \delta^{a_{5} x} \right) j^{y}_{-1} \bigg] \mathcal{O}^{x}_{\Delta_{4}+\Delta_{5}-1,+}(z_{5},\bar{z}_{5})
\end{split}
\end{equation}

This matches with the corresponding result obtained  in \cite{Ebert:2020nqf}.

%%%%%%%%%%%%%%%%%%%%%%%%%%%%%%%%%%%%%%%%%%%%%%%%%%%%%%%%%%%%%%%%
\section{Mixed Helicity OPE from $4$-pt MHV gluon amplitude}
\label{pmope}

In this section, using the $4$-point Mellin amplitude, we will obtain the first subleading correction to the leading OPE between a positive helicity and a negative helicity gluon primary. 

\subsection{$4$-point Mellin amplitude}

Consider the color-dressed $4$-point MHV gluon amplitude $\mathcal{A}(1^{-a_{1}} , 2^{+a_{2}}, 3^{+a_{3}}, 4^{-a_{4}}  ) $. This is given by
\begin{equation}
\label{ymgl4pt1}
\begin{split}
& \mathcal{A}(1^{-a_{1}} , 2^{+a_{2}}, 3^{+a_{3}}, 4^{-a_{4}}  ) = (i g)^{2} \left( f^{a_{1}a_{2}x} f^{x a_{3} a_{4}} +  f^{a_{1}a_{3}x} f^{x a_{2}a_{4}} \ \frac{s_{34}}{s_{24}}  \right) A(1^{-} , 2^{+}, 3^{+}, 4^{-} ) 
\end{split}
\end{equation}

where 
\begin{equation}
\label{PTmhv4pt}
\begin{split}
& A(1^{-} , 2^{+}, 3^{+}, 4^{-} ) =  - \frac{\langle 1 4 \rangle^{3}}{\langle 1 2 \rangle \langle 2 3 \rangle \langle 3 4 \rangle }
\end{split}
\end{equation}

Using the $(\omega_{i},z_{i},\bar{z}_{i})$ parametrisation of null momenta,  we get from \eqref{ymgl4pt1} 
\begin{equation}
\label{ymgl4pt2}
\begin{split}
& \mathcal{A}(1^{-a_{1}} , 2^{+a_{2}}, 3^{+a_{3}}, 4^{-a_{4}} ) \\
&=  (i g)^{2} \left( f^{a_{1}a_{2}x} f^{x a_{3}a_{4}} +  f^{a_{1}a_{3}x} f^{x a_{2}a_{4}} \ \frac{\epsilon_{3}\omega_{3}}{\epsilon_{2}\omega_{2}} \ \frac{z_{34}\bar{z}_{34}}{z_{24}\bar{z}_{24}} \right) \frac{\omega_{1}\omega_{4}}{\omega_{2} \omega_{3}} \ \frac{z_{14}^{3}}{z_{12}z_{23}z_{34}}
\end{split}
\end{equation}

%%%%%%%%%%%%%%%%%%%%%%%%%%%%%%%%%%%%%%%%%%%%%%%%%%%%%%%%%%%%%%%%
%\subsection{Mellin transform} 

The modified Mellin transform of \eqref{ymgl4pt2} is then 
\begin{equation}
\label{mym4pt}
\begin{split}
 & \mathcal{M}(1^{- a_{1}}, 2^{+a_{2}} ,3^{+a_{3}}, 4^{-a_{4}}) \\
 & = \left \langle  \mathcal{O}^{a_{1}}_{\Delta_{1},-}(1)  \mathcal{O}^{a_{2}}_{\Delta_{2},+}(2)  \mathcal{O}^{a_{3}}_{\Delta_{3},+}(3)   \mathcal{O}^{a_{4}}_{\Delta_{4},-}(4)  \right\rangle  \\
 &  = \int_{0}^{\infty}  \prod_{i=1}^{4} d\omega_{i} \hspace{0.05cm} \omega_{i}^{\Delta_{i}-1}  \hspace{0.05cm} e^{- i\sum\limits_{i=1}^{4} \epsilon_{i}\omega_{i}u_{i}} \hspace{0.05cm}  \mathcal{A}(1^{-a_{1}} , 2^{+a_{2}}, 3^{+a_{3}}, 4^{-a_{4}} ) \hspace{0.05cm} \delta^{(4)} \left( \sum_{i=1}^{4} \epsilon_{i} \omega_{i}q_{i}^{\mu} \right)%, \quad \Delta_i = 1 + i\lambda_i
\end{split}
\end{equation}

where $\Delta_{i}=1+i\lambda_{i}$. Now we shall be interested in extracting the celestial OPE between $\mathcal{O}^{a_{3}}_{\Delta_{3},+}(3)  $ and $\mathcal{O}^{a_{4}}_{\Delta_{4},-}(4) $ from the Mellin amplitude \eqref{mym4pt}. We will also take both of them to be outgoing, i.e., $\epsilon_{3}=\epsilon_{4}=1$. Then let us make the following change of variables in \eqref{mym4pt}. 
\begin{equation}
\begin{split}
 \omega_{3} = \omega_{P} \hspace{0.04cm} t , \quad \omega_{4} = \omega_{P}(1-t)
\end{split}
\end{equation} 

The momentum conservation imposing delta function can be written as
\begin{equation}
\label{del4pt}
\begin{split}
  \delta^{(4)} \left(  \sum_{i =1}^{4} \varepsilon_{i}\omega_{i} q_{i}^{\mu}\right)  = & \frac{1}{ 4 \epsilon_{1} \omega_{1} \epsilon_{2}\omega_{2} z_{12}^{2}} \prod_{i=1}^{2} \delta\left( \omega_{i} -\omega_{i}^{*}\right)  \delta\left( \bar{z}_{14}- \frac{\omega_{P}t }{\epsilon_{1}\omega_{1}} \frac{z_{23}}{z_{12}} \bar{z}_{34}\right)\delta\left( \bar{z}_{24} +  \frac{\omega_{P}t}{\epsilon_{2}\omega_{2}} \frac{z_{13}}{z_{12}}\bar{z}_{34}\right)
\end{split}
\end{equation} 

where\footnote{In this section for simplicity of notation, we will often use the same symbols to denote similar quantities which also occur in Section \ref{ope5pt}.  In all such cases, to avoid confusion, we urge the reader to note the relevant definitions which have been explicitly provided in the respective sections.}
\begin{equation}
\label{omegaist}
\begin{split}
&\omega^{*}_{i} =  \epsilon_{i} \omega_{P} \left( \sigma_{i,1} + z_{34} \hspace{0.04cm} t \hspace{0.04cm} \sigma_{i,2}\right), \quad i=1,2
\end{split}
\end{equation} 

and 
\begin{equation}
\label{sigmadef}
\begin{split}
\sigma_{1,1} = \frac{z_{24}}{z_{12}}, \quad \sigma_{2,1} = - \frac{z_{14}}{z_{12}}, \quad \sigma_{1,2} = - \sigma_{2,2} = - \frac{1}{z_{12}}
\end{split}
\end{equation} 

Applying the delta function \eqref{del4pt}, the $\omega_{1},\omega_{2}$ integrals can be trivially done.  Then using \eqref{ymgl4pt2} and performing the integral over $\omega_{P}$,  the Mellin amplitude can be expressed as
\begin{equation}
\label{mym4pt1}
\begin{split}
& \mathcal{M}(1^{-a_{1}}, 2^{+a_{2}} ,3^{+a_{3}}, 4^{-a_{4}})= - i   \hspace{0.05cm} \mathcal{N} \int_{0}^{1} dt \hspace{0.1cm} t^{i\lambda_{3}-1}(1-t)^{i\lambda_{4}+1} \ \prod_{i=1}^{3} \Theta(\epsilon_{i}(\sigma_{i,1} +  z_{34} \hspace{0.03cm} t \hspace{0.03cm} \sigma_{i,2} )) \ \mathcal{I}(t) 
\end{split}
\end{equation}

where the theta functions above impose $(\frac{\omega_{i}^{*}}{\omega_{P}} )\ge 0$. The prefactor $\mathcal{N}$ and the integrand $\mathcal{I}(t)$ are given by
\begin{equation}
\label{Ngldef4pt}
\begin{split}
\mathcal{N} &=  i  \hspace{0.04cm} \epsilon_{1}\epsilon_{2} \    \frac{z_{14}^{3}}{z_{12}^{3}z_{24}} \ (\epsilon_{1}\sigma_{1,1})^{i\lambda_{1}} (\epsilon_{2}\sigma_{2,1})^{i\lambda_{2}-2}  \ \frac{\Gamma(i\Lambda)}{( i\hspace{0.04cm} \mathcal{U}_{1})^{i\Lambda}}  
\end{split}
\end{equation}

with $\Lambda = \sum\limits_{i=1}^{4}\lambda_{i}, \hspace{0.1cm} \mathcal{U}_{1} = \sum\limits_{i=1}^{2} \sigma_{i,1} u_{i4}$ and
\begin{equation}
\label{Igldef4pt}
\begin{split}
\mathcal{I} (t) =  & \frac{1}{z_{34}} \left(1-\frac{z_{34}}{z_{24}}\right)^{-1} \bigg[ f^{a_{1}a_{2}x} f^{x a_{3}a_{4}} -   \frac{z_{12}z_{34}}{z_{14}z_{24}} \left(1-\frac{z_{34}}{z_{14}}\right)^{-1}  f^{a_{1}a_{3}x} f^{x a_{2}a_{4}}\bigg]  \times \\
& \left( 1+ z_{34} \hspace{0.03cm} t \hspace{0.03cm} \frac{\sigma_{1,2}}{\sigma_{1,1}}   \right)^{i\lambda_{1}}  \times \left( 1+ z_{34} \hspace{0.03cm} t \hspace{0.03cm} \frac{\sigma_{2,2}}{\sigma_{2,1}}  \right)^{i\lambda_{2}-2}  \left[ 1+ \frac{t}{\mathcal{U}_{1}} \left( z_{34} \ \mathcal{U}_{2} + u_{34}\right)\right]^{-i\Lambda}  \times \\
&   \delta\left( \bar{z}_{14}- \frac{t }{\sigma_{1,1}} \left( 1+ z_{34} \ t \ \frac{\sigma_{1,2}}{\sigma_{1,1}} \right)^{-1} \frac{z_{23}}{z_{12}} \bar{z}_{34}\right)\delta\left( \bar{z}_{24} +  \frac{t}{\sigma_{2,1}}  \left( 1+ z_{34} \ t \ \frac{\sigma_{2,2}}{\sigma_{2,1}} \right)^{-1}\frac{z_{13}}{z_{12}}\bar{z}_{34}\right)
\end{split}
\end{equation}

In \eqref{mym4pt1} we have again set the Yang-Mills coupling $g=2$. Now let us also note the expression of the $3$-point Mellin amplitude in terms of which the $4$-point Mellin amplitude factorises  in the $(3^{+},4^{-})$ OPE channel. This can be easily derived and is given by
\begin{equation}
\label{mym3pt}
\begin{split}
 \mathcal{M}(1^{- a_{1}}, 2^{+a_{2}} ,4^{- x}) & = \left \langle  \mathcal{O}^{a_{1}}_{\Delta_{1},-}(1)  \mathcal{O}^{a_{2}}_{\Delta_{2},+}(2)  \mathcal{O}^{ x}_{\Delta_{3}+\Delta_{4}-1,-}(4)  \right\rangle   \\
 & = f^{a_{1}a_{2}x} \hspace{0.04cm} \mathcal{N} \hspace{0.04cm}  \delta(\bar{z}_{14}) \delta(\bar{z}_{24}) \ \prod_{i=1}^{3} \Theta(\epsilon_{i}(\sigma_{i,1} ))
\end{split}
\end{equation}

with $\mathcal{N}$ defined in \eqref{Ngldef4pt}.

%%%%%%%%%%%%%%%%%%%%%%%%%%%%%%%%%%%%%%%%%%%%%%%%%%%%%%%%%%%%%%%%
\subsection{OPE decomposition of $4$-point Mellin amplitude} 

We can now extract the mixed helicity celestial OPE in the $(3^{+},4^{-})$ channel by setting $u_{34}=0$ in \eqref{mym4pt1}  and expanding the integrand $\mathcal{I}(t)$ around $z_{34} = 0, \bar{z}_{34} = 0$. As before, in order to avoid irrelevant contact terms we will simply set $z_{34}=0$ inside the argument of the theta functions in the Mellin integral in \eqref{mym4pt1}.  

\subsubsection{Leading term}

The leading term in \eqref{mym4pt1} as $z_{34} \rightarrow 0$ is
\begin{equation}
\label{ordzinv4pt}
\begin{split}
 \mathcal{M}(1^{- a_{1}}, 2^{+a_{2}} ,3^{+a_{3}}, 4^{-a_{4}}) & \approx - i  \hspace{0.05cm} \mathcal{N} \int_{0}^{1} dt \hspace{0.1cm} t^{i\lambda_{3}-1}(1-t)^{i\lambda_{4}+1}  \hspace{0.05cm} f^{a_{1}a_{2}x} f^{x a_{3}a_{4}} \hspace{0.04cm} \frac{ \delta(\bar{z}_{14}) \delta(\bar{z}_{24})}{z_{34}} \ \prod_{i=1}^{3} \Theta(\epsilon_{i}(\sigma_{i,1} )) \\
& = -   \frac{i f^{a_{3}a_{4}x}}{z_{34}} \hspace{0.05cm} B(\Delta_{3}-1, \Delta_{4}+1)  \hspace{0.05cm} \mathcal{M}(1^{- a_{1}}, 2^{+a_{2}} ,4^{- x}) 
\end{split}
\end{equation}

From the above result we get the leading celestial OPE to be 
\begin{equation}
\label{pmleadope}
\begin{split}
  \mathcal{O}^{ a_{3}}_{\Delta_{3},+}(z_{3},\bar{z}_{3})  \mathcal{O}^{a_{4}}_{\Delta_{4},-}(z_{4},\bar{z}_{4}) & \sim -   \frac{i f^{a_{3}a_{4}x}}{z_{34}} \hspace{0.05cm} B(\Delta_{3}-1, \Delta_{4}+1)  \hspace{0.05cm}  \mathcal{O}^{x}_{\Delta_{3}+\Delta_{4}-1,-}(z_{4},\bar{z}_{4})
\end{split}
\end{equation}

This agrees with the result obtained in \cite{Pate:2019lpp,Fan:2019emx}.

\subsubsection{Subleading terms: $\mathcal{O}(z_{34}^{0}\bar{z}_{34}^{0})$ } 

Using \eqref{Igldef4pt}, we find the $\mathcal{O}(z_{34}^{0}\bar{z}_{34}^{0})$ term from the $4$-pt Mellin amplitude to be 
\begin{equation}
\label{ord14pt}
\begin{split}
& \mathcal{M}(1^{- a_{1}}, 2^{+a_{2}} ,3^{+a_{3}}, 4^{-a_{4}})\Big|_{\mathcal{O}(z_{34}^{0}\bar{z}_{34}^{0})}\\
 & = - i  \hspace{0.05cm} B(\Delta_{3}-1, \Delta_{4}+1) \bigg[ \frac{i\lambda_{3}}{(i\lambda_{3}+i\lambda_{4}+2)} f^{a_{1}a_{2}x} f^{{x}a_{3}a_{4}}  \bigg( - \frac{ i\lambda_{1}}{z_{24}} - \frac{(i\lambda_{2}-2)}{z_{14}}  +   \frac{ i \Lambda}{\mathcal{U}_{1}} \hspace{0.05cm} \frac{ u_{12}}{z_{12}}   \bigg) \\
 & +  \left( - \frac{f^{a_{1}a_{4}x} f^{{x}a_{2}a_{3}}}{z_{24}} + \frac{ f^{a_{1}a_{3}x} f^{{x}a_{2}a_{4}}}{z_{14}}  \right)\bigg] \widetilde{\mathcal{M}}(1^{-}, 2^{+} ,4^{-})
\end{split}
\end{equation}

where
\begin{equation}
\label{3ptmellin}
\begin{split}
& \widetilde{\mathcal{M}}(1^{-}, 2^{+} ,4^{-}) = \mathcal{N} \hspace{0.05cm}  \delta(\bar{z}_{14}) \delta(\bar{z}_{24}) \ \prod_{i=1}^{3} \Theta(\epsilon_{i}(\sigma_{i,1} ))
 \end{split}
\end{equation}

is the modified Mellin transform of the color-stripped $3$-point amplitude $A(1^{- }, 2^{+} ,4^{-})$. 

\vspace{0.1cm}
Now  the R.H.S. of  \eqref{ord14pt} can be expressed in terms of a linear combination of $3$-point Mellin amplitudes which involve descendants created by the action of $J^{a}_{-1}$ and $j^{a}_{-1}$ respectively. In order to show this, let us first consider the correlator
\begin{equation}
\label{Ja3m13pt}
\begin{split}
& \left \langle  \mathcal{O}^{a_{1}}_{\Delta_{1},-}(1)  \mathcal{O}^{a_{2}}_{\Delta_{2},+}(2)    \left( J^{a_{3}}_{-1}\mathcal{O}^{a_{4}}_{\Delta_{3}+\Delta_{4},-}(4) \right)  \right\rangle  =  \mathcal{J}^{a_{3}}_{-1}(4) \mathcal{P}_{-1,-1}(4) \mathcal{M}(1^{- a_{1}}, 2^{+a_{2}} ,4^{-a_{4}}) 
 \end{split}
\end{equation}

where 
\begin{equation}
\label{Jmin13pt}
\begin{split}
  & \mathcal{J}^{a_{3}}_{-1}(4) \mathcal{P}_{-1,-1}(4) \mathcal{M}(1^{- a_{1}}, 2^{+a_{2}} ,4^{-a_{4}}) \\
  &  = \sum_{i=1}^{2}    \frac{ \epsilon_{i}\left(2\bar h_i -1 + u_{i4} \partial_{u_{i}} + \bar z_{i4} \partial_{\bar{z}_{i}}\right)}{2 \hspace{0.05cm}z_{i4}} \hspace{0.05cm}T^{a_{3}}_i P^{-1}_i  \hspace{0.05cm} \mathcal{P}_{-1,-1}(4) \mathcal{M}(1^{- a_{1}}, 2^{+a_{2}} ,4^{-a_{4}}) 
%& = - i \bigg[ \bigg(i \lambda_{1} f^{a_{3}a_{1}y} f^{ya_{2}a_{4}} - (i\lambda_{2}-2) f^{a_{3}a_{2}y} f^{a_{1}ya_{4}}\bigg) \frac{z_{12}}{z_{14}z_{24}} \\
%& - \frac{i \Lambda}{\mathcal{U}_{1}} \left( \frac{u_{14}}{z_{14}} f^{a_{3}a_{1}y} f^{ya_{2}a_{4}} + \frac{u_{24}}{z_{24}} f^{a_{3}a_{2}y} f^{a_{1} y a_{4}} \right) \bigg] \widetilde{\mathcal{M}}(1^{-}, 2^{+} ,4^{-})
\end{split}
\end{equation}
and 
\begin{equation}
\begin{split}
\mathcal{P}_{-1,-1}(4) \mathcal{M}(1^{- a_{1}}, 2^{+a_{2}} ,4^{-a_{4}}) = i \partial_{u_{4}}  \mathcal{M}(1^{- a_{1}}, 2^{+a_{2}} ,4^{-a_{4}}) 
\end{split}
\end{equation}

%$\mathcal{P}_{-1,-1} \mathcal{M}(1^{- a_{1}}, 2^{+a_{2}} ,4^{-a_{4}}) = i \partial_{u_{4}}  \mathcal{M}(1^{- a_{1}}, 2^{+a_{2}} ,4^{-a_{4}}) $. 
Note that since we are using here the modified Mellin transform, the differential operator representation of $J^{a_{3}}_{-1}$ explicitly includes the $u_{i}$ coordinates. This is obtained from \eqref{1} by making the replacement 
\begin{equation}
\begin{split}
2\bar{h}_{i} \rightarrow 2\bar{h}_{i}+ u_{i4} \partial_{u_{i}}. 
\end{split}
\end{equation}

In order to evaluate \eqref{Jmin13pt}, it is useful to note that since the derivatives with respect to $\bar{z}_{i}$  act only on the anti-holomorphic delta functions in the $3$-point Mellin amplitude, we can use the identity
\begin{equation}
\label{delbarderv}
\begin{split}
& \bar{z}_{i4} \hspace{0.05cm} \partial_{\bar{z}_{i}} \left(\delta(\bar{z}_{14}) \delta(\bar{z}_{24}) \right) = - \delta(\bar{z}_{14}) \delta(\bar{z}_{24}), \quad i=1,2 
\end{split}
\end{equation}

Then it can be easily shown that 
\begin{equation}
\label{Jmin13pt1}
\begin{split}
   \mathcal{J}^{a_{3}}_{-1}(4) \mathcal{P}_{-1,-1}(4) \mathcal{M}(1^{- a_{1}}, 2^{+a_{2}} ,4^{-a_{4}}) & = - i \bigg[ \bigg(i \lambda_{1} f^{a_{3}a_{1}y} f^{ya_{2}a_{4}} - (i\lambda_{2}-2) f^{a_{3}a_{2}y} f^{a_{1}ya_{4}}\bigg) \frac{z_{12}}{z_{14}z_{24}} \\
   & - \frac{i \Lambda}{\mathcal{U}_{1}} \left( \frac{u_{14}}{z_{14}} f^{a_{3}a_{1}y} f^{ya_{2}a_{4}} + \frac{u_{24}}{z_{24}} f^{a_{3}a_{2}y} f^{a_{1} y a_{4}} \right) \bigg] \widetilde{\mathcal{M}}(1^{-}, 2^{+} ,4^{-})
\end{split}
\end{equation}

Now let us consider the following $3$-point correlator
\begin{equation}
\label{jm13pt}
\begin{split}
& \left \langle  \mathcal{O}^{a_{1}}_{\Delta_{1},-}(1)  \mathcal{O}^{a_{2}}_{\Delta_{2},+}(2)     \left( j^{a_{3}}_{-1}\mathcal{O}^{a_{4}}_{\Delta_{3}+\Delta_{4}-1,-}(4) \right)  \right\rangle  =  \mathscr{J}^{a_{3}}_{-1}(4) \mathcal{M}(1^{- a_{1}}, 2^{+a_{2}} ,4^{-a_{4}}) 
 \end{split}
\end{equation}

Using \eqref{jamp} this is simply given by
\begin{equation}
\label{jm13pt1}
\begin{split}
&  \mathscr{J}^{a_{3}}_{-1} (4)\mathcal{M}(1^{- a_{1}}, 2^{+a_{2}} ,4^{-a_{4}}) = i  \left(  \frac{ f^{a_{3}a_{1}x} f^{{x}a_{2}a_{4}}}{z_{14}}  + \frac{f^{a_{3} a_{2}x} f^{a_{1} x a_{4}} }{z_{24}}   \right) \widetilde{\mathcal{M}}(1^{-}, 2^{+} ,4^{-})
\end{split}
\end{equation}

Then using \eqref{Jmin13pt1} and \eqref{jm13pt1} we get the following relation
\begin{equation}
\label{ord14pt2}
\begin{split}
&  f^{a_{1}a_{2}x} f^{{x}a_{3}a_{4}}    \bigg( - \frac{ i\lambda_{1}}{z_{24}} - \frac{(i\lambda_{2}-2)}{z_{14}}  +   \frac{ i \Lambda}{\mathcal{U}_{1}} \hspace{0.05cm} \frac{ u_{12}}{z_{12}}   \bigg)  \widetilde{\mathcal{M}}(1^{-}, 2^{+} ,4^{-}) \\
& = i  \mathcal{J}^{a_{3}}_{-1}(4) \mathcal{P}_{-1,-1} (4)\mathcal{M}(1^{- a_{1}}, 2^{+a_{2}} ,4^{-a_{4}}) + i (2+ i\lambda_{3}+i\lambda_{4}) \mathscr{J}^{a_{3}}_{-1}(4) \mathcal{M}(1^{- a_{1}}, 2^{+a_{2}} ,4^{-a_{4}})
\end{split}
\end{equation}

Gathering the above results, \eqref{ord14pt} can finally be written as
\begin{equation}
\label{ord14ptfinal}
\begin{split}
& \mathcal{M}(1^{- a_{1}}, 2^{+a_{2}} ,3^{+a_{3}}, 4^{-a_{4}})\Big|_{\mathcal{O}(z_{34}^{0}\bar{z}_{34}^{0})}\\
& = - i   \hspace{0.05cm} B(\Delta_{3}-1, \Delta_{4}+1)  \bigg[ \frac{\Delta_{3}-1}{(\Delta_{3}+\Delta_{4})}  \hspace{0.05cm} i \mathcal{J}^{a_{3}}_{-1}(4) \mathcal{P}_{-1,-1}(4)  +  i \hspace{0.05cm}\Delta_{3} \mathscr{J}^{a_{3}}_{-1}(4) \bigg] \mathcal{M}(1^{- a_{1}}, 2^{+a_{2}} ,4^{-a_{4}})
\end{split}
\end{equation}

Thus upto the first the subleading order, the celestial OPE between outgoing gluon primaries of opposite helicities is given by
\begin{equation}
\label{glpmopeord1}
\begin{split}
&  \mathcal{O}^{ a_{3}}_{\Delta_{3},+}(z_{3},\bar{z}_{3})  \mathcal{O}^{a_{4}}_{\Delta_{4},-}(z_{4},\bar{z}_{4}) \\
& \sim -  i \hspace{0.05cm}B(\Delta_{3}-1,\Delta_{4}+1)\bigg[ \frac{f^{a_{3} a_{4}x}}{z_{34}} + \frac{i (\Delta_{3}-1)}{(\Delta_{3}+\Delta_{4})} \hspace{0.04cm}   \delta^{a_{4}x }  J^{a_{3}}_{-1}P_{-1,-1} + i  \hspace{0.04cm} \Delta_{3}  \hspace{0.05cm} \delta^{a_{4}x }\hspace{0.05cm} j^{a_{3}}_{-1} \bigg] \mathcal{O}^{x}_{\Delta_{3}+\Delta_{4}-1,-}(z_{4},\bar{z}_{4})
\end{split}
\end{equation}
where $P_{-1,-1}\mathcal{O}^{x}_{\Delta_{3}+\Delta_{4}-1,-}= \mathcal{O}^{x}_{\Delta_{3}+\Delta_{4},-} $. The above OPE manifestly satisfies both the leading as well as the subleading conformal soft gluon theorem.

%%%%%%%%%%%%%%%%%%%%%%%%%%%%%%%%%%%%%%%%%%%%%%%%%%%%%%%%%%%%%%%%
%\subsubsection{Subleading terms: $\mathcal{O}(\bar{z}_{45})$}

\end{appendices}
%%%%%%%%%%%%%%%%%%%%%%%%%%%%%%%%%%%%%%%%%%%%%%%%%%%%%%%%%%%%%%


\begin{thebibliography}{99}
  
% \bibitem{Weinberg:1965nx} 
%  S.~Weinberg,
 % ``Infrared photons and gravitons,''
%  Phys.\ Rev.\  {\bf 140}, B516 (1965).
%  doi:10.1103/PhysRev.140.B516

 \bibitem{Strominger:2013lka} 
A.~Strominger,
``Asymptotic Symmetries of Yang-Mills Theory,''
JHEP {\bf 1407}, 151 (2014)
doi:10.1007/JHEP07(2014)151
[arXiv:1308.0589 [hep-th]].

\bibitem{Strominger:2013jfa} 
  A.~Strominger,
  ``On BMS Invariance of Gravitational Scattering,''
  JHEP {\bf 1407}, 152 (2014)
  doi:10.1007/JHEP07(2014)152
  [arXiv:1312.2229 [hep-th]].
   
\bibitem{Pate:2019lpp} 
  M.~Pate, A.~M.~Raclariu, A.~Strominger and E.~Y.~Yuan,
  ``Celestial Operator Products of Gluons and Gravitons,''
  arXiv:1910.07424 [hep-th].
    
\bibitem{Law:2019glh} 
  Y.~T.~A.~Law and M.~Zlotnikov,
  ``Poincare Constraints on Celestial Amplitudes,''
  arXiv:1910.04356 [hep-th].  
  
 \bibitem{Banerjee:2020zlg}
S.~Banerjee, S.~Ghosh and P.~Paul,
``MHV Graviton Scattering Amplitudes and Current Algebra on the Celestial Sphere,''
[arXiv:2008.04330 [hep-th]].
  
\bibitem{Himwich:2020rro}
E.~Himwich, S.~A.~Narayanan, M.~Pate, N.~Paul and A.~Strominger,
``The Soft $\mathcal{S}$-Matrix in Gravity,''
JHEP \textbf{09}, 129 (2020)
doi:10.1007/JHEP09(2020)129
[arXiv:2005.13433 [hep-th]].

 \bibitem{Fotopoulos:2019vac} 
  A.~Fotopoulos, S.~Stieberger, T.~R.~Taylor and B.~Zhu,
  ``Extended BMS Algebra of Celestial CFT,''
  arXiv:1912.10973 [hep-th].
  
\bibitem{Banerjee:2020kaa}
S.~Banerjee, S.~Ghosh and R.~Gonzo,
``BMS symmetry of celestial OPE,''
JHEP \textbf{04}, 130 (2020)
doi:10.1007/JHEP04(2020)130
[arXiv:2002.00975 [hep-th]].

\bibitem{Ebert:2020nqf}
S.~Ebert, A.~Sharma and D.~Wang,
``Descendants in celestial CFT and emergent multi-collinear factorization,''
[arXiv:2009.07881 [hep-th]].
  
\bibitem{Cachazo:2014fwa} 
 F.~Cachazo and A.~Strominger,
  ``Evidence for a New Soft Graviton Theorem,''
  arXiv:1404.4091 [hep-th].

\bibitem{Strominger:2017zoo}
A.~Strominger,
``Lectures on the Infrared Structure of Gravity and Gauge Theory,''
[arXiv:1703.05448 [hep-th]].
  
  \bibitem{Kapec:2014opa} 
  D.~Kapec, V.~Lysov, S.~Pasterski and A.~Strominger,
  ``Semiclassical Virasoro symmetry of the quantum gravity $ \mathcal{S}$-matrix,''
  JHEP {\bf 1408}, 058 (2014)
  doi:10.1007/JHEP08(2014)058
  [arXiv:1406.3312 [hep-th]].
  
  \bibitem{Kapec:2016jld}
  D.~Kapec, P.~Mitra, A.~M.~Raclariu and A.~Strominger,
  ``2D Stress Tensor for 4D Gravity,''
  Phys.\ Rev.\ Lett.\  {\bf 119}, no. 12, 121601 (2017)
  doi:10.1103/PhysRevLett.119.121601
  [arXiv:1609.00282 [hep-th]].
  
  \bibitem{He:2017fsb} 
  T.~He, D.~Kapec, A.~M.~Raclariu and A.~Strominger,
  ``Loop-Corrected Virasoro Symmetry of 4D Quantum Gravity,''
  JHEP {\bf 1708}, 050 (2017)
  doi:10.1007/JHEP08(2017)050
  [arXiv:1701.00496 [hep-th]].
 
   \bibitem{Ball:2019atb} 
  A.~Ball, E.~Himwich, S.~A.~Narayanan, S.~Pasterski and A.~Strominger,
  ``Uplifting AdS3/CFT2 to Flat Space Holography,''
 arXiv:1905.09809 [hep-th].
 
  \bibitem{Kapec:2017gsg} 
D.~Kapec and P.~Mitra,
``A $d$-Dimensional Stress Tensor for Mink$_{d+2}$ Gravity,''
arXiv:1711.04371 [hep-th]. 
 
  \bibitem{Banerjee:2019aoy}
S.~Banerjee, P.~Pandey and P.~Paul,
``Conformal properties of soft-operators - 1 : Use of null-states,''
Phys. Rev. D \textbf{101}, no.10, 106014 (2020)
doi:10.1103/PhysRevD.101.106014
[arXiv:1902.02309 [hep-th]].

\bibitem{Banerjee:2019tam}
S.~Banerjee and P.~Pandey,
``Conformal properties of soft-operators. Part II. Use of null-states,''
JHEP \textbf{02}, 067 (2020)
doi:10.1007/JHEP02(2020)067
[arXiv:1906.01650 [hep-th]].

\bibitem{Donnay:2020guq}
L.~Donnay, S.~Pasterski and A.~Puhm,
``Asymptotic Symmetries and Celestial CFT,''
[arXiv:2005.08990 [hep-th]].

\bibitem{Campiglia:2014yka} 
  M.~Campiglia and A.~Laddha,
  ``Asymptotic symmetries and subleading soft graviton theorem,''
  Phys.\ Rev.\ D {\bf 90}, no. 12, 124028 (2014)
  doi:10.1103/PhysRevD.90.124028
  [arXiv:1408.2228 [hep-th]].
    
 \bibitem{Campiglia:2020qvc}
M.~Campiglia and J.~Peraza,
``Generalized BMS charge algebra,''
Phys. Rev. D \textbf{101}, no.10, 104039 (2020)
doi:10.1103/PhysRevD.101.104039
[arXiv:2002.06691 [gr-qc]].

 \bibitem{Compere:2018ylh}
G.~Comp\`ere, A.~Fiorucci and R.~Ruzziconi,
``Superboost transitions, refraction memory and super-Lorentz charge algebra,''
JHEP \textbf{11}, 200 (2018)
[erratum: JHEP \textbf{04}, 172 (2020)]
doi:10.1007/JHEP11(2018)200
[arXiv:1810.00377 [hep-th]].
  
  \bibitem{Barnich:2009se} 
  G.~Barnich and C.~Troessaert,
  ``Symmetries of asymptotically flat 4 dimensional spacetimes at null infinity revisited,''
 Phys.\ Rev.\ Lett.\  {\bf 105}, 111103 (2010)
  doi:10.1103/PhysRevLett.105.111103
  [arXiv:0909.2617 [gr-qc]]. 
  
   \bibitem{Barnich:2011ct} 
  G.~Barnich and C.~Troessaert,
  ``Supertranslations call for superrotations,''
  PoS CNCFG {\bf 2010}, 010 (2010)
  [Ann.\ U.\ Craiova Phys.\  {\bf 21}, S11 (2011)]
  [arXiv:1102.4632 [gr-qc]]. 
  
  \bibitem{Bagchi:2016bcd} 
  A.~Bagchi, R.~Basu, A.~Kakkar and A.~Mehra,
  ``Flat Holography: Aspects of the dual field theory,''
  JHEP {\bf 1612}, 147 (2016)
  doi:10.1007/JHEP12(2016)147
 [arXiv:1609.06203 [hep-th]].
  
   \bibitem{Sachs:1962zza} 
  R.~Sachs,
  ``Asymptotic symmetries in gravitational theory,''
  Phys.\ Rev.\  {\bf 128}, 2851 (1962).
  doi:10.1103/PhysRev.128.2851

   \bibitem{Bondi:1962px} 
  H.~Bondi, M.~G.~J.~van der Burg and A.~W.~K.~Metzner,
  ``Gravitational waves in general relativity. 7. Waves from axisymmetric isolated systems,''
  Proc.\ Roy.\ Soc.\ Lond.\ A {\bf 269}, 21 (1962).
  doi:10.1098/rspa.1962.0161
  
%\bibitem{Sachs:1962wk} 
  R.~K.~Sachs,
  ``Gravitational waves in general relativity. 8. Waves in asymptotically flat space-times,''
  Proc.\ Roy.\ Soc.\ Lond.\ A {\bf 270}, 103 (1962).
  doi:10.1098/rspa.1962.0206
      
  \bibitem{He} 
  T.~He, V.~Lysov, P.~Mitra and A.~Strominger,
  ``BMS supertranslations and Weinberg's soft graviton theorem,''
  JHEP {\bf 1505}, 151 (2015)
  doi:10.1007/JHEP05(2015)151
  [arXiv:1401.7026 [hep-th]].
  
  \bibitem{Strominger:2014pwa} 
 A.~Strominger and A.~Zhiboedov,
 ``Gravitational Memory, BMS Supertranslations and Soft Theorems,''
 JHEP {\bf 1601}, 086 (2016)
doi:10.1007/JHEP01(2016)086
[arXiv:1411.5745 [hep-th]].

  \bibitem{Pasterski:2016qvg} 
  S.~Pasterski, S.~H.~Shao and A.~Strominger,
  ``Flat Space Amplitudes and Conformal Symmetry of the Celestial Sphere,''
  Phys.\ Rev.\ D {\bf 96}, no. 6, 065026 (2017)
  doi:10.1103/PhysRevD.96.065026
  [arXiv:1701.00049 [hep-th]].

\bibitem{Pasterski:2017kqt} 
  S.~Pasterski and S.~H.~Shao,
  ``Conformal basis for flat space amplitudes,''
  Phys.\ Rev.\ D {\bf 96}, no. 6, 065022 (2017)
  doi:10.1103/PhysRevD.96.065022
  [arXiv:1705.01027 [hep-th]].  
   
  \bibitem{Cheung:2016iub} 
  C.~Cheung, A.~de la Fuente and R.~Sundrum,
  ``4D scattering amplitudes and asymptotic symmetries from 2D CFT,''
  JHEP {\bf 1701}, 112 (2017)
  doi:10.1007/JHEP01(2017)112
  [arXiv:1609.00732 [hep-th]].
      
  \bibitem{deBoer:2003vf} 
  J.~de Boer and S.~N.~Solodukhin,
  ``A Holographic reduction of Minkowski space-time,''
  Nucl.\ Phys.\ B {\bf 665}, 545 (2003)
  doi:10.1016/S0550-3213(03)00494-2
  [hep-th/0303006]. 
  
  \bibitem{Banerjee:2018gce} 
  S.~Banerjee,
  ``Null Infinity and Unitary Representation of The Poincare Group,''
  JHEP {\bf 1901}, 205 (2019)
  doi:10.1007/JHEP01(2019)205
  [arXiv:1801.10171 [hep-th]].
  
  \bibitem{Banerjee:2019prz}
S.~Banerjee, S.~Ghosh, P.~Pandey and A.~P.~Saha,
``Modified celestial amplitude in Einstein gravity,''
JHEP \textbf{03}, 125 (2020)
doi:10.1007/JHEP03(2020)125
[arXiv:1909.03075 [hep-th]].
  
  \bibitem{Banerjee:2018fgd} 
  S.~Banerjee,
  ``Symmetries of free massless particles and soft theorems,''
  Gen.\ Rel.\ Grav.\  {\bf 51}, no. 9, 128 (2019)
  doi:10.1007/s10714-019-2609-z
  [arXiv:1804.06646 [hep-th]].
  
 \bibitem{Pasterski:2017ylz} 
  S.~Pasterski, S.~H.~Shao and A.~Strominger,
  ``Gluon Amplitudes as 2d Conformal Correlators,''
  Phys.\ Rev.\ D {\bf 96}, no. 8, 085006 (2017)
  doi:10.1103/PhysRevD.96.085006
  [arXiv:1706.03917 [hep-th]].
  
  \bibitem{Schreiber:2017jsr} 
  A.~Schreiber, A.~Volovich and M.~Zlotnikov,
  ``Tree-level gluon amplitudes on the celestial sphere,''
  arXiv:1711.08435 [hep-th]. 
  
  \bibitem{Cardona:2017keg} 
  C.~Cardona and Y.~t.~Huang,
  ``S-matrix singularities and CFT correlation functions,''
  JHEP {\bf 1708}, 133 (2017)
  doi:10.1007/JHEP08(2017)133
  [arXiv:1702.03283 [hep-th]].
 
   \bibitem{Lam:2017ofc} 
  H.~T.~Lam and S.~H.~Shao,
  ``Conformal Basis, Optical Theorem, and the Bulk Point Singularity,''
  arXiv:1711.06138 [hep-th].
    
  \bibitem{Banerjee:2017jeg} 
  N.~Banerjee, S.~Banerjee, S.~Atul Bhatkar and S.~Jain,
  ``Conformal Structure of Massless Scalar Amplitudes Beyond Tree level,''
  arXiv:1711.06690 [hep-th].
  
   \bibitem{Stieberger:2018edy} 
  S.~Stieberger and T.~R.~Taylor,
  ``Strings on Celestial Sphere,''
  Nucl.\ Phys.\ B {\bf 935}, 388 (2018)
  doi:10.1016/j.nuclphysb.2018.08.019
  [arXiv:1806.05688 [hep-th]]. 
  
  \bibitem{Stieberger:2018onx} 
  S.~Stieberger and T.~R.~Taylor,
  ``Symmetries of Celestial Amplitudes,''
  Phys.\ Lett.\ B {\bf 793}, 141 (2019)
  doi:10.1016/j.physletb.2019.03.063
  [arXiv:1812.01080 [hep-th]].
                        
  \bibitem{Fotopoulos:2019tpe} 
  A.~Fotopoulos and T.~R.~Taylor,
  ``Primary Fields in Celestial CFT,''
  arXiv:1906.10149 [hep-th].
  
   \bibitem{Albayrak:2020saa}
S.~Albayrak, C.~Chowdhury and S.~Kharel,
``On loop celestial amplitudes for gauge theory and gravity,''
[arXiv:2007.09338 [hep-th]].

%\cite{Casali:2020vuy}
\bibitem{Casali:2020vuy}
E.~Casali and A.~Puhm,
``A Double Copy for Celestial Amplitudes,''
[arXiv:2007.15027 [hep-th]].


%\cite{Law:2020xcf}
\bibitem{Law:2020xcf}
Y.~T.~A.~Law and M.~Zlotnikov,
``Relativistic partial waves for celestial amplitudes,''
[arXiv:2008.02331 [hep-th]].

%\cite{Gonzalez:2020tpi}
\bibitem{Gonzalez:2020tpi}
H.~A.~Gonz\'alez, A.~Puhm and F.~Rojas,
``Loops on the Celestial Sphere,''
[arXiv:2009.07290 [hep-th]].

%\cite{Muck:2020wtx}
\bibitem{Muck:2020wtx}
L.~Iacobacci and W.~M\"uck,
``Conformal Primary Basis for Dirac Spinors,''
[arXiv:2009.02938 [hep-th]].

%\cite{Narayanan:2020amh}
\bibitem{Narayanan:2020amh}
S.~A.~Narayanan,
``Massive Celestial Fermions,''
[arXiv:2009.03883 [hep-th]].
  
  \bibitem{Donnay:2018neh} 
  L.~Donnay, A.~Puhm and A.~Strominger,
  ``Conformally Soft Photons and Gravitons,''
  JHEP {\bf 1901}, 184 (2019)
  doi:10.1007/JHEP01(2019)184
  [arXiv:1810.05219 [hep-th]].
      
 \bibitem{Pate:2019mfs} 
  M.~Pate, A.~M.~Raclariu and A.~Strominger,
  ``Conformally Soft Theorem in Gauge Theory,''
  arXiv:1904.10831 [hep-th].
  
  \bibitem{Fan:2019emx} 
  W.~Fan, A.~Fotopoulos and T.~R.~Taylor,
  ``Soft Limits of Yang-Mills Amplitudes and Conformal Correlators,''
  JHEP {\bf 1905}, 121 (2019)
  doi:10.1007/JHEP05(2019)121
  [arXiv:1903.01676 [hep-th]]. 
  
  \bibitem{Nandan:2019jas} 
  D.~Nandan, A.~Schreiber, A.~Volovich and M.~Zlotnikov,
  ``Celestial Amplitudes: Conformal Partial Waves and Soft Limits,''
  arXiv:1904.10940 [hep-th].
     
  \bibitem{Adamo:2019ipt} 
  T.~Adamo, L.~Mason and A.~Sharma,
  ``Celestial amplitudes and conformal soft theorems,''
  arXiv:1905.09224 [hep-th]. 
    
  \bibitem{Puhm:2019zbl} 
  A.~Puhm,
  ``Conformally Soft Theorem In Gravity,''
 arXiv:1905.09799 [hep-th].
 
 \bibitem{Guevara:2019ypd} 
  A.~Guevara,
  ``Notes on Conformal Soft Theorems and Recursion Relations in Gravity,''
  arXiv:1906.07810 [hep-th].
         
  \bibitem{Bern:1998sv} 
  Z.~Bern, L.~J.~Dixon, M.~Perelstein and J.~S.~Rozowsky,
  ``Multileg one loop gravity amplitudes from gauge theory,''
  Nucl.\ Phys.\ B {\bf 546}, 423 (1999)
  doi:10.1016/S0550-3213(99)00029-2
  [hep-th/9811140].
  
%  \bibitem{White:2011yy} 
%  C.~D.~White,
%  ``Factorization Properties of Soft Graviton Amplitudes,''
%  JHEP {\bf 1105}, 060 (2011)
%  doi:10.1007/JHEP05(2011)060
%  [arXiv:1103.2981 [hep-th]]. 
  
%  \bibitem{Akhoury:2011kq} 
%  R.~Akhoury, R.~Saotome and G.~Sterman,
%  ``Collinear and Soft Divergences in Perturbative Quantum Gravity,''
%  Phys.\ Rev.\ D {\bf 84}, 104040 (2011)
%  doi:10.1103/PhysRevD.84.104040
%  [arXiv:1109.0270 [hep-th]].
  
  \bibitem{Nandan:2016ohb} 
  D.~Nandan, J.~Plefka and W.~Wormsbecher,
  ``Collinear limits beyond the leading order from the scattering equations,''
  JHEP {\bf 1702}, 038 (2017)
  doi:10.1007/JHEP02(2017)038
  [arXiv:1608.04730 [hep-th]].
      
  
%\bibitem{Strominger:2013lka} 
%  A.~Strominger,
%  ``Asymptotic Symmetries of Yang-Mills Theory,''
%  JHEP {\bf 1407}, 151 (2014)
%  doi:10.1007/JHEP07(2014)151
%  [arXiv:1308.0589 [hep-th]].

   
%\bibitem{Partha}
%S.~Banerjee, S.~Ghosh, P.~Pandey, P.~Paul, and A.~P.~Saha,
%\textit{Unpublished work}.


%%% Celestial OPE %%%%

% \bibitem{Weinberg:1965nx} 
%  S.~Weinberg,
%  ``Infrared photons and gravitons,''
%  Phys.\ Rev.\  {\bf 140}, B516 (1965).
%  doi:10.1103/PhysRev.140.B516
  
   \bibitem{He:2014cra} 
  T.~He, P.~Mitra, A.~P.~Porfyriadis and A.~Strominger,
  ``New Symmetries of Massless QED,''
  JHEP {\bf 1410}, 112 (2014)
 doi:10.1007/JHEP10(2014)112
  [arXiv:1407.3789 [hep-th]]. 
  
    \bibitem{Kapec:2014zla} 
  D.~Kapec, V.~Lysov and A.~Strominger,
  ``Asymptotic Symmetries of Massless QED in Even Dimensions,''
  Adv.\ Theor.\ Math.\ Phys.\  {\bf 21}, 1747 (2017)
  doi:10.4310/ATMP.2017.v21.n7.a6
  [arXiv:1412.2763 [hep-th]].
  
  \bibitem{Kapec:2015ena} 
  D.~Kapec, M.~Pate and A.~Strominger,
  ``New Symmetries of QED,''
  arXiv:1506.02906 [hep-th].
    
   \bibitem{Campiglia:2015qka} 
  M.~Campiglia and A.~Laddha,
 ``Asymptotic symmetries of QED and Weinberg's soft photon theorem,''
  JHEP {\bf 1507}, 115 (2015)
  doi:10.1007/JHEP07(2015)115
  [arXiv:1505.05346 [hep-th]].
  
  \bibitem{Nande:2017dba}
A.~Nande, M.~Pate and A.~Strominger,
``Soft Factorization in QED from 2D Kac-Moody Symmetry,''
JHEP \textbf{02}, 079 (2018)
doi:10.1007/JHEP02(2018)079
[arXiv:1705.00608 [hep-th]].

 \bibitem{He:2020ifr}
T.~He and P.~Mitra,
``Covariant Phase Space and Soft Factorization in Non-abelian Gauge Theories,''
[arXiv:2009.14334 [hep-th]].

 \bibitem{Lysov:2014csa}
V.~Lysov, S.~Pasterski and A.~Strominger,
``Low\textquoteright{}s Subleading Soft Theorem as a Symmetry of QED,''
Phys. Rev. Lett. \textbf{113}, no.11, 111601 (2014)
doi:10.1103/PhysRevLett.113.111601
[arXiv:1407.3814 [hep-th]].

 \bibitem{Campiglia:2016hvg}
M.~Campiglia and A.~Laddha,
``Subleading soft photons and large gauge transformations,''
JHEP \textbf{11}, 012 (2016)
doi:10.1007/JHEP11(2016)012
[arXiv:1605.09677 [hep-th]].

 \bibitem{Conde:2016csj}
E.~Conde and P.~Mao,
``Remarks on asymptotic symmetries and the subleading soft photon theorem,''
Phys. Rev. D \textbf{95}, no.2, 021701 (2017)
doi:10.1103/PhysRevD.95.021701
[arXiv:1605.09731 [hep-th]].
  
   \bibitem{Himwich:2019dug}
E.~Himwich and A.~Strominger,
``Celestial current algebra from Low\textquoteright{}s subleading soft theorem,''
Phys. Rev. D \textbf{100}, no.6, 065001 (2019)
doi:10.1103/PhysRevD.100.065001
[arXiv:1901.01622 [hep-th]].
  
  %  \bibitem{Campiglia:2019wxe} 
%  M.~Campiglia and A.~Laddha,
%  ``Loop Corrected Soft Photon Theorem as a Ward Identity,''
%  arXiv:1903.09133 [hep-th].

%\cite{He:2020ifr}

%\cite{McLoughlin:2016uwa}
\bibitem{Fan:2020xjj}
W.~Fan, A.~Fotopoulos, S.~Stieberger and T.~R.~Taylor,
``On Sugawara construction on Celestial Sphere,''
JHEP \textbf{09}, 139 (2020)
doi:10.1007/JHEP09(2020)139
[arXiv:2005.10666 [hep-th]].

\bibitem{McLoughlin:2016uwa}
T.~McLoughlin and D.~Nandan,
``Multi-Soft gluon limits and extended current algebras at null-infinity,''
JHEP \textbf{08}, 124 (2017)
doi:10.1007/JHEP08(2017)124
[arXiv:1610.03841 [hep-th]].



%%%%%%%%%%%%%%%%%%%%%%%%%%%%%%%%%%%%%%%%%%%%%%%%%%%%%%%%%%
  
  %%% Soft theorems gravity %%%
  
   
  %%%%%%%%%%%%%%%%%%%%%%%%%%%%%%%%%%%%%%%%%%%%%%%%%%%%%%%%%%
 %%% Celestial Amplitudes %%%%
  
   
%%%%%%%%%%%%%%%%%%%%%%%%%%%%%%%%%%%%%%%%%%%%%%%%%%%%%%%%%%
  %%% NEW %%%
  
  
%\cite{Fotopoulos:2020bqj}
%\bibitem{Fotopoulos:2020bqj}
%A.~Fotopoulos, S.~Stieberger, T.~R.~Taylor and B.~Zhu,
%``Extended Super BMS Algebra of Celestial CFT,''
%JHEP \textbf{09}, 198 (2020)
%doi:10.1007/JHEP09(2020)198
%[arXiv:2007.03785 [hep-th]].

%\cite{Albayrak:2020saa}

  
 

%%%%%%%%%%%%%%%%%%%%%%%%%%%%%%%%%%%%%%%%%%%%%%%%%%%%%%%%%%
  
  %%% BMS, Asymptotic Symmetries %%%
  
  
%%%%%%%%%%%%%%%%%%%%%%%%%%%%%%%%%%%%%%%%%%%%%%%%%%%%%%%%%%

   %%% Subleading collinear limit %%%
  
%  \bibitem{Nandan:2016ohb} 
%  D.~Nandan, J.~Plefka and W.~Wormsbecher,
 % ``Collinear limits beyond the leading order from the scattering equations,''
%  JHEP {\bf 1702}, 038 (2017)
%  doi:10.1007/JHEP02(2017)038
%  [arXiv:1608.04730 [hep-th]].
      
%%%%%%%%%%%%%%%%%%%%%%%%%%%%%%%%%%%%%%%%%%%%%%%%%%%%%%%%%%
      
%%% KZ equation %%%  

%\cite{Knizhnik:1984nr}
\bibitem{Knizhnik:1984nr}
V.~G.~Knizhnik and A.~B.~Zamolodchikov,
``Current Algebra and Wess-Zumino Model in Two-Dimensions,''
Nucl. Phys. B \textbf{247}, 83-103 (1984)
doi:10.1016/0550-3213(84)90374-2

%\cite{Ginsparg:1988ui}
\bibitem{Ginsparg:1988ui}
P.~H.~Ginsparg,
``APPLIED CONFORMAL FIELD THEORY,''
[arXiv:hep-th/9108028 [hep-th]].
  
  %%%%%%%%%%%%%%%%%%%%%%%%%%%%%%%%%%%%%%%%%%%%%%%%%%%%%%%%%%  

%%% Others %%%

%\cite{Bern:1998sv}
\bibitem{Bern:1998sv}
Z.~Bern, L.~J.~Dixon, M.~Perelstein and J.~S.~Rozowsky,
``Multileg one loop gravity amplitudes from gauge theory,''
Nucl. Phys. B \textbf{546}, 423-479 (1999)
doi:10.1016/S0550-3213(99)00029-2
[arXiv:hep-th/9811140 [hep-th]].

\bibitem{DelDuca:1999rs}
V.~Del Duca, L.~J.~Dixon and F.~Maltoni,
``New color decompositions for gauge amplitudes at tree and loop level,''
Nucl. Phys. B \textbf{571}, 51-70 (2000),
doi:10.1016/S0550-3213(99)00809-3,
[arXiv:hep-ph/9910563 [hep-ph]].

%\cite{Bern:2008qj}
\bibitem{Bern:2008qj}
Z.~Bern, J.~J.~M.~Carrasco and H.~Johansson,
``New Relations for Gauge-Theory Amplitudes,''
Phys. Rev. D \textbf{78}, 085011 (2008)
doi:10.1103/PhysRevD.78.085011
[arXiv:0805.3993 [hep-ph]].

%\cite{Parke:1986gb}
\bibitem{Parke:1986gb}
S.~J.~Parke and T.~R.~Taylor,
``An Amplitude for $n$ Gluon Scattering,''
Phys. Rev. Lett. \textbf{56}, 2459 (1986)
doi:10.1103/PhysRevLett.56.2459

%******************************************Twistor*****************************

%\cite{Witten:2003nn}
\bibitem{Witten:2003nn}
E.~Witten,
``Perturbative gauge theory as a string theory in twistor space,''
Commun. Math. Phys. \textbf{252}, 189-258 (2004)
doi:10.1007/s00220-004-1187-3
[arXiv:hep-th/0312171 [hep-th]].

\bibitem{Nair:1988bq}
V.~P.~Nair,
``A Current Algebra for Some Gauge Theory Amplitudes,''
Phys. Lett. B \textbf{214}, 215-218 (1988)
doi:10.1016/0370-2693(88)91471-2

%\cite{Cachazo:2004kj}
\bibitem{Cachazo:2004kj}
F.~Cachazo, P.~Svrcek and E.~Witten,
``MHV vertices and tree amplitudes in gauge theory,''
JHEP \textbf{09}, 006 (2004)
doi:10.1088/1126-6708/2004/09/006
[arXiv:hep-th/0403047 [hep-th]].

%\cite{Cachazo:2004zb}
\bibitem{Cachazo:2004zb}
F.~Cachazo, P.~Svrcek and E.~Witten,
``Twistor space structure of one-loop amplitudes in gauge theory,''
JHEP \textbf{10}, 074 (2004)
doi:10.1088/1126-6708/2004/10/074
[arXiv:hep-th/0406177 [hep-th]].

%\cite{Cachazo:2004by}
\bibitem{Cachazo:2004by}
F.~Cachazo, P.~Svrcek and E.~Witten,
``Gauge theory amplitudes in twistor space and holomorphic anomaly,''
JHEP \textbf{10}, 077 (2004)
doi:10.1088/1126-6708/2004/10/077
[arXiv:hep-th/0409245 [hep-th]].

%\cite{Roiban:2004yf}
\bibitem{Roiban:2004yf}
R.~Roiban, M.~Spradlin and A.~Volovich,
``On the tree level S matrix of Yang-Mills theory,''
Phys. Rev. D \textbf{70}, 026009 (2004)
doi:10.1103/PhysRevD.70.026009
[arXiv:hep-th/0403190 [hep-th]].

%\cite{Cachazo:2013hca}
\bibitem{Cachazo:2013hca}
F.~Cachazo, S.~He and E.~Y.~Yuan,
``Scattering of Massless Particles in Arbitrary Dimensions,''
Phys. Rev. Lett. \textbf{113}, no.17, 171601 (2014)
doi:10.1103/PhysRevLett.113.171601
[arXiv:1307.2199 [hep-th]].

%\cite{Cachazo:2013iea}
\bibitem{Cachazo:2013iea}
F.~Cachazo, S.~He and E.~Y.~Yuan,
``Scattering of Massless Particles: Scalars, Gluons and Gravitons,''
JHEP \textbf{07}, 033 (2014)
doi:10.1007/JHEP07(2014)033
[arXiv:1309.0885 [hep-th]].

%\cite{Adamo:2014yya}
\bibitem{Adamo:2014yya}
T.~Adamo, E.~Casali and D.~Skinner,
``Perturbative gravity at null infinity,''
Class. Quant. Grav. \textbf{31}, no.22, 225008 (2014)
doi:10.1088/0264-9381/31/22/225008
[arXiv:1405.5122 [hep-th]].

%\cite{Geyer:2014lca}
\bibitem{Geyer:2014lca}
Y.~Geyer, A.~E.~Lipstein and L.~Mason,
``Ambitwistor strings at null infinity and (subleading) soft limits,''
Class. Quant. Grav. \textbf{32}, no.5, 055003 (2015)
doi:10.1088/0264-9381/32/5/055003
[arXiv:1406.1462 [hep-th]].

%\cite{Adamo:2015fwa}
\bibitem{Adamo:2015fwa}
T.~Adamo and E.~Casali,
``Perturbative gauge theory at null infinity,''
Phys. Rev. D \textbf{91}, no.12, 125022 (2015)
doi:10.1103/PhysRevD.91.125022
[arXiv:1504.02304 [hep-th]].



 % \bibitem{Hodges:2011wm}
% A.~Hodges,
%``New expressions for gravitational scattering amplitudes,''
%JHEP \textbf{07}, 075 (2013)
%%
%[arXiv:1108.2227 [hep-th]].

%\bibitem{Hodges:2012ym}
%A.~Hodges,
%``A simple formula for gravitational MHV amplitudes,''
%[arXiv:1204.1930 [hep-th]].

% \bibitem{Partha2} 
%  Partha Paul (unpublished work)
     
 %  \bibitem{Bern:1998sv} 
  %Z.~Bern, L.~J.~Dixon, M.~Perelstein and J.~S.~Rozowsky,
  %``Multileg one loop gravity amplitudes from gauge theory,''
  %Nucl.\ Phys.\ B {\bf 546}, 423 (1999)
  %doi:10.1016/S0550-3213(99)00029-2
  %[hep-th/9811140].
  
%  \bibitem{White:2011yy} 
%  C.~D.~White,
%  ``Factorization Properties of Soft Graviton Amplitudes,''
%  JHEP {\bf 1105}, 060 (2011)
%  doi:10.1007/JHEP05(2011)060
%  [arXiv:1103.2981 [hep-th]]. 
  
%  \bibitem{Akhoury:2011kq} 
%  R.~Akhoury, R.~Saotome and G.~Sterman,
%  ``Collinear and Soft Divergences in Perturbative Quantum Gravity,''
%  Phys.\ Rev.\ D {\bf 84}, 104040 (2011)
%  doi:10.1103/PhysRevD.84.104040
%  [arXiv:1109.0270 [hep-th]].


%  \bibitem{Kapec:2015vwa} 
%  D.~Kapec, V.~Lysov, S.~Pasterski and A.~Strominger,
%  ``Higher-Dimensional Supertranslations and Weinberg's Soft Graviton Theorem,''
%  Annals of Mathematical Sciences and Applications, Volume 2 (2017),
%  pp 69-94
%  doi:10.4310/AMSA.2017.v2.n1.a2
%  [arXiv:1502.07644 [gr-qc]].
  
%  \bibitem{Pate:2017fgt} 
%  M.~Pate, A.~M.~Raclariu and A.~Strominger,
%  ``Gravitational Memory in Higher Dimensions,''
%  JHEP {\bf 1806}, 138 (2018)
%  doi:10.1007/JHEP06(2018)138
%  [arXiv:1712.01204 [hep-th]].

%\bibitem{Partha}

%S.~Banerjee, S.~Ghosh, P.~Pandey, P.~Paul, and A.~P.~Saha,
%\textit{Unpublished work}.


   
  
%  \bibitem{Kapec:2016jld}

%   \bibitem{Bondi:1962px} 
%  H.~Bondi, M.~G.~J.~van der Burg and A.~W.~K.~Metzner,
%  ``Gravitational waves in general relativity. 7. Waves from axisymmetric isolated systems,''
%  Proc.\ Roy.\ Soc.\ Lond.\ A {\bf 269}, 21 (1962).
%  doi:10.1098/rspa.1962.0161
  
%\bibitem{Sachs:1962wk} 
%  R.~K.~Sachs,
%  ``Gravitational waves in general relativity. 8. Waves in asymptotically flat space-times,''
%  Proc.\ Roy.\ Soc.\ Lond.\ A {\bf 270}, 103 (1962).
%  doi:10.1098/rspa.1962.0206

%\bibitem{Sachs:1962zza} 
%  R.~Sachs,
%  ``Asymptotic symmetries in gravitational theory,''
%  Phys.\ Rev.\  {\bf 128}, 2851 (1962).
%  doi:10.1103/PhysRev.128.2851 
    
 %   \bibitem{Barnich:2009se} 
%  G.~Barnich and C.~Troessaert,
%  ``Symmetries of asymptotically flat 4 dimensional spacetimes at null infinity revisited,''
% Phys.\ Rev.\ Lett.\  {\bf 105}, 111103 (2010)
%  doi:10.1103/PhysRevLett.105.111103
%  [arXiv:0909.2617 [gr-qc]]. 
    

%  \bibitem{Banerjee:2019aoy} 
%  S.~Banerjee, P.~Pandey and P.~Paul,
%  ``Conformal properties of soft-operators - 1 : Use of null-states,''
%  arXiv:1902.02309 [hep-th].
  
% \bibitem{Weinberg:1964ew} 
%  S.~Weinberg,
%  ``Photons and Gravitons in s Matrix Theory: Derivation of Charge Conservation and Equality of Gravitational and Inertial Mass,''
%  Phys.\ Rev.\  {\bf 135}, B1049 (1964).
%  doi:10.1103/PhysRev.135.B1049
  
     
%  \bibitem{Sahoo:2018lxl} 
%  B.~Sahoo and A.~Sen,
 % ``Classical and Quantum Results on Logarithmic Terms in the Soft Theorem in Four Dimensions,''
%  JHEP {\bf 1902}, 086 (2019)
%  doi:10.1007/JHEP02(2019)086
%  [arXiv:1808.03288 [hep-th]].
  
    

    
%  \bibitem{Penedones:2015aga} 
%  J.~Penedones, E.~Trevisani and M.~Yamazaki,
%  ``Recursion Relations for Conformal Blocks,''
%  JHEP {\bf 1609}, 070 (2016)
%  doi:10.1007/JHEP09(2016)070
%  [arXiv:1509.00428 [hep-th]].
  
      
%  \bibitem{Erdmenger:1997wy} 
 % J.~Erdmenger and H.~Osborn,
 % ``Conformally covariant differential operators: Symmetric tensor fields,''
%  Class.\ Quant.\ Grav.\  {\bf 15}, 273 (1998)
%  doi:10.1088/0264-9381/15/2/003
%  [gr-qc/9708040].
  
%  \bibitem{Dolan:2001ih} 
%  L.~Dolan, C.~R.~Nappi and E.~Witten,
 % ``Conformal operators for partially massless states,''
%  JHEP {\bf 0110}, 016 (2001)
%  doi:10.1088/1126-6708/2001/10/016
%  [hep-th/0109096].
    
%  \bibitem{Beccaria:2015vaa} 
%  M.~Beccaria and A.~A.~Tseytlin,
 % ``On higher spin partition functions,''
%  J.\ Phys.\ A {\bf 48}, no. 27, 275401 (2015)
%  doi:10.1088/1751-8113/48/27/275401
%  [arXiv:1503.08143 [hep-th]]
  
%  \bibitem{Barnich:2011ct} 
 % G.~Barnich and C.~Troessaert,
%  ``Supertranslations call for superrotations,''
%  PoS CNCFG {\bf 2010}, 010 (2010)
%  [Ann.\ U.\ Craiova Phys.\  {\bf 21}, S11 (2011)]
%  [arXiv:1102.4632 [gr-qc]]
  
%  \bibitem{Giveon:1998ns} 
%  A.~Giveon, D.~Kutasov and N.~Seiberg,
%  ``Comments on string theory on AdS(3),''
%  Adv.\ Theor.\ Math.\ Phys.\  {\bf 2}, 733 (1998)
%  doi:10.4310/ATMP.1998.v2.n4.a3
%  [hep-th/9806194].
  
% \bibitem{Kutasov:1999xu} 
%  D.~Kutasov and N.~Seiberg,
%  ``More comments on string theory on AdS(3),''
%  JHEP {\bf 9904}, 008 (1999)
%  doi:10.1088/1126-6708/1999/04/008
%  [hep-th/9903219].
  
%   \bibitem{Geyer:2014lca} 
%  Y.~Geyer, A.~E.~Lipstein and L.~Mason,
%  ``Ambitwistor strings at null infinity and (subleading) soft limits,''
%  Class.\ Quant.\ Grav.\  {\bf 32}, no. 5, 055003 (2015)
%  doi:10.1088/0264-9381/32/5/055003
 % [arXiv:1406.1462 [hep-th]].
  
%  \bibitem{Lipstein:2015rxa} 
 % A.~E.~Lipstein,
 % ``Soft Theorems from Conformal Field Theory,''
 % JHEP {\bf 1506}, 166 (2015)
%  doi:10.1007/JHEP06(2015)166
%  [arXiv:1504.01364 [hep-th]].
        
%  \bibitem{Kapec:2017gsg} 
%  D.~Kapec and P.~Mitra,
%  ``A $d$-Dimensional Stress Tensor for Mink$_{d+2}$ Gravity,''
%  arXiv:1711.04371 [hep-th]. 
             
  
%  \bibitem{Fotopoulos:2019tpe} 
%  A.~Fotopoulos and T.~R.~Taylor,
 % ``Primary Fields in Celestial CFT,''
%  arXiv:1906.10149 [hep-th].
    
  
  %\bibitem{Fan:2019emx} 
  %W.~Fan, A.~Fotopoulos and T.~R.~Taylor,
  %``Soft Limits of Yang-Mills Amplitudes and Conformal Correlators,''
  %JHEP {\bf 1905}, 121 (2019)
  %doi:10.1007/JHEP05(2019)121
  %[arXiv:1903.01676 [hep-th]].
  
   
%   \bibitem{Ball:2019atb} 
%  A.~Ball, E.~Himwich, S.~A.~Narayanan, S.~Pasterski and A.~Strominger,
%  ``Uplifting AdS3/CFT2 to Flat Space Holography,''
%  arXiv:1905.09809 [hep-th].
  
  
  
%\bibitem{osborn}
%  Hugh Osborn, 
% "Lectures on conformal field theories in more than two dimensions",  
%[http://www.damtp.cam.ac.uk/user/ho/CFTNotes.pdf]  
      
% \bibitem{Peeters:2007wn} 
%  K.~Peeters,
%  ``Introducing Cadabra: A Symbolic computer algebra system for field theory problems,''
%  hep-th/0701238.
  
%\bibitem{Peeters:2006kp} 
%  K.~Peeters,
 % ``A Field-theory motivated approach to symbolic computer algebra,''
 % Comput.\ Phys.\ Commun.\  {\bf 176}, 550 (2007)
%  doi:10.1016/j.cpc.2007.01.003
%  [cs/0608005 [cs.SC]].
    
          
 % \bibitem{Bondi:1962px} 
  %H.~Bondi, M.~G.~J.~van der Burg and A.~W.~K.~Metzner,
  %``Gravitational waves in general relativity. 7. Waves from axisymmetric isolated systems,''
  %Proc.\ Roy.\ Soc.\ Lond.\ A {\bf 269}, 21 (1962).
 % doi:10.1098/rspa.1962.0161
  
%\bibitem{Sachs:1962wk} 
 % R.~K.~Sachs,
%  ``Gravitational waves in general relativity. 8. Waves in asymptotically flat space-times,''
%  Proc.\ Roy.\ Soc.\ Lond.\ A {\bf 270}, 103 (1962).
%  doi:10.1098/rspa.1962.0206
        
%   \bibitem{Barnich:2009se} 
%  G.~Barnich and C.~Troessaert,
%  ``Symmetries of asymptotically flat 4 dimensional spacetimes at null infinity revisited,''
%  Phys.\ Rev.\ Lett.\  {\bf 105}, 111103 (2010)
%  doi:10.1103/PhysRevLett.105.111103
%  [arXiv:0909.2617 [gr-qc]].  
  
% \bibitem{Barnich:2011ct} 
%  G.~Barnich and C.~Troessaert,
%  ``Supertranslations call for superrotations,''
 % PoS CNCFG {\bf 2010}, 010 (2010)
%  [Ann.\ U.\ Craiova Phys.\  {\bf 21}, S11 (2011)]
%  [arXiv:1102.4632 [gr-qc]]. 
  
%\bibitem{Barnich:2011mi} 
%  G.~Barnich and C.~Troessaert,
%  ``BMS charge algebra,''
%  JHEP {\bf 1112}, 105 (2011)
%  doi:10.1007/JHEP12(2011)105
%  [arXiv:1106.0213 [hep-th]].  
  
% \bibitem{Barnich:2013axa} 
%  G.~Barnich and C.~Troessaert,
%  ``Comments on holographic current algebras and asymptotically flat four dimensional spacetimes at null infinity,''
%  JHEP {\bf 1311}, 003 (2013)
%  doi:10.1007/JHEP11(2013)003
%  [arXiv:1309.0794 [hep-th]]. 
  
%  \bibitem{Banks:2003vp} 
%  T.~Banks,
%  ``A Critique of pure string theory: Heterodox opinions of diverse dimensions,''
%  hep-th/0306074.
    
% 
%   \bibitem{bargman} 
%  V. Bargmann,
%  "Irreducible Unitary Representations of the Lorentz Group",
%Annals of Mathematics,
%Second Series, Vol. 48, No. 3 (Jul., 1947), pp. 568-640

% \bibitem{Gadde:2017sjg} 
%  A.~Gadde,
% ``In search of conformal theories,''
 % arXiv:1702.07362 [hep-th].  
  
%  \bibitem{Hogervorst:2017sfd} 
%  M.~Hogervorst and B.~C.~van Rees,
%  ``Crossing symmetry in alpha space,''
%  JHEP {\bf 1711}, 193 (2017)
%  doi:10.1007/JHEP11(2017)193
%  [arXiv:1702.08471 [hep-th]].
    
%  \bibitem{Simmons-Duffin:2017nub} 
%  D.~Simmons-Duffin, D.~Stanford and E.~Witten,
%  ``A spacetime derivation of the Lorentzian OPE inversion formula,''
%  arXiv:1711.03816 [hep-th].  
  
%  \bibitem{Ashtekar:1978zz} 
%  A.~Ashtekar and R.~O.~Hansen,
%  ``A unified treatment of null and spatial infinity in general relativity. I - Universal structure, asymptotic symmetries, and conserved quantities at spatial infinity,''
%  J.\ Math.\ Phys.\  {\bf 19}, 1542 (1978).
%  doi:10.1063/1.523863 
  
%  \bibitem{Ashtekar:1981sf} 
%  A.~Ashtekar,
%  ``Asymptotic Quantization of the Gravitational Field,''
%  Phys.\ Rev.\ Lett.\  {\bf 46}, 573 (1981).
%  doi:10.1103/PhysRevLett.46.573
  
%  \bibitem{Ashtekar:1981bq} 
%  A.~Ashtekar and M.~Streubel,
%  ``Symplectic Geometry of Radiative Modes and Conserved Quantities at Null Infinity,''
%  Proc.\ Roy.\ Soc.\ Lond.\ A {\bf 376}, 585 (1981).
%  doi:10.1098/rspa.1981.0109
  
%  \bibitem{Ashtekar:1987tt} 
%  A.~Ashtekar,
%  ``Asymptotic Quantization: Based On 1984 Naples Lectures,''
%  NAPLES, ITALY: BIBLIOPOLIS (1987) 107 P. (MONOGRAPHS AND TEXTBOOKS IN PHYSICAL SCIENCE, 2) 
  
%  \bibitem{Kapec:2016jld} 
%  D.~Kapec, P.~Mitra, A.~M.~Raclariu and A.~Strominger,
%  ``2D Stress Tensor for 4D Gravity,''
%  Phys.\ Rev.\ Lett.\  {\bf 119}, no. 12, 121601 (2017)
%  doi:10.1103/PhysRevLett.119.121601
%  [arXiv:1609.00282 [hep-th]].
     
% \bibitem{Kapec:2014opa} 
%  D.~Kapec, V.~Lysov, S.~Pasterski and A.~Strominger,
%  ``Semiclassical Virasoro symmetry of the quantum gravity $ \mathcal{S}$-matrix,''
%  JHEP {\bf 1408}, 058 (2014)
%  doi:10.1007/JHEP08(2014)058
%  [arXiv:1406.3312 [hep-th]].
  
%  \bibitem{Pasterski:2015tva} 
%  S.~Pasterski, A.~Strominger and A.~Zhiboedov,
%  ``New Gravitational Memories,''
%  JHEP {\bf 1612}, 053 (2016)
%  doi:10.1007/JHEP12(2016)053
%  [arXiv:1502.06120 [hep-th]].
 
%                  
  
%  \bibitem{Schwab:2014xua} 
%  B.~U.~W.~Schwab and A.~Volovich,
%  ``Subleading Soft Theorem in Arbitrary Dimensions from Scattering Equations,''
 % Phys.\ Rev.\ Lett.\  {\bf 113}, no. 10, 101601 (2014)
%  doi:10.1103/PhysRevLett.113.101601
%  [arXiv:1404.7749 [hep-th]].
  
 % \bibitem{Bern:2014oka} 
 % Z.~Bern, S.~Davies and J.~Nohle,
%  ``On Loop Corrections to Subleading Soft Behavior of Gluons and Gravitons,''
%  Phys.\ Rev.\ D {\bf 90}, no. 8, 085015 (2014)
%  doi:10.1103/PhysRevD.90.085015
%  [arXiv:1405.1015 [hep-th]].
  
%  \bibitem{Broedel:2014fsa} 
%  J.~Broedel, M.~de Leeuw, J.~Plefka and M.~Rosso,
%  ``Constraining subleading soft gluon and graviton theorems,''
%  Phys.\ Rev.\ D {\bf 90}, no. 6, 065024 (2014)
%  doi:10.1103/PhysRevD.90.065024
%  [arXiv:1406.6574 [hep-th]].
  
 % \bibitem{Avery:2015gxa} 
 % S.~G.~Avery and B.~U.~W.~Schwab,
 % ``Burg-Metzner-Sachs symmetry, string theory, and soft theorems,''
 % Phys.\ Rev.\ D {\bf 93}, 026003 (2016)
%  doi:10.1103/PhysRevD.93.026003
 % [arXiv:1506.05789 [hep-th]].
  
%   \bibitem{Sen:2017xjn} 
%  A.~Sen,
%  ``Soft Theorems in Superstring Theory,''
%  JHEP {\bf 1706}, 113 (2017)
%  doi:10.1007/JHEP06(2017)113
 % [arXiv:1702.03934 [hep-th]].
  
 % \bibitem{Sen:2017nim} 
 % A.~Sen,
 % ``Subleading Soft Graviton Theorem for Loop Amplitudes,''
 % JHEP {\bf 1711}, 123 (2017)
%  doi:10.1007/JHEP11(2017)123
%  [arXiv:1703.00024 [hep-th]].
  
%  \bibitem{Laddha:2017ygw} 
%  A.~Laddha and A.~Sen,
%  ``Sub-subleading Soft Graviton Theorem in Generic Theories of Quantum Gravity,''
%  JHEP {\bf 1710}, 065 (2017)
%  doi:10.1007/JHEP10(2017)065
%  [arXiv:1706.00759 [hep-th]].
  
%  \bibitem{Chakrabarti:2017ltl} 
%  S.~Chakrabarti, S.~P.~Kashyap, B.~Sahoo, A.~Sen and M.~Verma,
%  ``Subleading Soft Theorem for Multiple Soft Gravitons,''
%  arXiv:1707.06803 [hep-th].
  
%  \bibitem{Chakrabarti:2017zmh} 
%  S.~Chakrabarti, S.~P.~Kashyap, B.~Sahoo, A.~Sen and M.~Verma,
 % ``Testing Subleading Multiple Soft Graviton Theorem for CHY Prescription,''
%  arXiv:1709.07883 [hep-th].
  
%  \bibitem{Laddha:2017vfh} 
%  A.~Laddha and P.~Mitra,
%  ``Asymptotic Symmetries and Subleading Soft Photon Theorem in Effective Field Theories,''
%  arXiv:1709.03850 [hep-th].
  
%  \bibitem{Laddha:2018rle} 
%  A.~Laddha and A.~Sen,
%  ``Gravity Waves from Soft Theorem in General Dimensions,''
%  arXiv:1801.07719 [hep-th].
  
 % \bibitem{Campoleoni:2017mbt} 
 % A.~Campoleoni, D.~Francia and C.~Heissenberg,
 % ``On higher-spin supertranslations and superrotations,''
%  JHEP {\bf 1705}, 120 (2017)
 % doi:10.1007/JHEP05(2017)120
%  [arXiv:1703.01351 [hep-th]].
  
 % \bibitem{Hawking:2016msc} 
 % S.~W.~Hawking, M.~J.~Perry and A.~Strominger,
%  ``Soft Hair on Black Holes,''
%  Phys.\ Rev.\ Lett.\  {\bf 116}, no. 23, 231301 (2016)
 % doi:10.1103/PhysRevLett.116.231301
%  [arXiv:1601.00921 [hep-th]].
  
  
%  \bibitem{Barnich:2014kra} 
%  G.~Barnich and B.~Oblak,
%  ``Notes on the BMS group in three dimensions: I. Induced representations,''
%  JHEP {\bf 1406}, 129 (2014)
%  doi:10.1007/JHEP06(2014)129
%  [arXiv:1403.5803 [hep-th]].
  
%  \bibitem{Barnich:2015uva} 
%  G.~Barnich and B.~Oblak,
 % ``Notes on the BMS group in three dimensions: II. Coadjoint representation,''
%  JHEP {\bf 1503}, 033 (2015)
%  doi:10.1007/JHEP03(2015)033
%  [arXiv:1502.00010 [hep-th]].
  
% \bibitem{Campoleoni:2016vsh} 
%  A.~Campoleoni, H.~A.~Gonzalez, B.~Oblak and M.~Riegler,
 % ``BMS Modules in Three Dimensions,''
%  Int.\ J.\ Mod.\ Phys.\ A {\bf 31}, no. 12, 1650068 (2016)
 % doi:10.1142/S0217751X16500688, 10.1142/9789813144101_0011
%  [arXiv:1603.03812 [hep-th]].
  
 %\bibitem{Penrose:1967wn} 
%  R.~Penrose,
%  ``Twistor algebra,''
%  J.\ Math.\ Phys.\  {\bf 8}, 345 (1967).
 % doi:10.1063/1.1705200  
  
% \bibitem{Weinberg:1995mt} 
%%  S.~Weinberg,
%  ``The Quantum theory of fields. Vol. 1: Foundations,''  



\end{thebibliography}
\end{document}